\documentclass[twocolumn]{aastex631}

\usepackage{graphicx}
\usepackage{amsmath}
\usepackage{float}

\begin{document}

\title{
    Learning and Predicting the Nonlinear Variability of X-ray Binaries with the Koopman Operator
}

\author[0009-0009-6455-3804]{Eric Miao}\affiliation{Columbia Astrophysics Laboratory, Columbia University, New York, NY 10027, USA}

\author[0000-0002-9856-989X]{Ruo-Yu Shang}\affiliation{Department of Physics and Astronomy, Barnard College, Columbia University, NY 10027, USA}

\author[0000-0002-9709-5389]{Kaya Mori}
\affiliation{Columbia Astrophysics Laboratory, Columbia University, New York, NY 10027, USA}

\author[0000-0002-3223-0754]{Reshmi Mukherjee}
\affiliation{Department of Physics and Astronomy, Barnard College, Columbia University, NY 10027, USA}

\correspondingauthor{Eric Miao}
\email{em3928@columbia.edu}

\begin{abstract}

    X-ray variability in compact-object binaries encodes the nonlinear dynamics of corona-jet interactions and accretion disk instabilities. Standard timing techniques characterize periodic and quasi-periodic variability well, but do not model underlying nonlinear dynamics or forecast their evolution. We apply Koopman operator theory and a data-driven approximation, extended dynamic mode decomposition (EDMD), to X-ray light curves for the first time. Koopman theory represents nonlinear evolution as an infinite-dimensional linear operator $\mathcal{K}$, whose eigendecomposition separates a complex system into independently evolving linear modes. We derive that each Koopman eigenfunction contributes a Lorentzian peak to the power spectrum, giving quasi-periodic oscillations a dynamical interpretation in which process noise damps modes and broadens their peaks. In both chaotic Duffing oscillator simulations and $\sim$30 yr of RXTE ASM and MAXI monitoring of the X-ray binary 4U 1705-44, the slowest-varying eigenfunction partitions state space into low- and high-flux regimes, changing sign days to weeks before transitions become visible in the light curve. Iterating $\mathcal{K}$ additionally yields flux forecasts on days-to-weeks horizons.
The strengths of this framework are its generality across linear and nonlinear systems, its intrinsic interpretability through decomposed modes carrying explicit dynamical meaning, and its predictive power from propagating learned dynamics forward. These results establish Koopman operator theory as a new frontier of astrophysical timing and help advance interpretable machine learning for scientific discovery and understanding.

\end{abstract}


\section{Introduction}
\label{sec:intro}

X-ray binaries, such as Cyg X-3, MAXI J1820+070, and GRS 1915+105, are stellar systems where a compact object (a black hole or a neutron star) accretes matter from a companion star, typically forming an accretion disk \citep{lewin1997x,done2007modelling}. 
These objects exhibit varying emissions through various mechanisms, including accreting gas from the companion and releasing gravitational energy in X-rays, up-scattering disk photons in corona regions to produce hard X-rays, and colliding relativistic jets with ambient materials to produce radio, IR/optical, X-ray, and gamma-ray emissions \citep{shakura1973black,frank2002accretion,mirabel1999sources,fender2001powerful,fender2004towards}.
Since the central compact objects in the binary systems are only a few solar masses, dynamical processes unfold on humanly accessible timescales, making them ideal laboratories for accretion and jet physics.

X-ray binaries exhibit distinct states with different spectral and temporal profiles during outbursts, reflecting changes in the geometry and dynamics of the systems \citep{zdziarski2004radiative, trushkin2017giant, sobolewska2003spectral, neilsen2012accretion}.
The states of X-ray binaries are defined by the observed spectral and temporal features.
The Low/Hard State (LHS) shows spectra dominated by hard power-law emission from the corona with strong variability and accompanied by compact steady jets, while the High/Soft State (HSS) shows spectra dominated by thermal emission from the accretion disk with lower variability and jets quenched or absent \citep{zdziarski2004radiative}.
Intermediate states are the transitional phases between LHS and HSS and are characterized by quasi-periodic oscillations (QPOs) \citep{motta2016quasi, ingram2019review}.

X-ray binaries provide a scaled-down analog of AGN state changes, helping unify black hole accretion physics across mass scales.
The transition states of X-ray binaries offer crucial insight into corona–jet interactions and the physics of accretion instabilities, and multi-wavelength campaigns ($\gamma$-ray, X-ray, optical/IR, radio) during the transitions can help disentangle cause vs.\ effect since each band traces a different physical region.
For example, X-ray observations can reveal spectral and timing evolution (disk temperature, power law, QPOs), while $\gamma$-ray observations can detect transient high-energy flares during jet ejections.

However, the timescales of these states and phase transitions are often highly aperiodic, spanning from months to seconds, making it challenging to plan multi-wavelength observations with pointing instruments.
An algorithm that models these state changes would enable observers to capture the beginning and end of stable-variability states, allowing detections of early stages of elevated flares and multi-wavelength observations of spectral state changes.

In this paper, we introduce Koopman operator theory \citep{koopman1931hamiltonian}, which transforms a nonlinear dynamical system to a linear one by lifting the original state space to a high-dimensional observable space. 
This paper serves as a pilot study for a series of studies of Koopman operator theory applications in time-domain high-energy astrophysics by demonstrating its potential for interpreting and predicting nonlinear dynamic systems.

We emphasize interpretability and understanding alongside prediction, aligning our work with interpretable machine learning, a growing field that prioritizes model understanding, scientific discovery, safety, and trustworthiness \citep{rudin2019stop,allen2024interpretable}. 
In scientific applications, interpretability is highly desirable, as models only support scientific discovery when their internal quantities can be examined, audited, and related to the observable behavior of the system \citep{allen2024interpretable}. 
Koopman operator theory is intrinsically interpretable in this sense: as we will demonstrate, its learned quantities are linear modes with explicit dynamical meaning, where eigenvalues govern temporal evolution and separate the timescales of the dynamics, Koopman modes describe the physical structures, and eigenfunction values locate the system's current dynamical regime. 
Predictions then follow from propagating these learned coordinates forward in time, and are therefore highly interpretable as well.

We first explain the shortcomings of traditional timing methods and existing nonlinear analysis methods.
Then, we introduce Extended Dynamic Mode Decomposition (EDMD), a data-driven method that allows a finite-dimensional approximation of the Koopman operator and subsequent eigendecomposition.
Using the simulated Duffing oscillator as an example, we demonstrate how the eigenmodes of the Koopman operator can partition the state space of a nonlinear system in a predictive way. We show how predictions can be made with Koopman eigenfunctions in addition to the direct evolution of the present state. 
Using the Duffing oscillator, we also demonstrate how the Koopman operator's eigenvalues are related to QPOs and how measurement noise affects prediction accuracy. 
We then discuss all the caveats of data processing, including our treatment of noise, data gaps, outliers, and measurement uncertainties in telescope data. 
Finally, we compute approximations to the Koopman operators of the X-ray binary 4U 1705-44 and demonstrate the potential of Koopman operator theory to predict the future behavior of nonlinear astrophysical sources in a highly interpretable way.

\section{Review of previous methods}
\label{sec:review}

Conventional timing analysis revolves around Fourier analysis, which decomposes a light curve (time series of photon counts) into frequency components, enabling the identification and characterization of periodic, quasi-periodic, and broadband noise variability \citep{merrifield1994estimating, belloni2002unified}. 
Typically, a Fourier power spectral density (PSD) is computed from the light curve, transforming time-domain data into the frequency domain, revealing QPO signals as broadened peaks described by Lorentzian profiles \citep{pottschmidt2003long}, and linking timing phenomena to accretion physics.
For example, \citet{miller2001high} and \citet{stella1998lense,stella1999correlations,stella1999khz} demonstrated how QPO frequencies can be related to physical parameters such as disk flux, black hole mass and spin, or accretion state.
Fourier-based methods are excellent for identifying persistent periodic/quasi-periodic features, and there are robust statistical frameworks for noise characterization and significance tests \citep{van1989fourier}. 
However, since Fourier methods decompose complex signals into a linear superposition of sinusoids, predictive ability is also limited to periodic signals. 
Complex (nonlinear or even chaotic) variability, such as aperiodic flaring, switches between high and low states, or transient periodicities, is smeared in PSDs and manifests as broadband noise.

Another method of signal decomposition, Singular Spectrum Analysis (SSA), has attracted increasing interest from the astrophysics community \citep{rico2025singular,thekkeppattu2023singular,fatheddin2024singular,poore2024comparative}. 
SSA is a nonparametric, data-driven technique for decomposing astrophysical light curves into interpretable components such as trends, oscillations, and noise.
SSA operates by transforming a light curve into a lag-time embedding matrix (also known as the Hankel matrix) and then using Singular Value Decomposition (SVD) to extract components.
SSA is advantageous in handling non-stationary sources, removing stochastic noise, filling observation gaps, and removing systematic instrumental effects. 
Although SSA is useful for interpolating data between observations, it is severely limited for long-term forecasts as its additive decomposition of light curves into trend, oscillations, and noise is directly in conflict with the aperiodic, nonlinear nature of chaotic systems.
As a result, the linear approximation of SSA diverges from the nonlinear path of the chaotic system over a time scale known as the Lyapunov time.

There has been work seeking to directly address nonlinearity in observational astrophysics, with efforts centered around identifying and characterizing nonlinear and chaotic dynamics; specifically, distinguishing deterministic or chaotic processes from stochastic processes.
X-ray flux fluctuations in X-ray binaries are often characterized by a lognormal distribution \citep{uttley2005non}, where the logarithm of the X-ray flux follows a normal (Gaussian) distribution, that is, the amplitude of the fluctuations is proportional to the current mean flux.
The lognormal variability of X-ray binaries is supported by the discovery of a linear relationship between the Root Mean Square (rms) variability and the mean X-ray flux of the sources \citep{uttley2001flux}, and this suggests that the short-term variabilities of X-ray binaries are connected to their long-term variabilities, leading to nonlinearity.
The rms-flux relation can be explained by a multiplicative process in which long timescale variability (e.g., a fluctuation in the outer regions of the disk) modulates shorter timescale fluctuations as it propagates inward to the inner regions \citep{uttley2017rms}.
While the rms-flux relation could be an effective tool for determining the nonlinearity of a dynamic system, its robustness has been questioned \citep{scargle2020studies}, and it does not offer the ability to model the dynamics of the system.

Another interesting example of nonlinear astrophysical analysis is \cite{phillipson2018chaotic}, who analyzed nearly 20 years of X-ray monitoring data of the low-mass X-ray binary 4U 1705-44.
By comparing phase-space embeddings of the light curves and topological metrics, such as relative rotation rates, they showed that 4U 1705-44 and the analytical Duffing oscillator likely share the same underlying dynamical template. 
This chaotic system provides a physically motivated framework for understanding long-term X-ray variability in this and similar accreting neutron star binaries.
\citet{phillipson2018chaotic} established 4U 1705-44 as one of the most thoroughly characterized nonlinear X-ray sources; combined with a very long continuous monitoring baseline ($\sim$30 years of RXTE ASM and MAXI data), 4U 1705-44 is an ideal pilot source for validating a new nonlinear systems method. 
However, while a simple model like the Duffing oscillator provides a clear understanding, its simplicity also makes it difficult to generalize to more complex systems. 

Many have attempted to analyze astronomical time-series data using neural network algorithms.
A common approach is to derive statistical features from time series and to use neural-network-based methods to separate objects into a few classes \citep{mahabal2017deep, monsalves2024application}.
Neural network models have also been used to interpolate and predict time-series data of astrophysical sources \citep{peng2024kilonova,wei2021deep}.
Neural networks are effective in detecting faint signals from noisy measurements, for example \cite{krastev2020real} uses convolutional neural networks for real-time detection of binary neutron-star gravitational-wave signals in time series, distinguishing them from noise and black-hole signals.
Although neural network algorithms are powerful for modeling complex dynamics, their ``black-box'' nature prevents the extraction of physically meaningful dynamical states, a fundamental limitation of neural networks that hinders both interpretability and reliable extrapolation beyond training data.

Most similar to our method are recent applications of dynamic mode decomposition (DMD) in astronomy: to galactic phase-space spirals \citep{darling2019eigenfunctions}, to sunspot data \citep{albidah2021proper}, and, in its higher-order form, to light curves of RR Lyrae stars \citep{trevisan2023case,mekhael2024koopman}.
These studies use DMD to describe structure already present in the data, decomposing an observed signal into modes and summarizing them with the recovered eigenvalues.
We use the Koopman decomposition to interpret variability as well, but also as a predictive model: we advance the operator forward to forecast unobserved behavior, and we identify slow-varying eigenfunctions whose values are both physically interpretable and predictive of state changes.

\section{Koopman operator theory}
\label{sec:koopman_theory}

Linear systems, where the dynamics are linearly related to the current state, can be fully characterized by spectral decomposition. In turn, there exists a general toolkit for prediction, decomposition, and estimation using eigenvalues and eigenvectors. 

In contrast, for nonlinear systems, where superposition fails, there is currently no overarching mathematical framework for their explicit and general characterization \citep{brunton_modern_2022}. While particular nonlinear systems admit exact solutions, there is no general toolkit for the prediction, estimation, and control of arbitrary nonlinear dynamics.

Deterministic nonlinear dynamics, such as chaotic dynamics, are famously difficult to predict because of their sensitivity to initial conditions, a problem worsened by noisy measurements and by the fact that the governing equations of most systems of interest, such as accretion instability or jet flow, are unknown. However, a dynamical system framework explicitly models future state evolution from the current system state, making prediction a fundamental objective despite these challenges.

A nonlinear dynamical system can be represented as
\begin{equation}
\vec{s}_{t+1} = \vec{F}_{\Delta{t}}\left(\vec{s}_{t}\right),
\end{equation}
where $\vec{s}_{t}$ is the full state of the system (e.g., the accretion flow parameters such as density, temperature, and magnetic fields), $t = t_k$ is a discrete time with timestep $\Delta{t}$, and $\vec{F}_{\Delta{t}}$ is the nonlinear discrete time propagator that governs the future evolution of the full state. In 1931, Bernard O. Koopman \citep{koopman1931hamiltonian} showed that it is possible to represent this nonlinear system in terms of an infinite-dimensional linear operator $\mathcal{K}$ acting on a Hilbert space of observable functions of the state of the system $g\left(\vec{s}_{t}\right)$. That is,
\begin{equation}
\mathcal{K} g\left(\vec{s}_{t}\right)
= g\left(\vec{F}_{\Delta t}\left(\vec{s}_{t}\right)\right)
= g\left(\vec{s}_{t+1}\right).
\end{equation}

The so-called Koopman operator $\mathcal{K}$ is an infinite-dimensional linear operator whose spectral decomposition provides a complete description of the behavior of the nonlinear system \citep{budivsic2012applied,brunton_modern_2022}, bringing all the benefits of linearity to a nonlinear system.
However, its infinite dimensionality poses challenges for real-world problems at the engineering level. 
Applied Koopman theory approximates the evolution of observable functions $g$ in a \textit{Koopman-invariant} subspace, spanned by a set of basis functions $\{\psi_1, \psi_2, \cdots, \psi_p\}$ that remains in the subspace after the application of the Koopman operator. 
\begin{equation}
g = \alpha_1 \psi_1 + \alpha_2 \psi_2 + \cdots + \alpha_p \psi_p
\end{equation}
\begin{equation}
\mathcal{K}g = \beta_1 \psi_1 + \beta_2 \psi_2 + \cdots + \beta_p \psi_p.
\end{equation}
In this coordinate system, given by values of $\psi_j(\vec{s}_t)$, the Koopman operator is a finite matrix that globally linearizes the dynamics. Therefore, finding an invariant subspace and a finite-dimensional Koopman operator is a key objective in Koopman analysis, as it enables harnessing the power of the Koopman mode decomposition for modeling and prediction. 

\subsection{Koopman Mode Decomposition}
An obvious set of Koopman-invariant basis functions is the eigenfunctions of the Koopman operator themselves.
Recall that we work with discrete time, indexed by $t$, with a timestep $\Delta t$.
For a discrete-time Koopman operator $\mathcal{K}$, eigenfunction $\varphi(\vec{s})$, and the corresponding eigenvalue $\lambda$,
\begin{equation}
\varphi(\vec{s}_{t+1}) = \mathcal{K}\varphi(\vec{s}_t) = \lambda \varphi(\vec{s}_t).
\end{equation}
Given an observable $g_{i}(\vec{s}_t)$ of a state $\vec{s}_t$, which in our case is the light curve flux of an astrophysical system, Koopman mode decomposition seeks to decompose the nonlinear dynamics of $g$ into a linear sum of independently evolving eigenmodes. This linearity enables model reduction, model simplification, and prediction that eigendecomposition typically provides for linear systems. 

The eigenfunctions $\varphi_{j}(\vec{s})$ form the basis of the Hilbert space of observables, or the measurement functions \citep{brunton_HAVOK_2017}. So, we can write the observable $g_{i}$ as a linear superposition of eigenfunctions:
\begin{equation}
g_i(\vec{s}_t)
= \mathcal{K}^{t} g_{i}(\vec{s}_0)
= \sum_{j=1}^{\infty} \lambda_j^t \varphi_j(\vec{s}_0) v_{ij}.
\end{equation}
For multiple measurements of a system, for example, with $p$ light curves from differing energy bands, a general observable vector $\vec{g}$ can be arranged:
\begin{equation}
\vec{g}(\vec{s}) = \begin{bmatrix} g_1(\vec{s}) \\ g_2(\vec{s}) \\ \vdots \\ g_p(\vec{s}) \end{bmatrix} = \sum_{j=1}^{\infty} \varphi_j(\vec{s}) \vec{v}_j
\end{equation}
where $\vec{v}_{j}$ is the \textit{Koopman mode} associated with the eigenfunction $\varphi_{j}$. This set of triplets, $\{(\lambda_j, \varphi_j, \vec{v}_j)\}_{j=1}^{\infty}$, is known as the \textit{Koopman mode decomposition}, or \textit{KMD} \citep{mezic_kmd_2005}.
This mode decomposition greatly simplifies complex dynamics into interpretable components, but the challenge is approximating this decomposition given limited, noisy, discrete measurements of a nonlinear system. 

\subsection{Extended Dynamic Mode Decomposition (EDMD)}
\label{sec:edmd}

Extended dynamic mode decomposition, or EDMD, \citep{williams2015data, brunton_modern_2022} is one of the leading data-driven methods of approximating the Koopman operator. The goal of EDMD is to approximate the Koopman mode decomposition and therefore reduce a system, from discrete measurements, into linearly evolving modes in eigenfunction space. We first describe time-delay embedding, then the lifting procedure, and finally the computation of the Koopman matrix.

Choosing appropriate measurements of a nonlinear dynamical system is critical for modeling and predicting it \citep{brunton_HAVOK_2017}. \textit{Observability} of a dynamical system requires certain conditions such that a full system state can be reconstructed from measurements of a system. Often, only partial observations are available, and many system variables are latent. 
This is the case with light curve data, where we only see a low-dimensional projection of a highly complex astrophysical system. One powerful way to ``enrich'' this single scalar measurement is with the time-delay vector $\vec{x}_t = [x_t, x_{t-1}, x_{t-2}, \ldots, x_{t-d+1}] \in \mathbb{R}^{1 \times d}$, created from time-shifted copies of a scalar measurement $x(t)$. 
The Takens embedding theorem \citep{takens_embedding_1981} states that, under certain conditions, this enriched measurement $\vec{x}_t$ contains sufficient information about the attractor of the entire dynamical system. 
In other words, it may be possible to reconstruct the entire attractor of the dynamical system that underlies a complex accretion flow with a single time series of light curve flux $x(t)$. 
Because this ``enriched'' observable is able to summarize pertinent information, we choose to use the time-delay vector $\vec{x}_t$ as our state observable for all analyses.

Given $n$ snapshot pairs $(\vec{x}_{t}, \vec{x}_{t+1}), \cdots, (\vec{x}_{t-n+1}, \vec{x}_{t-n+2})$, we arrange the delay vectors into data matrices where each row is an observation:
\begin{equation}
\begin{gathered}
\mathbf{X}_{\mathrm{past}} = \begin{bmatrix}
\text{---} & \vec{x}_{t} & \text{---} \\
\text{---} & \vec{x}_{t-1} & \text{---} \\
& \vdots & \\
\text{---} & \vec{x}_{t-n+1} & \text{---}
\end{bmatrix} \in \mathbb{R}^{n \times d}, \\
\mathbf{X}_{\mathrm{future}} = \begin{bmatrix}
\text{---} & \vec{x}_{t+1} & \text{---} \\
\text{---} & \vec{x}_{t} & \text{---} \\
& \vdots & \\
\text{---} & \vec{x}_{t-n+2} & \text{---}
\end{bmatrix} \in \mathbb{R}^{n \times d},
\end{gathered}
\end{equation}
where $\mathbf{X}_{\mathrm{future}}$ contains the time-shifted successors of each row in $\mathbf{X}_{\mathrm{past}}$.

EDMD approximates the Koopman operator by lifting these snapshot pairs into a higher-dimensional feature space using a dictionary of basis functions. A \textit{dictionary} $\mathcal{D} = \{\psi_1, \psi_2, \ldots, \psi_p\}$ is a collection of $p$ scalar-valued functions, where each $\psi_j: \mathbb{R}^{1 \times d} \to \mathbb{R}$ maps the entire delay vector to a scalar. We can then define a vector-valued function $\vec{\psi}(\vec{x}_{t})$ by applying the dictionary to a single delay vector, which outputs the $p$ dimensional \textit{lifted state} $\vec{y}_{t}$:
\begin{equation}
\label{eq:y_t}
\vec{y}_{t}=\vec{\psi}(\vec{x}_{t}) = \begin{bmatrix}
\psi_{1}(\vec{x}_{t}), &
\psi_{2}(\vec{x}_{t}), &
\cdots, &
\psi_{p}(\vec{x}_{t})
\end{bmatrix} \in \mathbb{R}^{1 \times p}.
\end{equation}
Applying the dictionary to all $n$ snapshot pairs produces the lifted data matrices:
\begin{equation}
\begin{gathered}
\mathbf{Y}_{\mathrm{past}} = \begin{bmatrix}
\text{---} & \vec{y}_{t} & \text{---} \\
\text{---} & \vec{y}_{t-1} & \text{---} \\
& \vdots & \\
\text{---} & \vec{y}_{t-n+1} & \text{---}
\end{bmatrix} \in \mathbb{R}^{n \times p}, \\
\mathbf{Y}_{\mathrm{future}} = \begin{bmatrix}
\text{---} & \vec{y}_{t+1} & \text{---} \\
\text{---} & \vec{y}_{t} & \text{---} \\
& \vdots & \\
\text{---} & \vec{y}_{t-n+2} & \text{---}
\end{bmatrix} \in \mathbb{R}^{n \times p}.
\end{gathered}
\label{eq:observable_function_matrix}
\end{equation}
The lifted data should span a Koopman-invariant subspace, leading to linear dynamics in the lifted space:
\begin{equation}
\mathbf{Y}_{\mathrm{future}} \approx \mathbf{Y}_{\mathrm{past}} \mathbf{K},
\label{eq:edmd}
\end{equation}
where $\mathbf{K} \in \mathbb{R}^{p \times p}$ is the finite-dimensional Koopman matrix. Note that we use bold letters ($\mathbf{K}$) to notate matrices, and calligraphic letters ($\mathcal{K}$) to notate operators.
The finite matrix $\mathbf{K}$ is the approximation of $\mathcal{K}$ in the basis of dictionary functions. This formulation assumes that the time series is stationary, so that a single time-independent matrix $\mathbf{K}$ describes the evolution over the entire training window.

We then solve for $\mathbf{K}$ using the least squares solution:
\begin{equation}
\mathbf{K} = \mathbf{Y}_{\mathrm{past}}^{\dagger} \mathbf{Y}_{\mathrm{future}},
\label{eq:koopman_operator_solution}
\end{equation}
where $\mathbf{Y}_{\mathrm{past}}^{\dagger}$ signifies the pseudoinverse of $\mathbf{Y}_{\mathrm{past}}$. 

Under the infinite sampling limit, the matrix approximation $\mathbf{K}$ converges to the Koopman operator $\mathcal{K}$ projected onto the subspace spanned by $\mathcal{D}$ \citep{korda_mezic_convergence_2018}. However, this is true under one crucial condition: this set of user-defined basis functions must span a Koopman-invariant subspace; otherwise $\mathbf{K}$ will have no resemblance to $\mathcal{K}$ and spectral decomposition will result in spurious eigenvalues and eigenvectors \citep{brunton_modern_2022}. Unfortunately, there is no guarantee that a certain basis choice will be Koopman-invariant; in turn, the choice of dictionary requires careful selection. Common choices in the dictionary include polynomials, Fourier series, and radial basis functions.

As our primary choice in this paper, we use a basis dictionary of time-delay vectors and Fourier functions as our lifting functions $\vec{\psi}$, transforming our ground state of time-delay vectors. As an example, for $m$ frequency harmonics and including both sines and cosines,
\begin{equation}
\resizebox{0.88\linewidth}{!}{$\displaystyle
\vec{y}_{t} = \begin{bmatrix}
x_{t}, & x_{t-1}, & \cdots, & x_{t-d+2}, & x_{t-d+1}, \\
\sin(x_{t}), & \cos(x_{t}), & \cdots, & \sin(x_{t-d+1}), & \cos(x_{t-d+1}), \\
\sin(2x_{t}), & \cos(2x_{t}), & \cdots, & \sin(2x_{t-d+1}), & \cos(2x_{t-d+1}), \\
& & \vdots & & \\
\sin(mx_{t}), & \cos(mx_{t}), & \cdots, & \sin(mx_{t-d+1}), & \cos(mx_{t-d+1})
\end{bmatrix}
$}
\label{eq:fourier_dict}
\end{equation}
where $d$ is the delay embedding dimension and $m$ is the number of frequencies, giving $p = d+2md$ basis functions $\psi_j$. The Fourier dictionary is a natural choice for systems with known or suspected periodic/quasiperiodic dynamics. 
Although we have seen success with the combination of a time-delay and Fourier basis, we emphasize again that this dictionary is a classic, heuristic choice among many possibilities. Dictionary selection can be carried out with time-series cross-validation: the training light curve is split into contiguous training and validation segments, $\mathbf{K}$ is computed on the training segment for each candidate dictionary, and the dictionary that minimizes the multi-step prediction error on the held-out validation segment is selected \citep{bergmeir2012use}. More modern approaches include learning the dictionary itself from data with neural networks \citep{li2017extended,lusch2018deep}, or searching for Koopman-invariant subspaces algebraically with invariance and accuracy guarantees \citep{haseli_symmetric_subspace_2022,haseli_recursive_2025}. Overall, dictionary selection continues to be an active area of research. 

While this is the first application of EDMD to X-ray astrophysics, it is a well-established, yet actively evolving, data-driven technique for modeling complex nonlinear dynamical systems across a wide variety of scientific and engineering disciplines.
In fluid dynamics, EDMD is widely used for turbulent flow analysis to decompose turbulent flows into simpler, more understandable components \citep{colbrook2023residual}.
In the energy system sector, EDMD is used to analyze the stability of power grids by identifying dominant oscillatory modes from data and to provide early warnings of impending instabilities, allowing proactive control actions to prevent blackouts \citep{susuki2016applied}.
In neuroscience, EDMD is used to analyze neuronal data to identify and track neural oscillations and brain states, and it can help understand brain function and diagnose neurological disorders by extracting meaningful dynamic modes from noisy, high-dimensional brain signals \citep{gallos2024data}.
Our focus in this paper is the predictive ability of the Koopman operator's spectral decomposition, as learned by EDMD. But, as shown in countless other disciplines, Koopman spectral theory and EDMD also provide estimation, uncertainty quantification, control, and new frameworks for understanding highly complex dynamical systems \citep{brunton_modern_2022}.

\subsection{EDMD spectral analysis and interpretation}
\label{sec:spectral_analysis}

Similarly to how the eigendecomposition of a linear system completely describes the dynamics of that system, the eigendecomposition of the finite Koopman matrix $\mathbf{K}$ characterizes the dynamics of a deterministic nonlinear system captured in the span of the dictionary \citep{brunton_modern_2022}. Given a $\mathbf{K}$ calculated by EDMD, we compute the eigen-decomposition $\mathbf{K}=U_{R} \Lambda U_{L}^{\top}$, where

\begin{equation}
\Lambda = \mathrm{diag}(\lambda_{1}, \lambda_{2}, \ldots, \lambda_{p}),
\end{equation}

\begin{equation}
U_{L} = 
\begin{bmatrix}
| & | &  & | &\\
\vec{l}_{1} & \vec{l}_{2} & \cdots & \vec{l}_{p} \\
| & | &  & | &\\
\end{bmatrix}
\end{equation}
and 
\begin{equation}
U_{R} = 
\begin{bmatrix}
| & | &  & | &\\
\vec{r}_{1} & \vec{r}_{2} & \cdots & \vec{r}_{p} \\
| & | &  & | &\\
\end{bmatrix}
\end{equation}
are the eigenvalues and the left and right eigenvectors of $\mathbf{K}$.

Using this eigendecomposition, we can then approximate the Koopman mode decomposition $\{(\lambda_j, \varphi_j, \vec{v}_j)\}_{j=1}^{\infty}$.
The $p$ eigenvalues of $\mathbf{K}$ approximate the Koopman spectrum in the following sense: in the limit of infinite data, $\mathbf{K}$ converges to the projection of $\mathcal{K}$ onto the $p$-dimensional subspace spanned by the dictionary, and accumulation points of its spectrum correspond to eigenvalues of $\mathcal{K}$ \citep{korda_mezic_convergence_2018}. Eigenvalues whose eigenfunctions are well represented in the span of $\mathcal{D}$ are approximated accurately, while others are missed or appear as spurious eigenvalues. The quality of an individual eigenvalue--eigenfunction pair can be tested from data by computing its residual, as in residual DMD \citep{colbrook2023residual}, or with recently developed invariance diagnostics and error bounds \citep{conradie2026trustworthy}. 
For our purposes, we are mainly interested in the slow-varying modes that capture regime transitions, rather than the full spectrum. 

Eigenfunctions are calculated as the projection of the data onto the right eigenvector,
\begin{equation}
\label{eq:eigenfunction_calculation}
\varphi_{i}(t) = \vec{y}_{t} \cdot \vec{r}_{i},
\end{equation}
since
\begin{equation}
\begin{split}
\varphi_j(\vec{x}_{t+1}) &= \vec{y}_{t+1} \cdot \vec{r}_j \approx (\vec{y}_t \mathbf{K}) \cdot \vec{r}_j \\
&= \vec{y}_t \cdot (\mathbf{K} \vec{r}_j) = \lambda_j \vec{y}_t \cdot \vec{r}_j = \lambda_j \varphi_j(\vec{x}_t).
\end{split}
\end{equation}
The Koopman modes $\vec{v}_{j}$ are approximated from the left eigenvectors. 
Interested readers can find a detailed derivation in \citet{williams2015data}.

In summary: eigenvalues of $\mathbf{K}$ approximate the eigenvalues of $\mathcal{K}$, the right eigenvectors of $\mathbf{K}$ generate approximate eigenfunctions, and the left eigenvectors of $\mathbf{K}$ generate approximate Koopman modes.

The Koopman mode decomposition is highly interpretable: the fitted model's internal quantities each answer a specific scientific question. The eigenvalues $\lambda_{j}$ set the timescales on which the system evolves; the eigenfunction values $\varphi(\vec{x})$ represent the state $\vec{x}$ as a coordinate in eigenfunction space, where similar states have similar coordinates, locating the dynamical regime in which the system currently resides; and the Koopman modes (the left eigenvectors $\vec{l}_{i}$) describe the time-independent structures that grow, decay, or oscillate at the timescale determined by the eigenvalues.
In this way, eigenfunctions can partition state space, in the sense that the value of each eigenfunction quantifies membership in a regime; an eigenfunction that exhibits rapid changes in value signifies a regime switch and may actually hold predictive power.
This will become clearer with an example in Section \ref{sec:partition}.

It is helpful to rewrite the eigenvalues $\lambda_{j}$ of the discrete operator $\mathcal{K}$ in terms of eigenvalues of the continuous-time Koopman operator $\mathcal{L}$, where
\begin{equation}
\frac{d}{dt}g(\vec{s}(t)) = \mathcal{L}g(\vec{s}(t))
\end{equation}
\begin{equation}
\mathcal{L}\varphi_j(\vec{s}) = \mu_j \varphi_j(\vec{s}).
\end{equation}
$\mathcal{K}$ and $\mathcal{L}$ share eigenfunctions $\varphi_{j}$, with eigenvalues related by $\lambda_{j}=e^{\mu_{j}\Delta t}$ \citep{williams2015data}. By decomposing the continuous time eigenvalue into real and imaginary parts $\mu_{j}=\rho_{j}+i\omega_{j}$, we can express the complete mode decomposition of an observable with explicit time evolution:
\begin{equation}
g_i(\vec{s}(t)) = \sum_{j=1}^{\infty} \lambda_j^t \varphi_j(\vec{s}(0)) v_{ij} = \sum_{j=1}^{\infty} \varphi_j(\vec{s}(0)) e^{(\rho_j + i \omega_j) t \Delta t} v_{ij}.
\end{equation}
The eigenvalues $\lambda_{j} = e^{\rho_j \Delta t} e^{i \omega_j \Delta t}$, which contain both a growth/decay term with time scale $1/\lvert \rho_{j} \rvert$ and an oscillating term with frequency $\omega_{j}$, govern the time evolution of the eigenfunction $\varphi_{j}(\vec{s})$.

In discrete time, the growth or decay of a mode is set by the modulus of its eigenvalue: $|\lambda_{j}| = e^{\rho_{j} \Delta t}$, so modes with $|\lambda_{j}|<1$ ($\rho_{j}<0$) decay, modes with $|\lambda_{j}|=1$ ($\rho_{j}=0$) persist, and modes with $|\lambda_{j}|>1$ ($\rho_{j}>0$) grow. The oscillation frequency is set by the phase angle, $\theta_{j} = \omega_{j} \Delta t$. For a dissipative system evolving on an attractor, we expect eigenvalues on or inside the unit circle, and this is what we observe in all systems analyzed in this paper.

As we show in Section \ref{sec:partition}, Koopman eigenfunctions corresponding to different time scales can partition the state space and be used to predict state changes at those timescales.

\subsection{Partitioning state space with Koopman eigenfunctions}
\label{sec:partition}

To illustrate how Koopman eigenfunctions can partition state space, we apply EDMD to discrete samples of a simulated system. 
One of the most common examples of a chaotic system is the forced Duffing oscillator, governed by the differential equation
\begin{equation}
\ddot{x} = -\delta \dot{x} -\alpha x -\beta x^{3} + \gamma \cos{\Omega t},
\end{equation}
where $\delta$, $\alpha$, $\beta$, $\gamma$, and $\Omega$ are the damping, linear stiffness, nonlinear stiffness, driving amplitude, and driving frequency constants, respectively.
The Koopman spectral properties of the Duffing oscillator have been studied extensively \citep{williams2016extending,li2017extended,mezic2026koopman}. 
We use a well-studied chaotic set of parameters: $\delta=0.2$, $\alpha=-1.2$, $\beta=0.9$, $\gamma=0.3$, and $\Omega=1.2$. 
With these parameters, the system exhibits deterministic but chaotic switching between two stable oscillatory states. We divide the simulated Duffing light curve into training and test segments, using training data to learn $\mathbf{K}$, and calculating the values of the eigenfunction for a streaming input of test data according to Eq. \ref{eq:eigenfunction_calculation}.

Recall from Section \ref{sec:spectral_analysis} that each continuous eigenvalue $\mu = \rho + i\omega$ combines a growth/decay rate $\rho$ and an oscillation frequency $\omega$, corresponding to the modulus $|\lambda| = e^{\rho \Delta t}$ and phase angle $\theta = \omega \Delta t$ of the discrete eigenvalue $\lambda$.

\begin{figure}
\centering
\includegraphics[width=0.9\linewidth]
{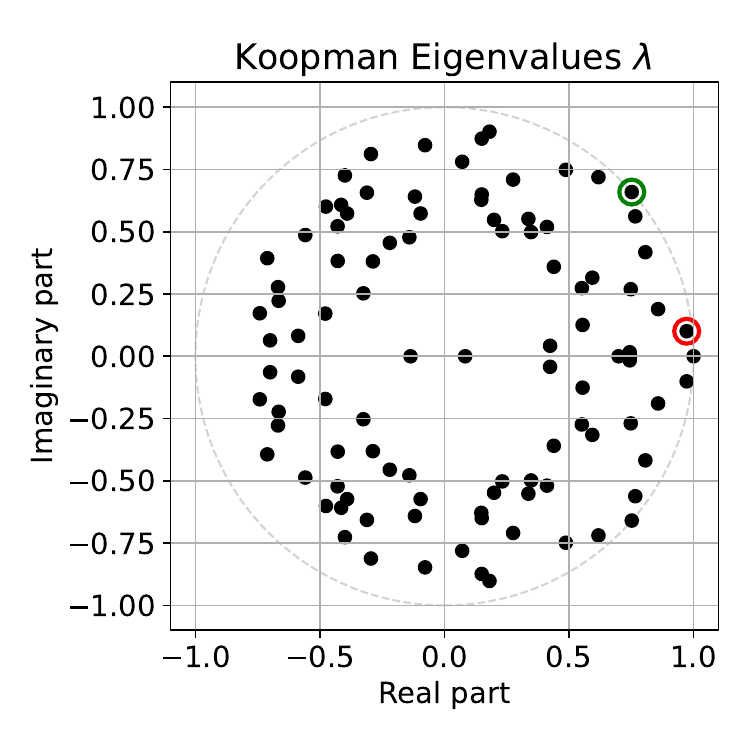}
\caption{
The eigenvalues of the Koopman matrix $\mathbf{K}$ of synthetic Duffing oscillator data. The $x$-axis values are the real parts of the eigenvalues, and the $y$-axis values are the imaginary parts of the eigenvalues.
The dashed curve marks the unit circle. We highlight two example eigenvalues: the red circle marks one of the eigenvalues with a small phase angle $\theta$ (a slow-varying mode), and the green circle marks one of the eigenvalues with a larger phase angle $\theta$ (a fast-oscillating mode). Their corresponding eigenfunctions are labeled $\varphi_{\mathrm{slow}}$ and $\varphi_{\mathrm{fast}}$.
}
\label{fig:eigenvalue_polar_plot}
\end{figure}

Crucially, we see that the values of the eigenfunction signify membership to a certain regime of a spatial mode, evolving with some temporal behavior governed by $\mu$. For example, eigenvalues where $\rho \approx 0$ and $\omega \approx 0$ govern nearly conserved, slow-varying spatial modes that decay and oscillate slowly.
Converting to discrete eigenvalues (eigenvalues of $\mathbf{K}$), this is equivalent to $|\lambda| \approx 1$, and the phase angle $\theta \approx 0$. 
In contrast, an eigenvalue with $|\omega| > 0$, or $|\theta| > 0$, governs the faster-oscillating dynamics.
Figure \ref{fig:eigenvalue_polar_plot} shows the eigenvalue spectrum of the discrete Koopman matrix $\mathbf{K}$ derived from the time-series data of the Duffing oscillator.
We highlight two example eigenvalues by a red circle and a green circle, representing a slow and a fast-varying dynamic mode.
Their corresponding eigenfunctions are labeled as $\varphi_{\mathrm{slow}}$ and $\varphi_{\mathrm{fast}}$.

\begin{figure*}
\raggedleft

\includegraphics[width=0.98\linewidth]
{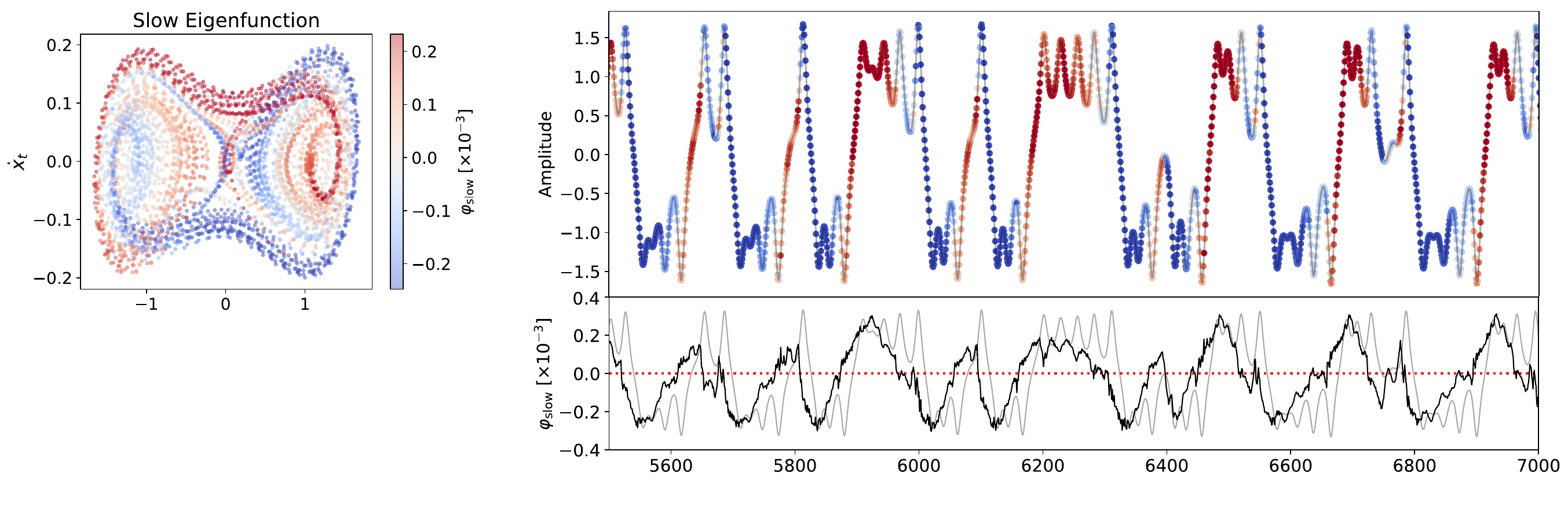}

\includegraphics[width=0.995\linewidth]
{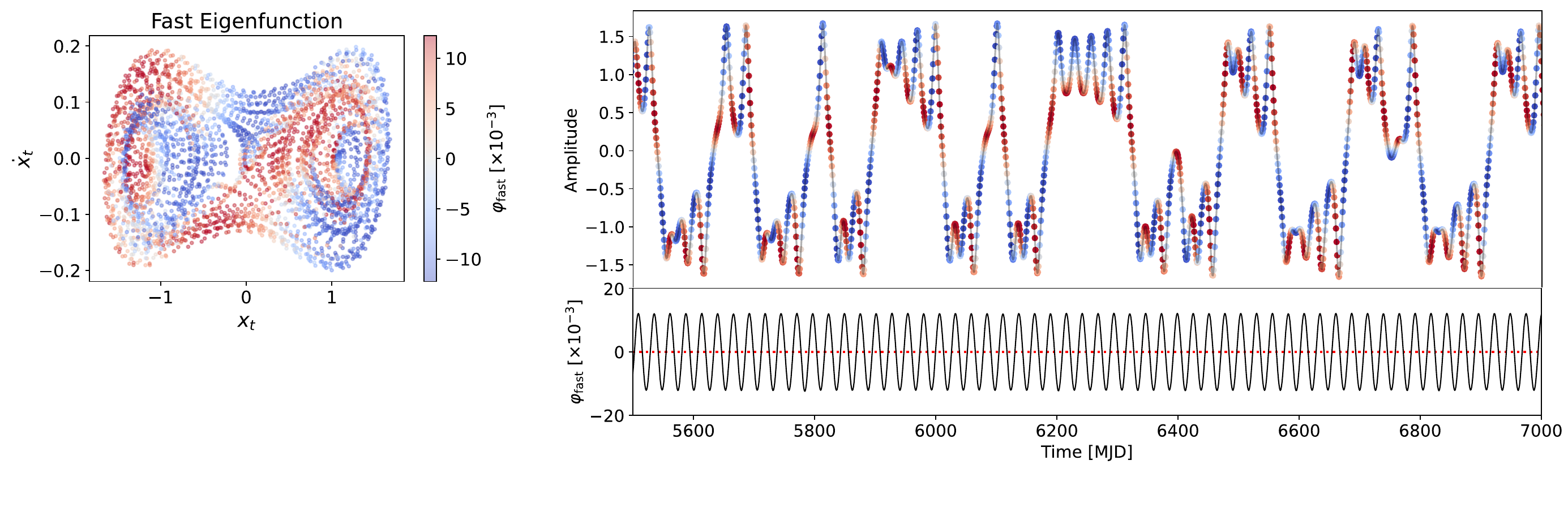}
\caption{Duffing oscillator light curve test data and phase space colored by eigenfunction amplitude. $\mathbf{K}$ was calculated with EDMD using the Fourier dictionary specified in Section \ref{sec:edmd} with $m=2$. 
The phase-space plots (left) show the flux $x_t$ against its time derivative $\dot{x}_t$, computed with backward finite differences of the light curve. The lower panel of each group plots the eigenfunction values $\varphi(t)$, with the dotted red line marking $\varphi = 0$. 
\textbf{Top:} Slow-varying mode $\varphi_{\mathrm{slow}}(t)$; blue ($\varphi_{\mathrm{slow}}(t) < 0$) and red ($\varphi_{\mathrm{slow}}(t) > 0$) partition the light curve into low-flux and high-flux regimes, respectively. $\varphi_{\mathrm{slow}}(t) = 0$ signifies a transition between regimes. The phase space plot (left) shows two distinct stable basins labeled by $\varphi_{\mathrm{slow}}$, with low-flux (blue) on the left and high-flux (red) on the right; long ``tails'' of opposing color connect the basins, signifying color switching before the system reaches the opposing basin. We additionally overlay the eigenfunction plot (bottom right) with the light curve flux at each point (in light gray) to demonstrate how sign changes of $\varphi_{\mathrm{slow}}(t)$ tend to precede the flux changes visible in the light curve.
\textbf{Bottom:} Fast-varying mode $\varphi_{\mathrm{fast}}(t)$ oscillates rapidly within each oscillation, providing no predictions for slow-varying regime changes. The phase-space plot shows no clear distinction between the left and right basins.
}
\label{fig:eigenfunction_light_curve_duffing}
\end{figure*}

As described in Section \ref{sec:spectral_analysis}, eigenfunctions $\varphi_j$ represent the original state $\vec{x}$ as a coordinate in an intrinsic coordinate space. The different areas of this coordinate space represent distinct regimes, quantified by the values of $\varphi_j(t)$. In this way, eigenfunction values partition state space into distinct regimes, as illustrated in Figure \ref{fig:eigenfunction_light_curve_duffing}. The top panel shows $\varphi_{\mathrm{slow}}(t)$ from the test light curve, and it clearly labels the slow-varying regimes of the light curve (high-flux vs low-flux states) through the eigenfunction's magnitude and sign. Crucially, sign changes in $\varphi_{\mathrm{slow}}(t)$ (color switches) \textit{precede} visible regime changes in the light curve flux $x_t$, sometimes by more than an entire stable oscillation. This is also visible when superimposing the flux over $\varphi_{\mathrm{slow}}(t)$. In other words, not only do eigenfunctions contain information about regime membership, but they hold \textit{predictive} power over transitions. This is evident in the phase space plot, where two distinct stable basins are connected by long ``tails'' of the opposing basin's color---the system's eigenfunction value changes before the system physically reaches the opposing basin. Eigenfunction amplitudes are actually highest (darkest color) right before regime changes, highlighting the usefulness of this slowly varying eigenfunction for predicting them.

In contrast, the fast-varying mode eigenfunction $\varphi_{\mathrm{fast}}(t)$ rapidly switches between red and blue, also labeling regimes, but at a short timescale as governed by the eigenvalue phase $\theta$ rather than the slower varying low and high flux regimes. The phase space plot displays no labeling of these stable regimes (left and right basins) nor predictive ``tails,'' but rather rapid color changes corresponding to each individual revolution in phase space. Therefore, the eigenfunction associated with this eigenvalue is not useful for predicting the longer-scale regime changes or flares of interest to us.

An impending regime change is thus signaled by the sign change of $\varphi_{\mathrm{slow}}(t)$, a single coordinate that encodes membership in the low-flux or high-flux basin and can be inspected directly in eigenfunction space (Figure \ref{fig:eigenfunction_light_curve_duffing}).

\subsection{Comparing Koopman operator methods to other approaches}
\label{sec:comparison}

Koopman operator theory approaches nonlinear dynamics by lifting the system into a higher-dimensional space of observables in which the evolution becomes approximately linear, enabling spectral analysis, modal decomposition, and physically interpretable forecasting. 
This contrasts with local state-space prediction methods, such as nearest-neighbor or local linear forecasting, which rely on short-range geometric similarity in delay-coordinate space and make predictions using nearby trajectories rather than a global dynamical representation \citep{farmer1987predicting,casdagli1989nonlinear,sugihara1990nonlinear}. 
While the local state-space methods are good for short-range linearized dynamics, their local linear assumptions limit their long-range forecasting ability and make global interpretation difficult.

On the other hand, recurrent neural networks (RNNs), including Long Short-Term Memory (LSTM) and Gated Recurrent Unit (GRU) \citep{graves2012long,gers2000learning,cho2014learning}, and reservoir computers \citep{jaeger2001echo,lukovsevivcius2009reservoir}, take a fundamentally different approach by directly learning nonlinear temporal dependencies from data through adaptive hidden states. 
Compared to Koopman approximations, RNNs can represent complicated nonlinear memory effects without requiring carefully designed observable dictionaries. 
However, this flexibility comes at the cost of interpretability \citep{krakovna2016increasing}. 
By contrast, Koopman methods provide explicit spectral objects such as eigenfunctions, eigenvalues, and eigenmodes that can reveal oscillatory structures, metastable states, and timescale separations in the dynamics \citep{rowley2009spectral,mezic_kmd_2005,budivsic2012applied}.
Modern deep-learning variants of Koopman theory increasingly mix these approaches by using neural networks to learn the lifting observables automatically, producing hybrid architectures such as Koopman autoencoders, DeepDMD, and neural Koopman operators that combine the representational power of RNNs with the interpretability and spectral structure of operator-theoretic methods.

Another interesting approach is sparse equation discovery: methods such as SINDy (Sparse Identification of Nonlinear Dynamical systems) seek explicit governing equations by identifying a sparse set of active nonlinear terms from a candidate function library \citep{brunton2016discovering}. 
The sparse equation discovery methods are highly interpretable and provide physically meaningful equations. 
Consequently, SINDy is especially attractive when the true dynamics is governed by a relatively compact set of nonlinear interactions that can be expressed in a suitable basis. 
The two frameworks are deeply connected mathematically because both rely on selecting informative observable libraries, and Koopman eigenfunctions can provide sparse intrinsic coordinates for nonlinear systems. 
Recent work increasingly combines these perspectives by using SINDy within Koopman-invariant subspaces, learning Koopman eigenfunctions with sparse regression, or constructing hybrid models that simultaneously recover interpretable equations and linear spectral structure \citep{brunton2016sparse,gao2026sparse}.

In Section \ref{sec:real_data}, we present a vanilla application of EDMD to the X-ray observations of a binary system. While this simple EDMD method might not outperform other state-of-the-art techniques in terms of forecasting nonlinear dynamics, we wish to demonstrate the potential of EDMD not only in predicting the futures of astrophysical systems but also in its capacity for spectral analysis and physical interpretation.

Finally, we caution that broadband variability is a challenge for Koopman methods as well. For chaotic or mixing dynamics, the Koopman operator generally has a continuous spectrum in addition to (or instead of) discrete eigenvalues. A continuous spectrum cannot be represented by a small set of eigenfunctions, and in some cases eigenfunctions may not exist at all \citep{lusch2018deep}. Data-driven methods that rigorously handle continuous spectra are an active area of research \citep{colbrook2024rigorous,sakata2024enhancing}. This matters here because systems that switch chaotically between metastable states, as 4U 1705-44 does between its low- and high-flux states, may have continuous or mixed spectra \citep[e.g.,][]{nagdi2026learning}. The discrete eigenvalues extracted by EDMD in Section \ref{sec:real_data} should therefore be interpreted as a finite approximation that captures the slow switching behavior, not as a complete description of the spectrum.

\section{Koopman's view of quasi-periodic oscillations}
\label{sec:connect_QPO}

In this section, we discuss how process noise can change the spectrum of Koopman eigenvalues and their connection to QPOs.
Interested readers can find a more detailed discussion on how process noise affects the spectrum of the Koopman operator in \cite{chekroun2020ruelle,vcrnjaric2020koopman,wanner2022robust}.

QPOs have attracted increasing interest because these quasi-periodicities may be linked to oscillatory physics of the central compact object and the relativistic jet structure \citep{camenzind1992lighthouse,caproni2017jet,sandrinelli2016quasi,sobacchi2016model,ingram2019review,belloni2002unified}.
However, these oscillatory dynamics are affected by stochastic processes due to magnetic turbulence, destroying the long-term periodicities of these oscillatory dynamics.

Consider a stochastic system in which the high-dimensional accretion flow state $\vec{s}$ (e.g., density, temperature, magnetic fields) evolves under a deterministic drift $\vec{f}(\vec{s})$ and a stochastic process noise of strength $\sqrt{2D}$ (e.g., turbulence), which is assumed to be uncorrelated for a simplified discussion. Let $\phi(\vec{s})$ be a projected observable function of the accretion flow state (e.g., the X-ray flux) that is governed by the stochastic Koopman generator $\mathcal{L}$; the precise formulation is given in Appendix \ref{sec:process_noise}.

Suppose that $\varphi_{k}(\vec{s})$ is an eigenfunction that satisfies $\mathcal{L}\varphi_{k}(\vec{s})=\mu_{k} \varphi_{k}(\vec{s})$. Defining the power spectral density $S(\omega)$ of an observable as the Fourier transform of its autocorrelation function (see Appendix \ref{sec:process_noise} for definitions and the full derivation), the contribution of the eigenfunction to the PSD has a Lorentzian form \citep{chekroun2020ruelle},
\begin{equation}
S_{k}(\omega) \propto
2 \mathrm{Re} \left( \frac{1}{i\omega-\mu_{k}} \right)
= \frac{-2 \mathrm{Re}(\mu_{k})}
{\left( \omega - \mathrm{Im}(\mu_{k}) \right)^{2} + \mathrm{Re}(\mu_{k})^{2}}.
\label{eq:qpo_spike}
\end{equation}

\begin{figure*}
\centering
\includegraphics[width=0.7\linewidth]{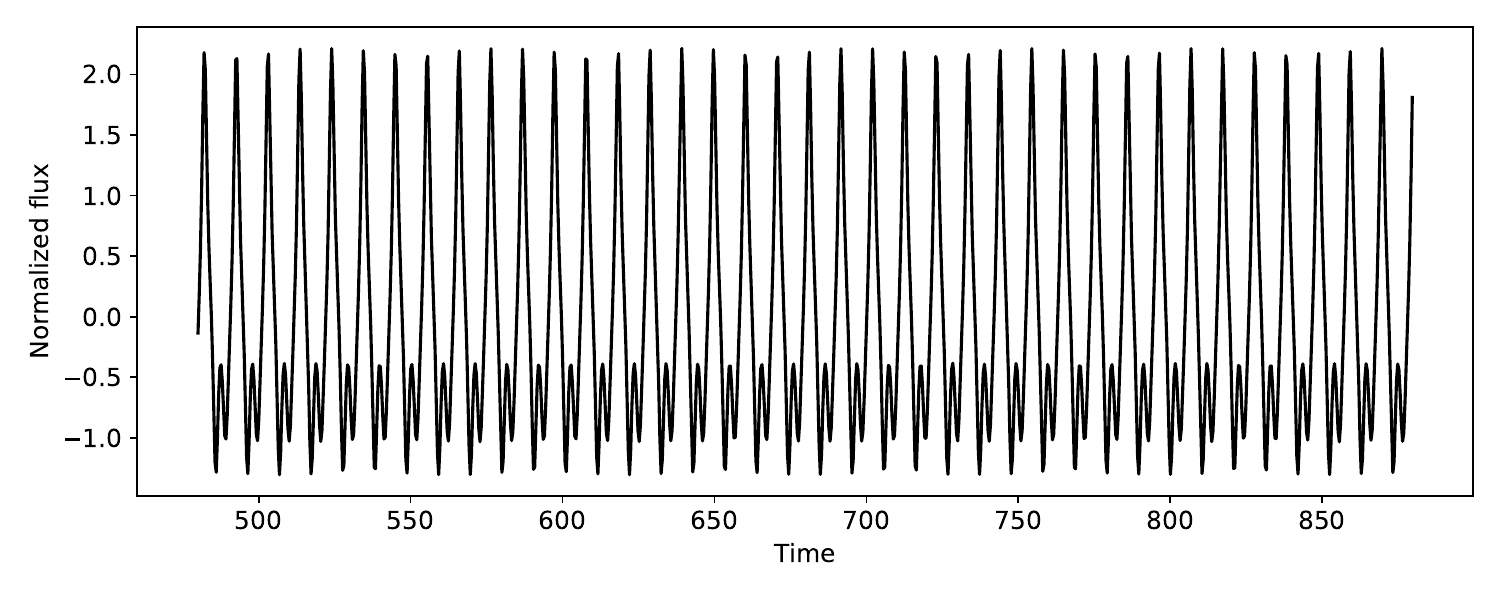}
\includegraphics[width=0.33\linewidth]{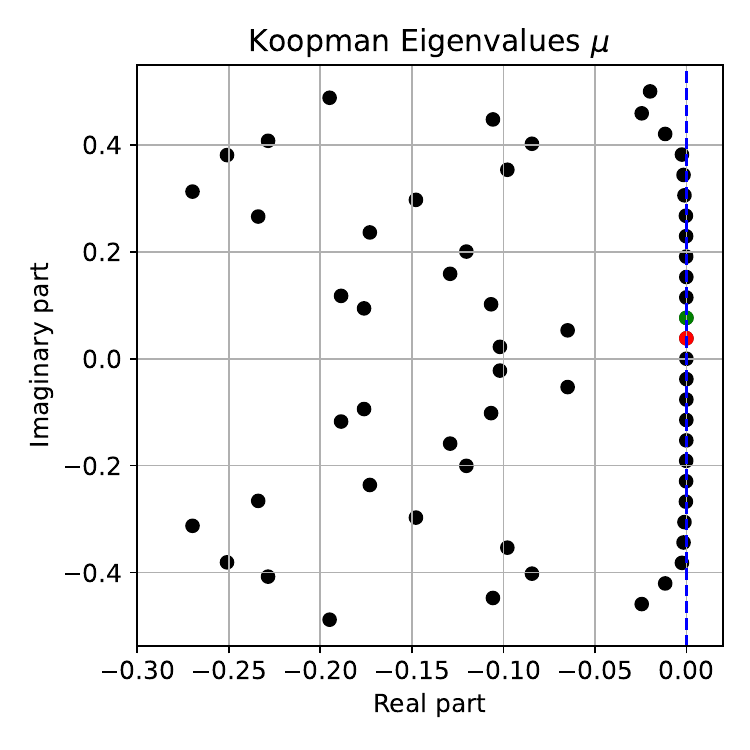}
\includegraphics[width=0.45\linewidth]{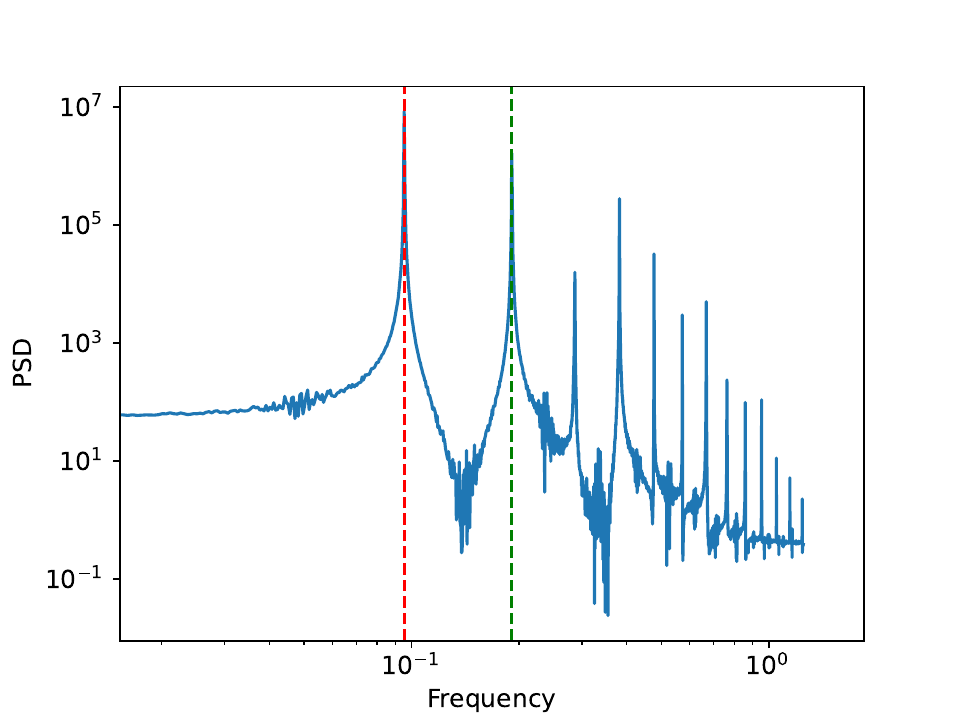}
\caption{
Koopman eigenvalues and PSD for a noise-free Duffing oscillator ($D=0$, $\delta=0.2$, $\alpha=-1.2$, $\beta=0.9$, $\gamma=1.0$, $\Omega=0.6$).
\textit{Top}: Light curve showing quasi-periodic behavior.
\textit{Bottom left}: Eigenvalues of $\mathcal{L}$ in the complex plane; $2\times 13$ eigenvalues lie near the imaginary axis (highlighted: red and green).
\textit{Bottom right}: PSD with 13 narrow QPOs; vertical lines mark frequencies corresponding to the highlighted eigenvalues.
}
\label{fig:connecting_qpo_koopman_noise0}
\end{figure*}

\begin{figure*}
\centering
\includegraphics[width=0.7\linewidth]{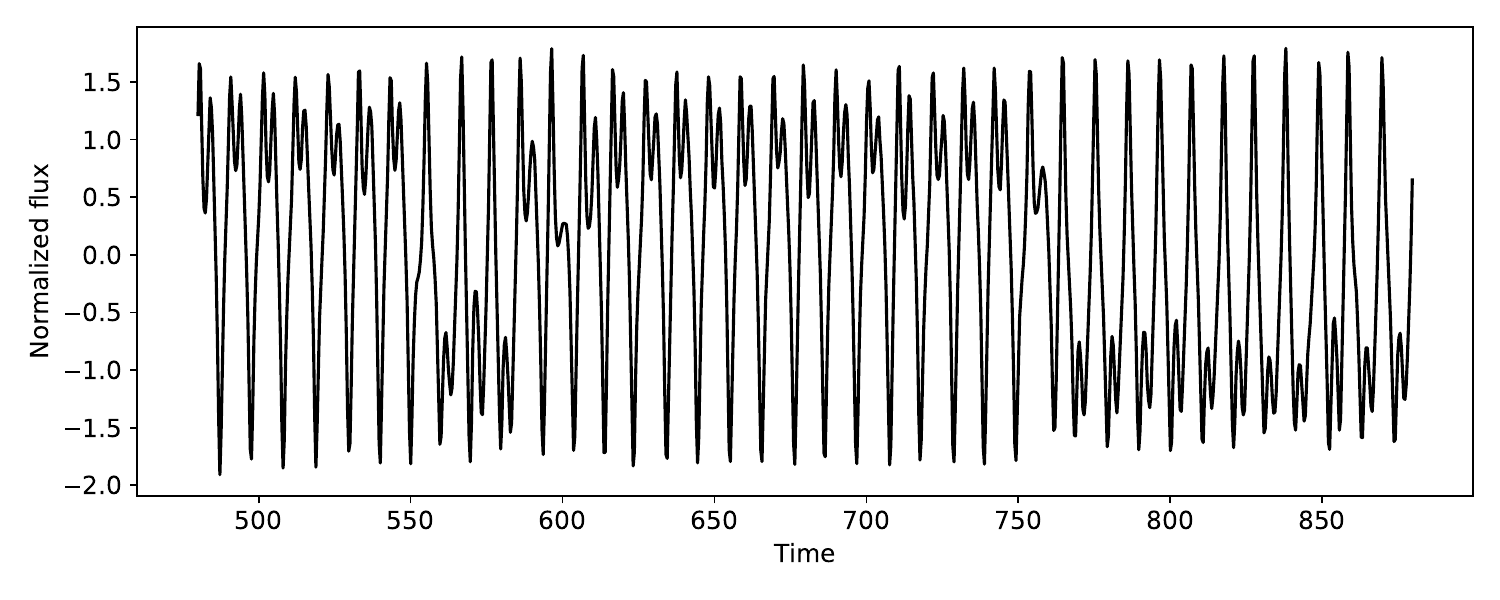}
\includegraphics[width=0.33\linewidth]{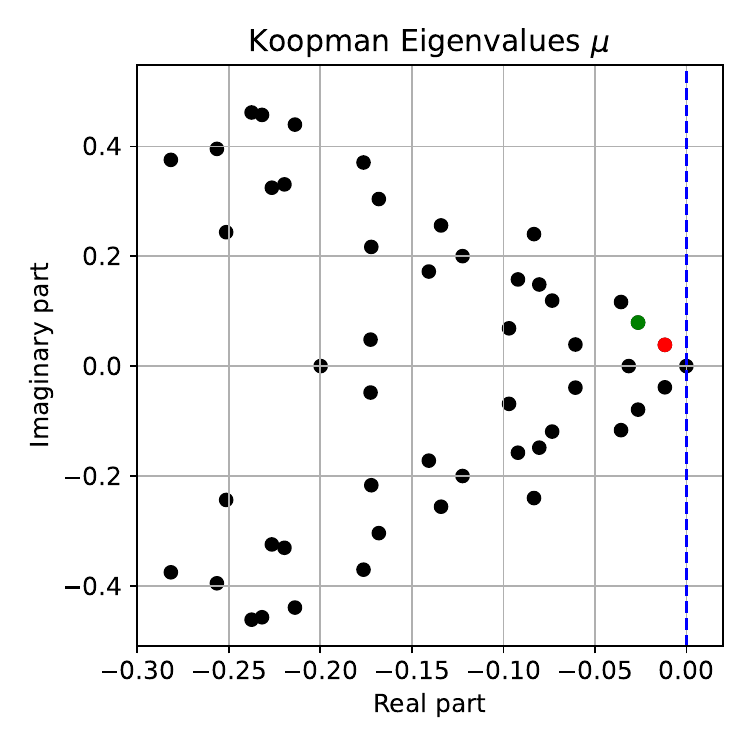}
\includegraphics[width=0.45\linewidth]{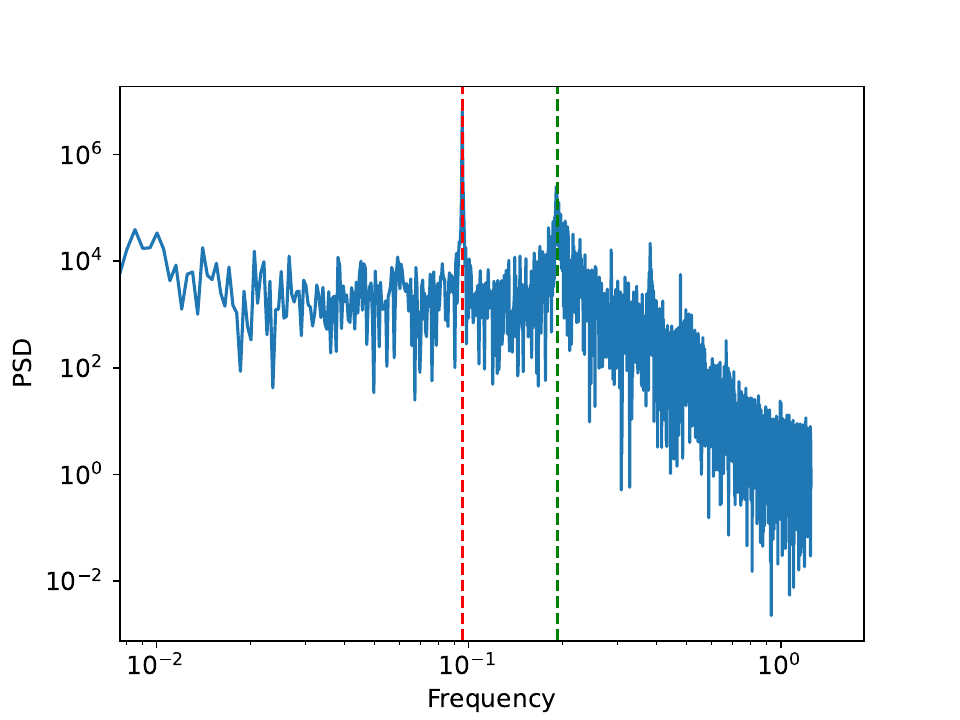}
\caption{
Same as Figure~\ref{fig:connecting_qpo_koopman_noise0}, but with process noise $D=0.2$.
Eigenvalues are shifted away from the imaginary axis, broadening the corresponding QPO peaks in the PSD.
The red-highlighted eigenvalue remains relatively close to zero, preserving a narrow QPO, while the green-highlighted eigenvalue shifts further, producing a broader peak.
}
\label{fig:connecting_qpo_koopman_noise20}
\end{figure*}

The eigenfunction $\varphi_{k}(\vec{s})$ contributes to a spike (QPO) at the frequency of $\omega=\mathrm{Im}(\mu_{k})$, and the width of the QPO is determined by $|\mathrm{Re}(\mu_{k})|$, which controls the decay timescale and is modified by the noise coefficient $D$.
The increase of the noise coefficient $D$ will lead to the shortening of QPO lifetimes and the broadening of the widths of QPOs, and multiple QPOs can merge into a noise-like bulk spectral distribution.
Each eigenvalue therefore predicts the frequency and width of a Lorentzian component of the PSD, so individual components of the fitted operator can be checked directly against the observed power spectrum (Figures~\ref{fig:connecting_qpo_koopman_noise0} and \ref{fig:connecting_qpo_koopman_noise20}).

Figures~\ref{fig:connecting_qpo_koopman_noise0} and \ref{fig:connecting_qpo_koopman_noise20} demonstrate how process noise reshapes the Koopman spectrum and the resulting PSD. We use a weakly chaotic set of Duffing oscillator parameters: $\delta=0.2$, $\alpha=-1.2$, $\beta=0.9$, $\gamma=1.0$, and $\Omega=0.6$, and we inject two levels of process noise with $D=0$ and $D=0.2$. We use the method described in Section \ref{sec:edmd} to compute the approximation of the Koopman operator $\mathcal{K}$ and its eigenvalues $\lambda_{k}$, and then the eigenvalues of the Koopman generator $\mathcal{L}$ are calculated as $\mu_{k}=\log(\lambda_{k})/\Delta t$.
Without noise (Figure~\ref{fig:connecting_qpo_koopman_noise0}), eigenvalues cluster near the imaginary axis, producing sharp, well-separated QPOs.
When noise is introduced (Figure~\ref{fig:connecting_qpo_koopman_noise20}), eigenvalues are pushed toward more negative real parts, which shortens coherence times and broadens the spectral peaks.
Notably, the modes respond differently: the red-marked mode remains relatively coherent, while the green-marked mode is more strongly damped by noise.
This is expected from the PSDs, since the red-marked peak is the tallest. However, peak height mixes the amplitude of a mode in the observable with the noise damping of the mode itself. The Koopman spectrum separates the two: the noise-induced damping of each mode is measured directly by the shift of $\mathrm{Re}(\mu_{k})$, which depends on the process noise and gradient structure of its eigenfunction rather than on its amplitude (Appendix \ref{sec:process_noise}; \citealt{chekroun2020ruelle}). Koopman analysis thus provides an amplitude-independent measure of robustness to process noise for each mode.

\section{Predicting the future of a dynamical system}
\label{sec:predicting}

Given a dynamical system,
\begin{equation}
\vec{s}_{t+1} = \vec{F}(\vec{s}_{t}),
\end{equation}
where $\vec{s}_{t}$ is the full state and $\vec{F}$ is the nonlinear propagator that evolves $\vec{s}_{t}$, we have shown how to represent future evolution with the linear Koopman operator $\mathcal{K}$.
To predict the future $n$ steps ahead of the present $t_0$, we can then iteratively apply $\mathcal{K}$ to observables $\vec{g}(\vec{s}_t)$
\begin{equation}
    \vec{g}\left(\vec{s}_{t_0+n}\right)
    = \mathcal{K}^{n} \vec{g}\left(\vec{s}_{t_0}\right),
\end{equation}
and predict the future state
\begin{equation}
\vec{s}_{t_0+n}
= \vec{g}^{-1} \left( \mathcal{K}^{n} \vec{g}\left(\vec{s}_{t_0}\right) \right),
\label{eq:future_flux_prediction}
\end{equation}
where $n$ is the number of steps forward in time, and $\vec{g}^{-1}$ is the inverse of the observable functions.  

In a similar fashion, we can use a learned Koopman matrix $\mathbf{K}$ from EDMD to propagate observed measurements into the future. 
Given the most recent observed state $\vec{x}_{t_0} \in \mathbb{R}^{d}$ (i.e., the time-delay vector of the X-ray flux), we can transform $\vec{x}_{t_0}$ into the lifted state using the function $\vec{\psi}$ and propagate in the lifted state:

\begin{equation}
\vec{y}_{t_0}=\vec{\psi}(\vec{x}_{t_0})
\end{equation}
\begin{equation}
    \vec{y}_{t_0+n}
    = \vec{y}_{t_0} \mathbf{K}^{n}.
\end{equation}
\begin{equation}
    \vec{x}_{t_0+n}=\vec{\psi}^{-1}(\vec{y}_{t_0+n}).
\end{equation}

Here, $\vec{\psi}^{-1}$ is the inverse of the lifting function (which extracts the time-delay vector from the lifted state). There are multiple ways to compute this reverse transformation. By encoding the observed state $\vec{x}_t$ itself in the dictionary (i.e., through an augmented Fourier dictionary $[\vec{x}_{t}, \sin(\vec{x}_{t}), \cos(\vec{x}_{t}), \ldots, \sin(m\vec{x}_{t}), \cos(m\vec{x}_{t})]$), $\vec{\psi}^{-1}$ simply extracts the elements of $\vec{y}_t$ that correspond to $\vec{x}_t$. We can also learn $\vec{\psi}^{-1}$ using a neural network decoder trained on past data, leading to a more accurate transformation.

In practice, data may contain significant measurement noise, rendering a perfectly Koopman-invariant subspace of dictionary functions intractable or even impossible. Thus, predictions will inevitably contain errors that accumulate with each subsequent step. 
Let $\vec{z}_{t} = \vec{x}_{t} + \vec{\epsilon}_{t}$ be the observed state corrupted by the measurement noise $\vec{\epsilon}_{t}$, where $\vec{\epsilon}_{t} \sim \mathcal{N}(\vec{0},\,\sigma^{2}\mathbf{I}_{d})$ is Gaussian noise, i.e., the errors on the $d$ components of $\vec{x}_{t}$ are assumed to be identically distributed and mutually uncorrelated for simplicity of the discussion.
The noise-corrupted Koopman operator gives an expected future of the system,
\begin{equation}
\tilde{\mathcal{K}} \vec{\psi}(\vec{z}_{t}) = \mathbb{E} \left[ \vec{\psi}(\vec{z}_{t+1}) \right].
\end{equation}
The presence of measurement noise in the data can make $\mathbb{E} \left[ \vec{\psi}(\vec{z}_{t+1}) \right]$ deviate from $\vec{\psi}\left(\vec{F}(\vec{z}_{t})\right)$,

\begin{align}
\label{eq:measurement_noise}
\mathbb{E}\left[\psi_i(\vec{z}_{t+1})\right] &={} 
\psi_i\left(\vec{F}(\vec{z}_{t})\right)
+\frac{1}{2}\sigma^{2} \sum_{jk} \delta_{jk} H^{\psi}_{ijk}\left(\vec{F}(\vec{z}_{t})\right)
\nonumber \\
&+\frac{1}{2}\sigma^{2} \sum_{jk} H^{\psi}_{ijk}\left(\vec{F}(\vec{z}_{t})\right)
\nonumber \sum_{uv} \delta_{uv} J^{F}_{ju}(\vec{z}_{t}) J^{F}_{kv}(\vec{z}_{t})
\nonumber \\
&+\frac{1}{2}\sigma^{2} \sum_{j} J^{\psi}_{ij}\left(\vec{F}(\vec{z}_{t})\right) \sum_{uv} \delta_{uv} H^{F}_{juv}(\vec{z}_{t}),
\end{align}

where $J^{\psi}$ and $H^{\psi}$ are the Jacobian and Hessian of the dictionary functions $\vec{\psi}$, and $J^{F}$ and $H^{F}$ are the Jacobian and Hessian of the propagator $\vec{F}$, as defined in Appendix \ref{sec:measurement_noise}.
Equation \ref{eq:measurement_noise} shows that the prediction error is scaled by the measurement noise variance $\sigma^{2}$, the curvature of the dictionary functions $\psi_i$ given by $\mathbf{H}^{\psi}(\vec{x})$, and the curvature of the trajectory given by the Hessian of the propagator $\mathbf{H}^{F}(\vec{x})$, with additional amplification where the dynamics are strongly stretching through $\mathbf{J}^{F}(\vec{x})$.
More details on the derivation of Equation \ref{eq:measurement_noise} can be found in Appendix \ref{sec:measurement_noise}.

\begin{figure*}
\centering
\includegraphics[width=0.49\linewidth]{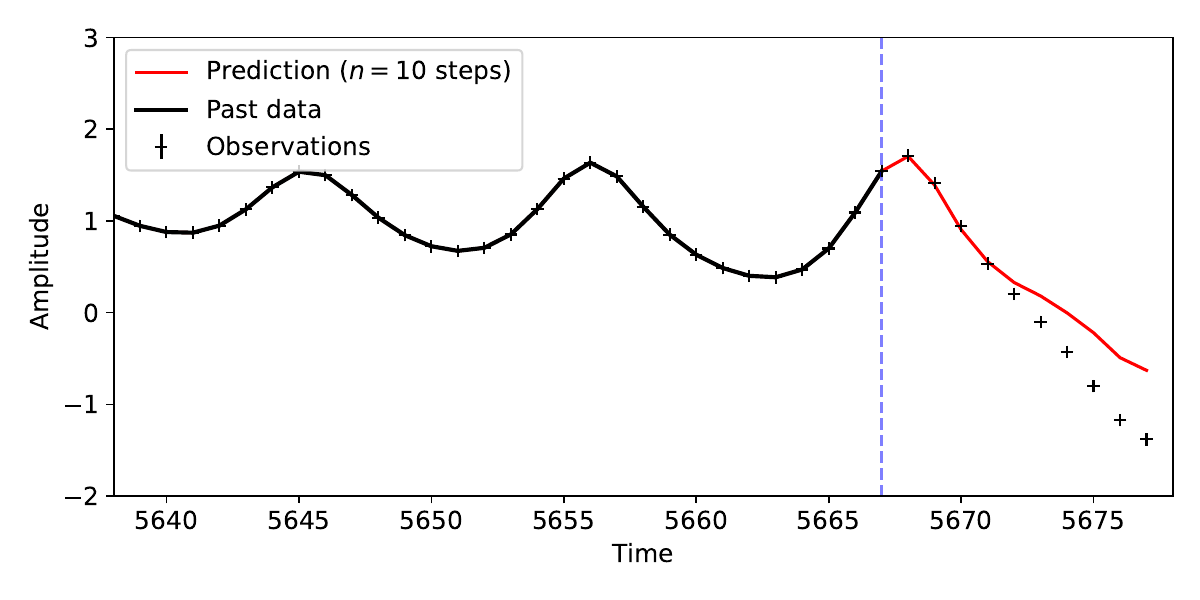}
\includegraphics[width=0.49\linewidth]{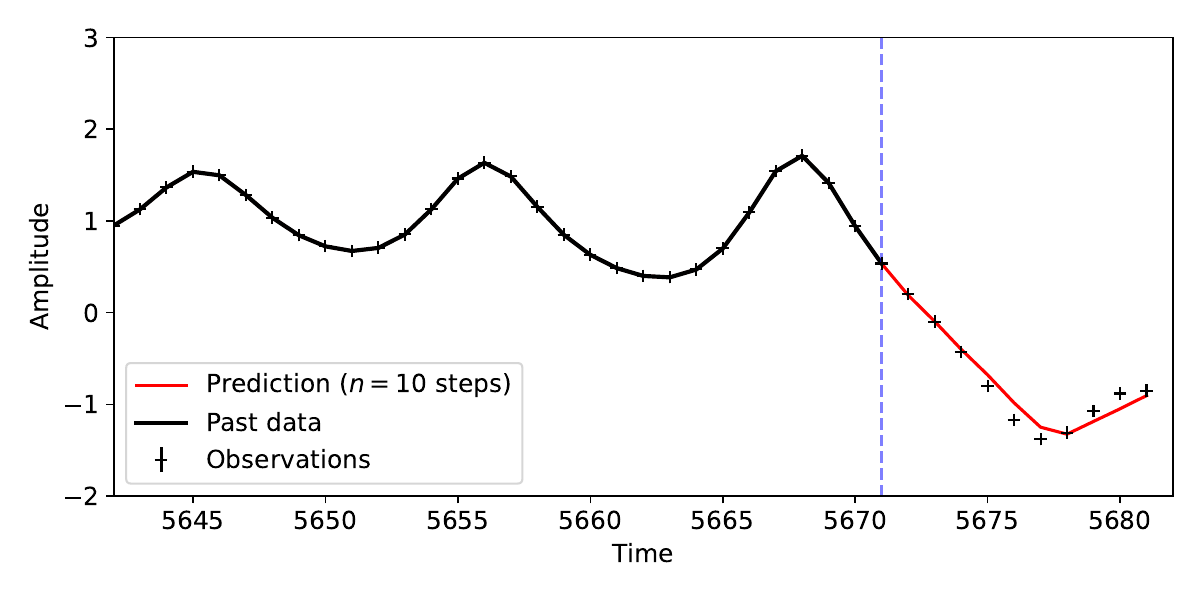} \\
\includegraphics[width=0.49\linewidth]{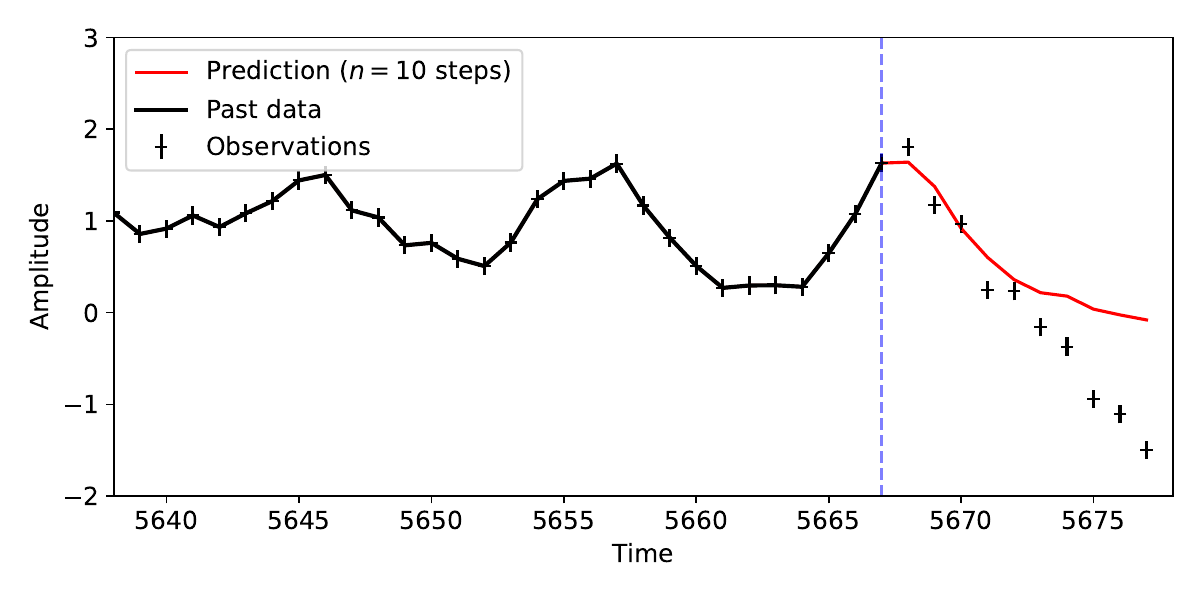}
\includegraphics[width=0.49\linewidth]{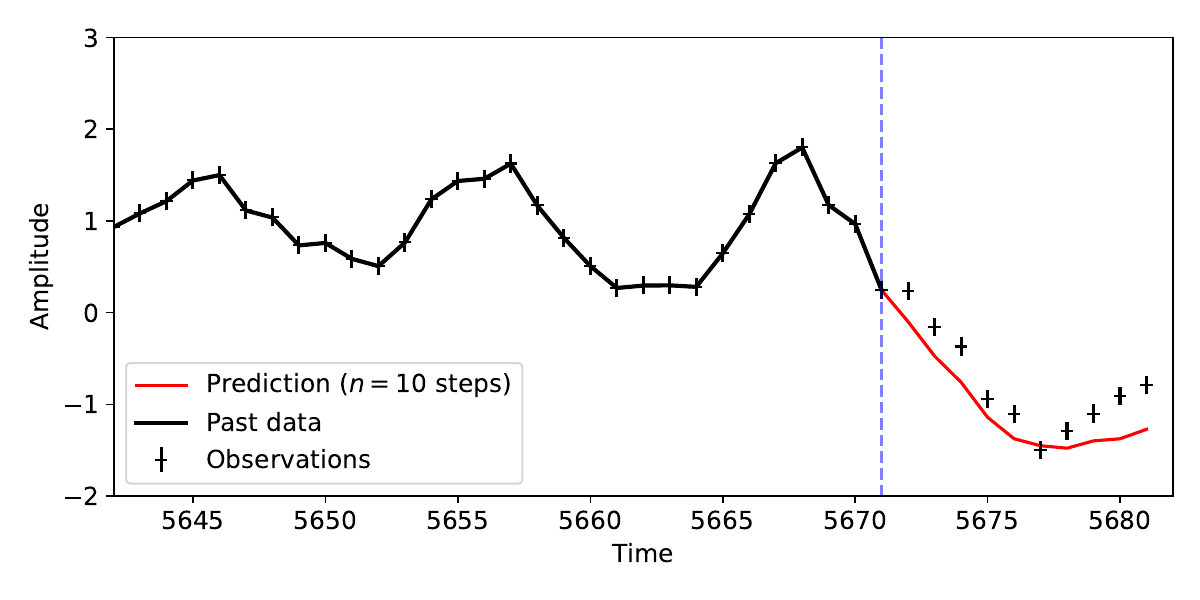} \\
\includegraphics[width=0.49\linewidth]{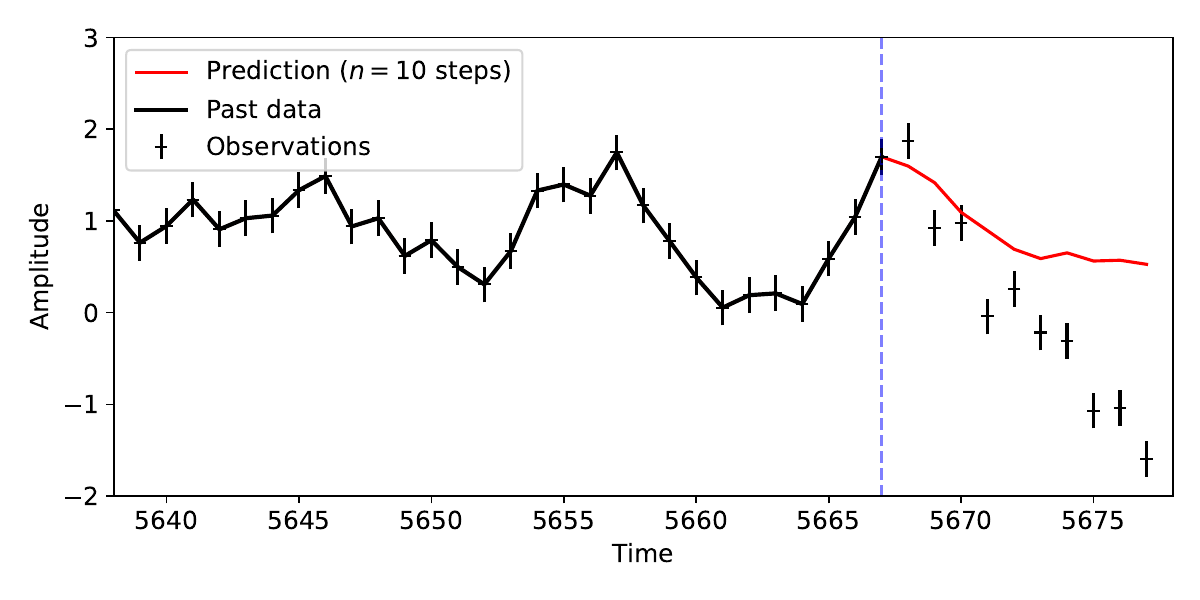}
\includegraphics[width=0.49\linewidth]{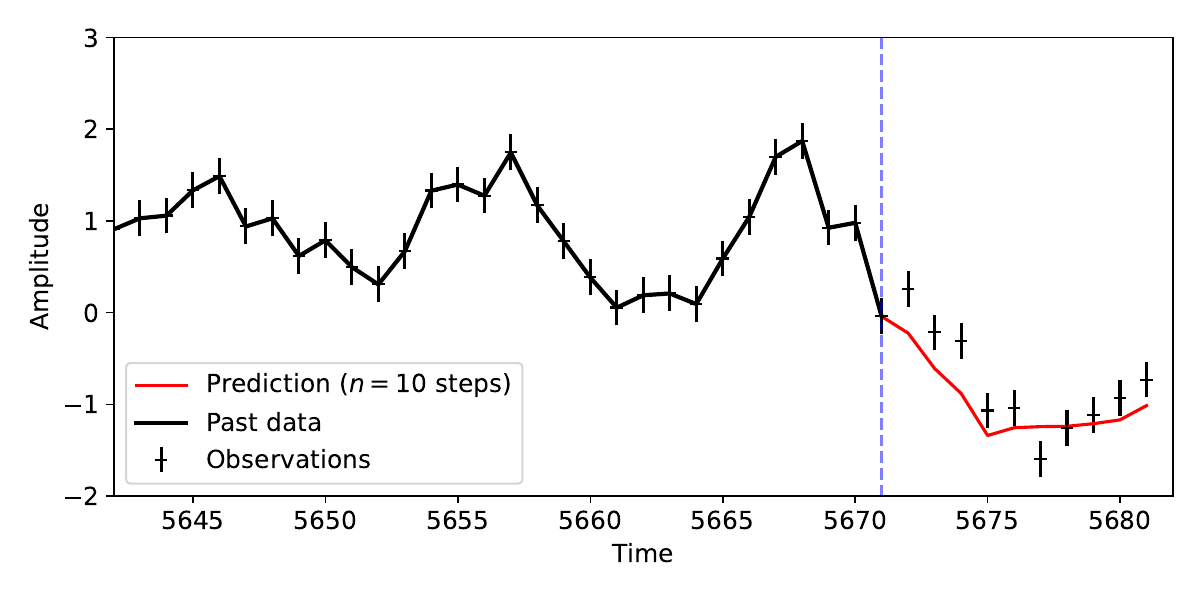} \\
\caption{
(Top panels) The Duffing oscillator time-series data injected with a measurement noise level of $\sigma=0$. The left panel shows the training data (solid black curve) stopped at the blue vertical line before the sharp turning point, and the right panel shows the training data stopped after the sharp turning point.
(Middle panels) The Duffing oscillator time-series data injected with a measurement noise level of $\sigma=0.1$.
(Bottom panels) The Duffing oscillator time-series data injected with a measurement noise level of $\sigma=0.2$.
The red curves show predictions for future time-series data. 
}
\label{fig:predict_light_curve_duffing}
\end{figure*}

To illustrate the effect of measurement noise, we injected Gaussian noise into simulated Duffing oscillator data, $x_{t} \leftarrow x_{t} + \epsilon$, where $\epsilon \sim \mathcal{N}(0,\,\sigma^{2})$ and $\sigma$ is set to three values $(0.0,0.1,0.2)$ for comparison.
For each level of simulated noise, $\mathbf{K}$ is learned from training data, and we predict 10 steps into the future using the procedures outlined above.
The left panels of Figure \ref{fig:predict_light_curve_duffing} show predictions from a high-curvature section of the light curve, while the right panels show predictions from a low-curvature section. Although predictions on both sides worsen as the noise level $\sigma$ increases, predictions on the right (low-curvature section) are significantly more accurate than predictions from the high-curvature point. 
This is expected from Equation \ref{eq:measurement_noise}, which tells us that the error is scaled by the noise variance $\sigma^{2}$ and amplified in high-curvature sections of the light curve through $\mathbf{H}^{F}(\vec{x})$. As a result, when seeking to make future predictions of a light curve, one must pay attention to the location of the present point in addition to the dictionary choice and data processing procedures.

\section{Real X-ray data application: 4U 1705-44}
\label{sec:real_data}

We selected the low-mass X-ray binary 4U 1705-44 to demonstrate how we can use EDMD to analyze past X-ray data and predict future activities. We will first describe the data processing procedures, including scaling and combining RXTE and MAXI data, rebinning, smoothing, and handling uncertainties. The goal is to construct a uniformly binned light curve that maximally captures interesting dynamics and minimizes measurement noise, such that $\mathbf{K}$ learns the underlying system dynamics. Then, we will demonstrate the ability of $\mathbf{K}$ to predictively partition the light curve into low and high flux states and make direct flux predictions, similar to what we did with the simulated Duffing oscillator in Sections \ref{sec:partition} and \ref{sec:predicting}.

Interestingly, while the Fourier dictionary with $m=3$ frequency harmonics and $d=30$ time-delay length was highly effective for flux predictions, calculating $\mathbf{K}$ using a long time-delay basis ($\vec{y}_t = [x_t, x_{t-1}, x_{t-2}, \ldots, x_{t-49}] \in \mathbb{R}^{1 \times 50}$) was significantly more effective at partitioning regimes and predicting state changes using slow-varying eigenfunctions. This differs from the Duffing oscillator simulation in Section \ref{sec:partition}, where a Fourier dictionary of the same form was most effective at predictive partitions. Therefore, not only is dictionary choice essential across different systems, but even within very similar systems, such as the forced Duffing oscillator and 4U 1705-44, separate dictionaries may be required depending on the task. In this section, we will use the time-delay basis for eigenfunction predictions and the Fourier-lifted time-delay basis for flux predictions. 

\subsection{Data selection}

The continuous long-term X-ray light curve for 4U 1705-44 was created by combining observations from RXTE ASM (All-Sky Monitor) and MAXI (Monitor of All-Sky X-ray Image).
RXTE ASM observed the source daily from December 1995 until the satellite's decommissioning in January 2012 and scanned $80\%$ of the sky every 90 minutes in the 2--10 keV energy band, with a sensitivity of 30 mCrab \citep{levine1996first}.
MAXI has operated on the International Space Station since August 2009, scanning nearly the entire sky every 96 minutes in the 2--20 keV energy band with a sensitivity of 20 mCrab \citep{matsuoka2009maxi}.
The combined data provide a continuous light curve spanning nearly 30 years (from January 6, 1996, through December 2025).
This light curve reveals over 50 high-amplitude transitions that occur on timescales of the order of hundreds of days.

Combining the light curves from the two instruments requires careful cross-calibration, as the detectors have distinct energy responses, effective areas, and count-rate sensitivities.
First, we align the energy bands by combining the MAXI 2--4 keV and 4--10 keV channels to match the 2--10~keV band measured by RXTE ASM.
We then perform cross-calibration using data from the overlap period during which both instruments observed the source simultaneously (MJD 55045--55434; see Appendix~\ref{sec:data_processing_appendix}).
During this interval, we applied Orthogonal Distance Regression (ODR) to derive a linear transformation from MAXI count rates to the RXTE rates, accounting for measurement uncertainties in both datasets.
The calibrated MAXI data extend the light curve from the end of the RXTE mission through MJD 61011 or December 2, 2025. While MAXI continues to observe the source, our dataset includes observations only up to that point. 
Further details of this procedure, including the fitted calibration parameters, are provided in the Appendix~\ref{sec:data_processing_appendix}.

We note that the character of the light curve changes around MJD 56000: the source pauses its usual switching and remains in the high-flux state for $\sim$2000 days, before returning to switching behavior shortly before MJD 58000. 
For a chaotic system switching between two metastable states, dwell times are broadly distributed and occasional long dwellings are expected, and the recurrence analysis of \citet{phillipson2018chaotic} found the series consistent with stationarity over their baseline. However, a true change in the underlying system cannot be excluded from a single realization. We also note that this epoch roughly coincides with the transition from RXTE ASM to the rescaled MAXI data (January 2012, MJD 55927); while we find no visible change in the noise properties of the light curve or in prediction performance across this boundary, imperfections in the cross-calibration could contribute to apparent changes in the series. If the underlying system does change over time, the single matrix $\mathbf{K}$ learned here represents a time-averaged operator and its forecasts would degrade accordingly. Koopman extensions to time-varying systems, such as nonautonomous operator families \citep{macesic2018koopman} and online DMD \citep{zhang2019online}, are natural directions for future work (Section \ref{sec:conclusion}).

\newpage

\subsection{Data rebinning and smoothing}
\label{sec:rebinning_smoothing_gp}

Due to inherent instrumental and observational limitations, raw data from RXTE ASM and MAXI are often sparse and uneven, and contain points with varying noise levels. 
EDMD requires an evenly sampled time series that best represents the underlying dynamical system.

To address the problem of uneven data, we rebin the raw data onto a uniform time grid using inverse-variance weighting. 
Given $n$ raw data points in a bin, each with a flux value $x_{i}$ and uncertainty $\Delta_{i}$, the weighted mean flux and its uncertainty are
\begin{equation}
\bar{x} = \frac{\sum_{i}^{n} w_{i}x_{i}}{\sum_{i}w_{i}}, \quad
\sigma_{\bar{x}} = \frac{1}{\sqrt{\sum_{i}w_{i}}}, \quad
\text{where } w_{i} = \frac{1}{\Delta_{i}^{2}}.
\end{equation}
The rebinned flux is then standardized to zero mean and unit variance, with uncertainties scaled by the same factor.

However, as large gaps still exist in the rebinned data, we use Gaussian processes (GP) to regress the data onto an even grid, interpolating through gaps and assigning appropriate uncertainties.
We divide the light curve into $W$ overlapping windows and fit each with a GP using a radial basis function kernel. Let $f_w(t)$ and $\sigma_w(t)$ denote the GP mean prediction and uncertainty from window $w$, respectively. Multiple window predictions are combined using weights $\pi_w(t)$ that incorporate both Gaussian decay from the window center (to ensure smooth transitions) and inverse-variance weighting. The smoothed flux prediction and $1\sigma$ uncertainty at time $t$ are then

\begin{equation}
\hat{x}(t) = \frac{\sum_{w=1}^W \pi_w(t) f_w(t)}{\sum_{w=1}^W \pi_w(t)}
\end{equation}
\begin{equation}
\text{SD}(\hat{x}(t)) = \frac{\sum_{w=1}^W \pi_w(t) \sigma_w(t)}{\sum_{w=1}^W \pi_w(t)}.
\end{equation}
The standard deviations are combined linearly rather than in quadrature because the errors of the overlapping windows are strongly correlated (see Appendix \ref{sec:data_processing_appendix}).

Note that the GP-smoothing step, which combines measurement errors from different observations, leads to autocorrelations in the errors on the final data.
Full details on window sizes, kernel parameters, and the weighting formula are provided in Appendix~\ref{sec:data_processing_appendix}.
Furthermore, the smoothed flux $\hat{x}_{t}$ is calculated from both past and future data points, which means that the GP-smoothed light curve has a large uncertainty at the last data point of the time series, since there is no future data beyond this point.

To further improve the data quality, we employ an iterative approach to remove outliers.
A data point is excluded when its standardized residual relative to the smoothed light curve exceeds a $4\sigma$ threshold.
After all outliers are removed, the data are refitted with a second GP to produce the final light curve. This light curve is finally used for training $\mathbf{K}$.

The combined dataset contains 97976 measurements. Rebinning onto a uniform grid yields 4345 binned points, of which 22 ($0.5\%$) are removed as outliers; the final GP-smoothed light curve is evaluated on a uniform grid of 5000 time steps.
For more details on data processing procedures, refer to Appendix \ref{sec:data_processing_appendix}.

We note that EDMD requires an evenly sampled, gap-free time series, so the processing described above (rebinning, outlier removal, and GP interpolation and smoothing) is necessary to produce usable input from telescope data. Nevertheless, this processing is sure to affect the estimated operator and its prediction performance. A detailed future study of data processing in simulation studies could be done to investigate these effects.

\subsection{Eigenfunction partitioning of 4U 1705-44}

\begin{figure*}
\centering
\includegraphics[width=0.9\linewidth]{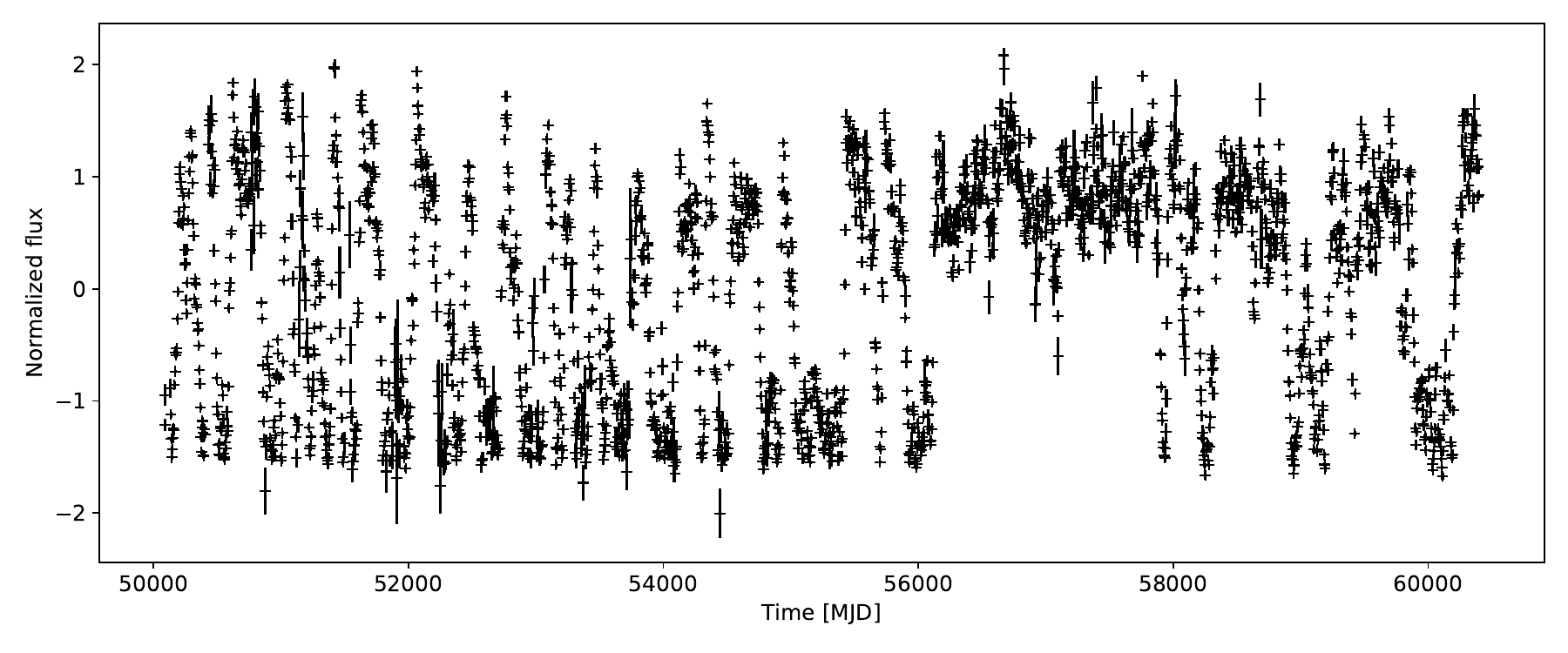}
\caption{
The combined light curve of 4U 1705-44 spanning $\sim$30 years (1996--2025), rebinned and standardized. Data processing details are provided in Appendix~\ref{sec:data_processing_appendix}.
}
\label{fig:full_light_curve_4u1705}
\end{figure*}

Figure \ref{fig:full_light_curve_4u1705} shows the full normalized light curve data of 4U 1705-44 observed by RXTE ASM and MAXI.
As pointed out by \cite{phillipson2018chaotic}, the light curve of 4U 1705-44 resembles a Duffing oscillator.
Indeed, we can see that the flux oscillates between two normalized states: a low-flux state at $\sim -1$ and a high-flux state at $\sim +1$.

\begin{figure}
\centering
\includegraphics[width=0.9\linewidth]
{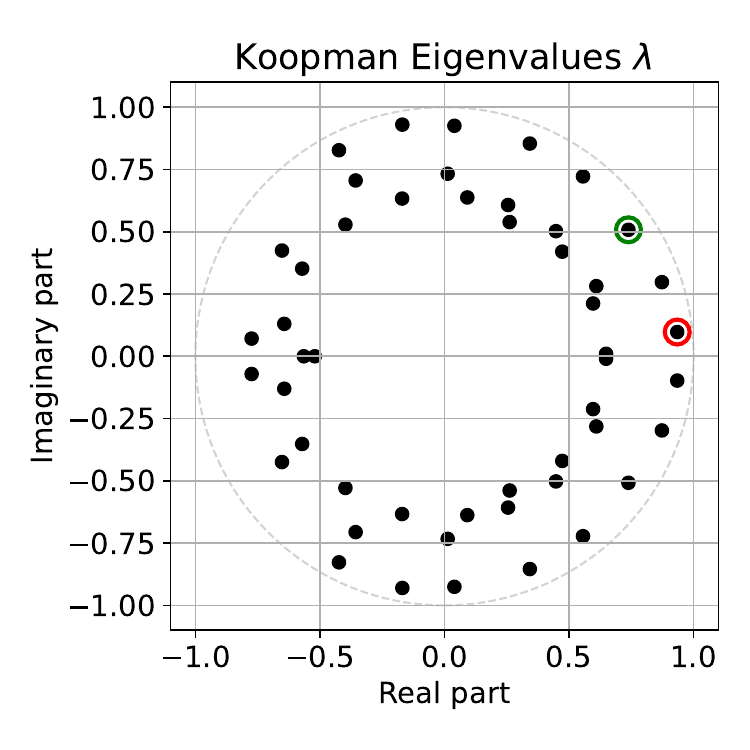}
\caption{The eigenvalues of the Koopman matrix $\mathbf{K}$ of 4U 1705-44 data, calculated with EDMD using a time-delay dictionary specified in Section \ref{sec:edmd} with length $d=50$. The $x$-axis values are the real parts of the eigenvalues, and the $y$-axis values are the imaginary parts of the eigenvalues.
The red circle highlights one of the eigenvalues with a small phase angle $\theta$, and the green circle highlights one of the eigenvalues with a larger phase angle $\theta$.
}
\label{fig:eigenvalue_polar_plot_4u1705}
\end{figure}

Following the procedure in Section \ref{sec:partition}, we see that the slowest-varying eigenfunction of $\mathbf{K}$, calculated from the 4U 1705-44 light curve, is able to \textit{predictively partition} the light curve into low-flux and high-flux states. To show this, we first train $\mathbf{K}$ on the first portion of the 4U 1705-44 data (up to MJD $\sim$57000).
Figure \ref{fig:eigenvalue_polar_plot_4u1705} shows the eigenvalue spectrum of the discrete Koopman matrix $\mathbf{K}$ derived from this segment of the 4U 1705-44 light curve, with the eigenvalues governing slow and fast-varying dynamics circled. 

\begin{figure*}[h]
\raggedleft

\includegraphics[width=0.98\linewidth]
{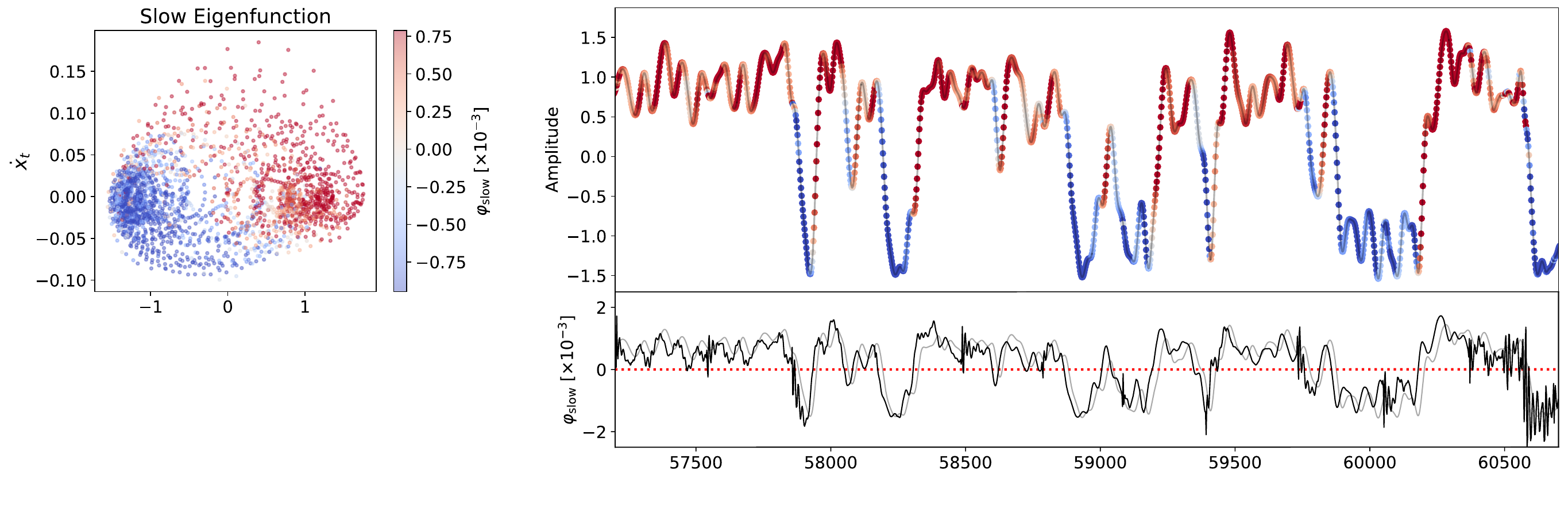}

\includegraphics[width=0.98\linewidth]
{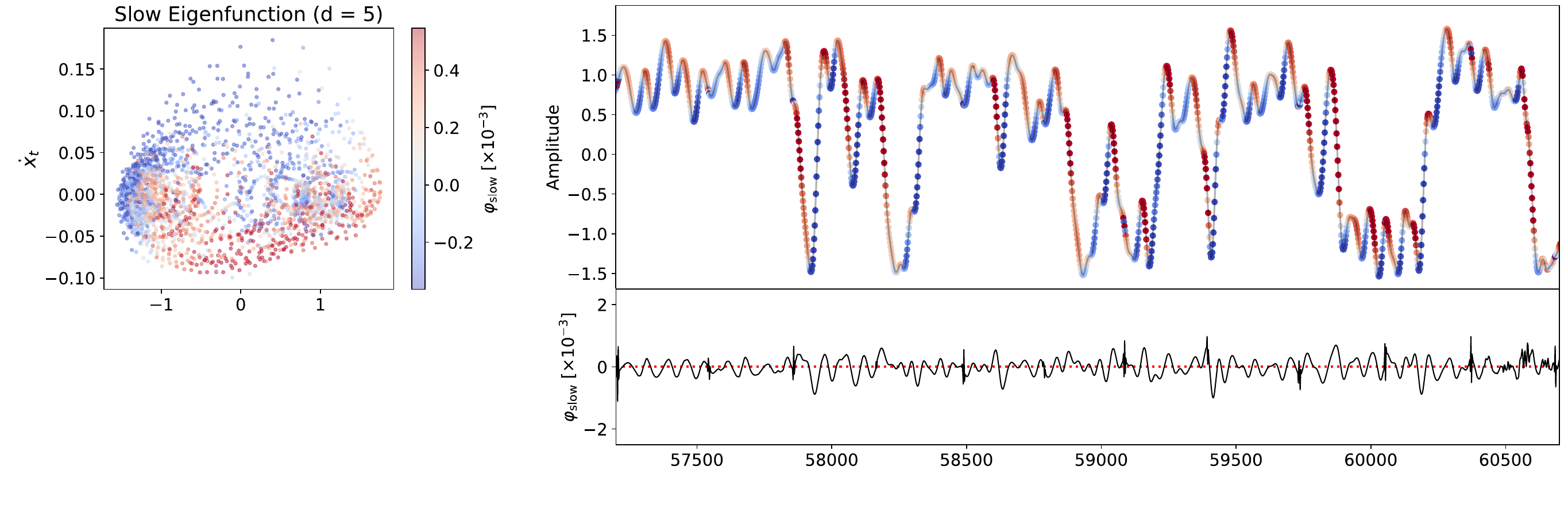}

\includegraphics[width=0.995\linewidth]
{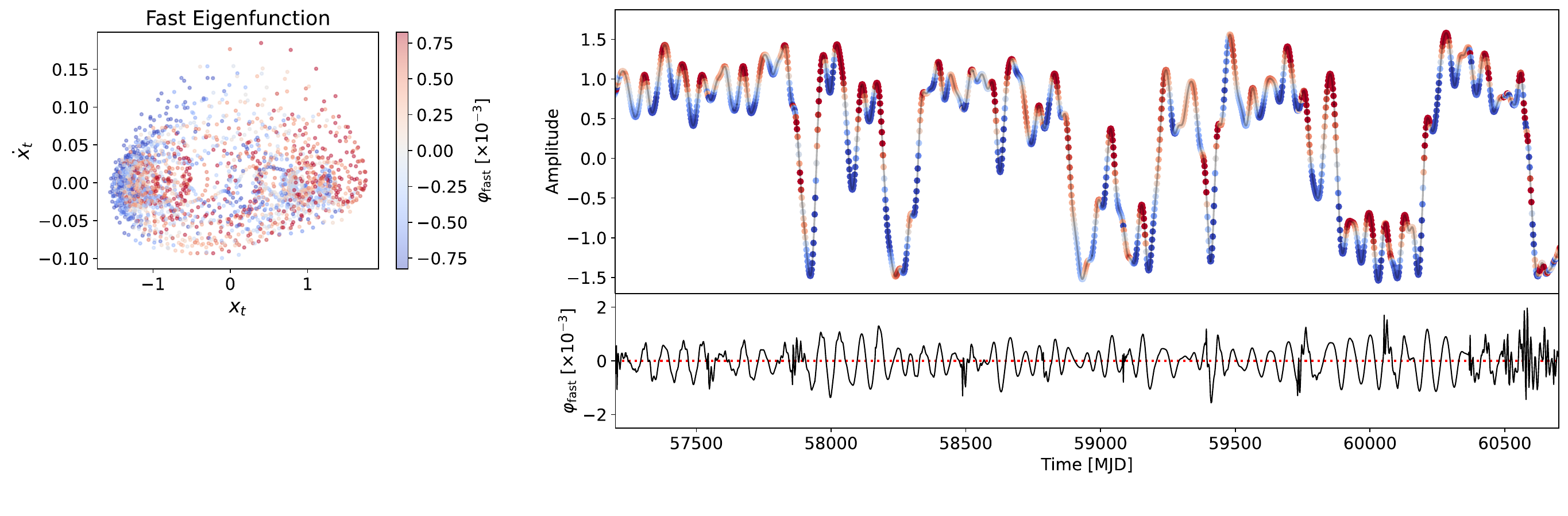}
\caption{
4U 1705-44 light curve and phase space colored by eigenfunction amplitude. $\mathbf{K}$ was calculated with EDMD using a time-delay dictionary specified in Section \ref{sec:edmd} with length $d=50$. The phase-space plots (left) show the flux $x_t$ against its time derivative $\dot{x}_t$, computed with backward finite differences of the light curve. The lower panel of each group plots the eigenfunction values $\varphi(t)$, with the dotted red line marking $\varphi = 0$.  \textbf{Top:} Slow-varying eigenfunction $\varphi_{\mathrm{slow}}(t)$; blue ($\varphi_{\mathrm{slow}}(t) < 0$) and red ($\varphi_{\mathrm{slow}}(t) > 0$) partition the light curve into low-flux and high-flux regimes, respectively. The light curve plot shows color switching before the flux reaches large variance, demonstrating predictive ability. The phase space plot (left) shows two distinct stable basins labeled by $\varphi_{\mathrm{slow}}$, with low-flux (blue) on the left and high-flux (red) on the right; as in Figure \ref{fig:eigenfunction_light_curve_duffing}, long ``tails'' of opposing color connect the basins, signifying color switching before the system begins switching to the opposite basin. We also overlaid the eigenfunction value plot (bottom right) with the light curve flux at each point (light gray) to demonstrate how sign changes of $\varphi_{\mathrm{slow}}(t)$ tend to precede the flux changes visible in the light curve. \textbf{Middle:} $\varphi_{\mathrm{slow}}(t)$ from $\mathbf{K}$ calculated with time-delay length $d=5$ observables. While state changes are predicted, false-positives (color changes) are rampant during stable sections (right), with colors also changing during each stable oscillation. As a result, the phase space plot (left) shows no labeling of stable regimes. \textbf{Bottom:} Fast-varying mode $\varphi_{\mathrm{fast}}(t)$ oscillates rapidly within each oscillation. The phase space plot also shows no clear regime labeling.
}
\label{fig:eigenfunction_light_curve_4u1705}
\end{figure*}

Then, we calculate eigenfunction values at each point of the ``test'' segment according to Eq. \ref{eq:eigenfunction_calculation}. 
As with the Duffing oscillator, the eigenfunction values partition state space into distinct regimes, as illustrated in the top panels of Figure \ref{fig:eigenfunction_light_curve_4u1705}.
The slow-varying eigenfunction $\varphi_{\mathrm{slow}}(t)$ corresponds to the long-timescale transitions between high-flux and low-flux states.

Crucially, as with the simulated Duffing data, sign changes in $\varphi_{\mathrm{slow}}(t)$ (color switches) \textit{precede} visible regime changes in the light curve flux. 
More importantly, the eigenfunction value switches sign days to weeks before the light curve exceeds the typical variance of stable oscillations, demonstrating that eigenfunctions hold significant power in detecting transitions in astrophysical data.
This is also evident in the phase-space plot (top left), where the eigenfunction colors clearly label two distinct basins on the left and right. These stable basins are connected by long ``tails'' of the opposing basin's color: the system's eigenfunction value changes before the system physically reaches the opposing basin.

We also demonstrate the importance of time-delay embedding as described in Section \ref{sec:edmd}. The middle panels of Figure \ref{fig:eigenfunction_light_curve_4u1705} show the slow-varying eigenfunction $\varphi_{\mathrm{slow}}(t)$ from a Koopman matrix $\mathbf{K}$ calculated using a very short $d=5$ time-delay basis rather than the $d=50$ basis used for the top panels. With only $d=5$ delay coordinates, each observable does not capture enough system history to determine whether flux increases are flares or stable oscillations. The eigenfunction $\varphi_{\mathrm{slow}}(t)$ therefore responds to both, producing significant false-positives during upwards trends in stable sections. With a longer delay length ($d=50$), each observable contains enough context to separate long and short timescales, and $\varphi_{\mathrm{slow}}(t)$ can predict flares with minimal false positives (top panel). 

Conversely, the fast-varying mode eigenfunction $\varphi_{\mathrm{fast}}(t)$ rapidly switches between red and blue in short timescales, labeling rapid oscillations rather than large-scale regime switches, and providing no usefulness for predicting longer-scale transitions. The phase space displays a homogeneous blend of colors: the basins are not clearly labeled.

\newpage

\subsection{Flux predictions of 4U 1705-44}

In addition to slow-varying eigenfunction values, we can make direct flux predictions by iteratively applying $\mathbf{K}$ to the most recent lifted observation $\vec{y}_{t_0}$, following the procedure in Section \ref{sec:predicting}. For each prediction point, we process and smooth only the data up to that point (Section \ref{sec:rebinning_smoothing_gp}), enforcing causality, and learn $\mathbf{K}$ from this processed data using the method described in Section \ref{sec:edmd}.

\begin{figure*}[h]
\centering
\includegraphics[width=0.49\linewidth]{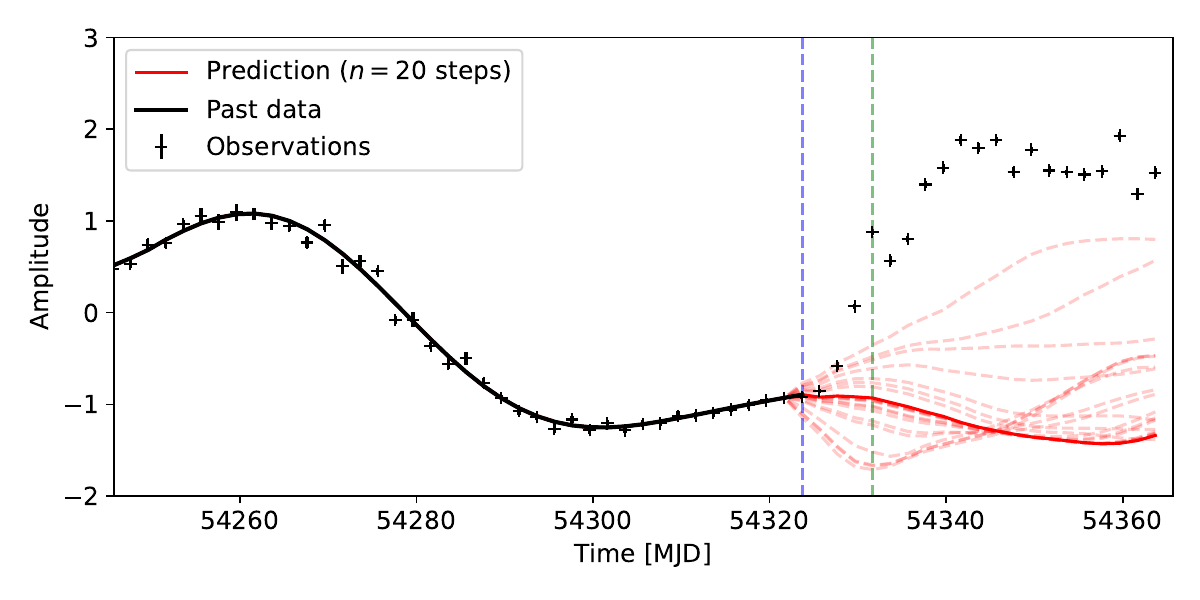}
\includegraphics[width=0.49\linewidth]{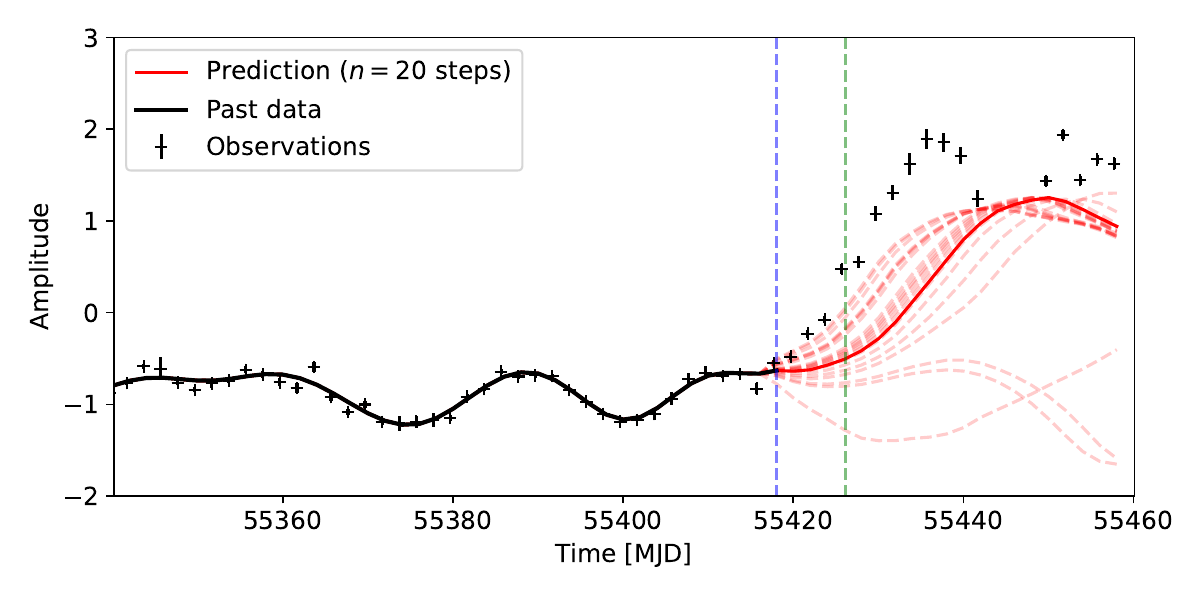}\\
\includegraphics[width=0.49\linewidth]{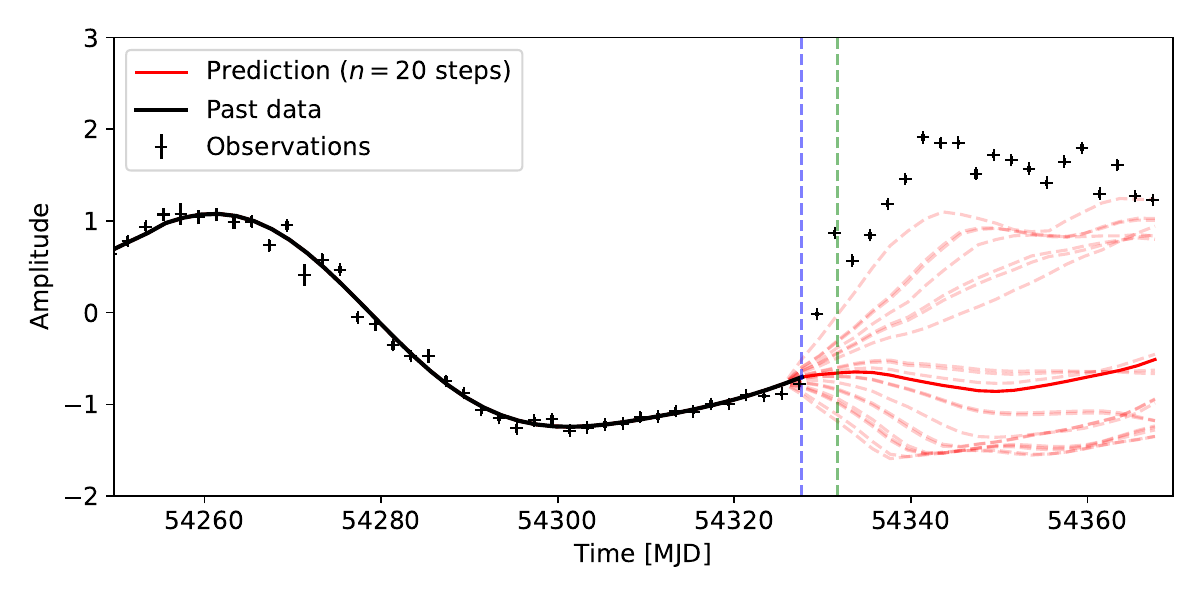}
\includegraphics[width=0.49\linewidth]{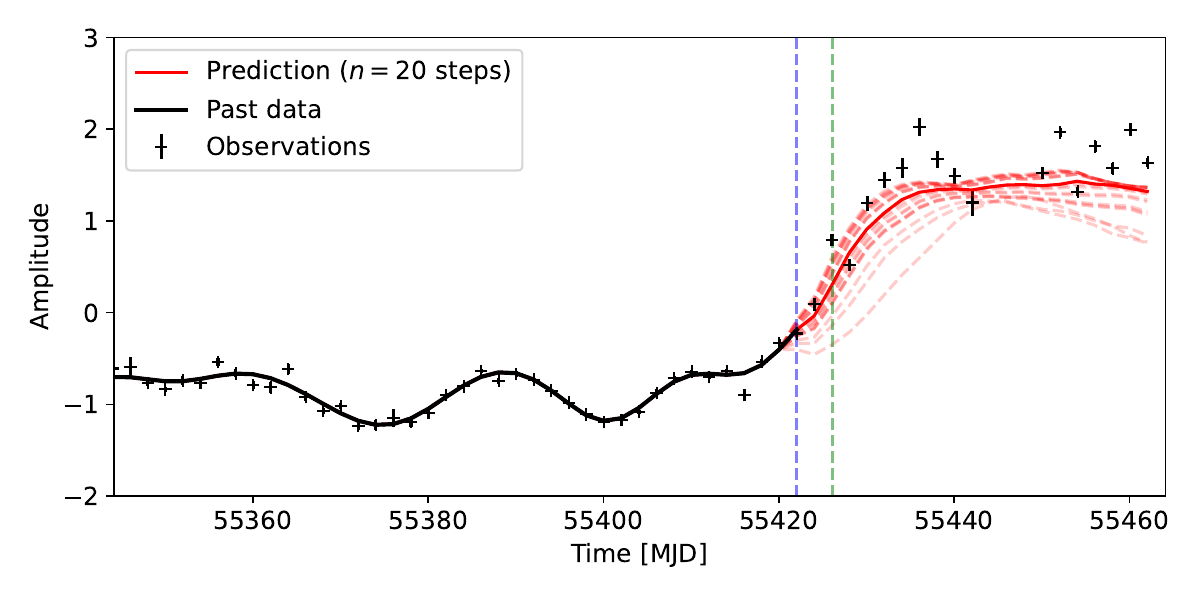}\\
\includegraphics[width=0.49\linewidth]{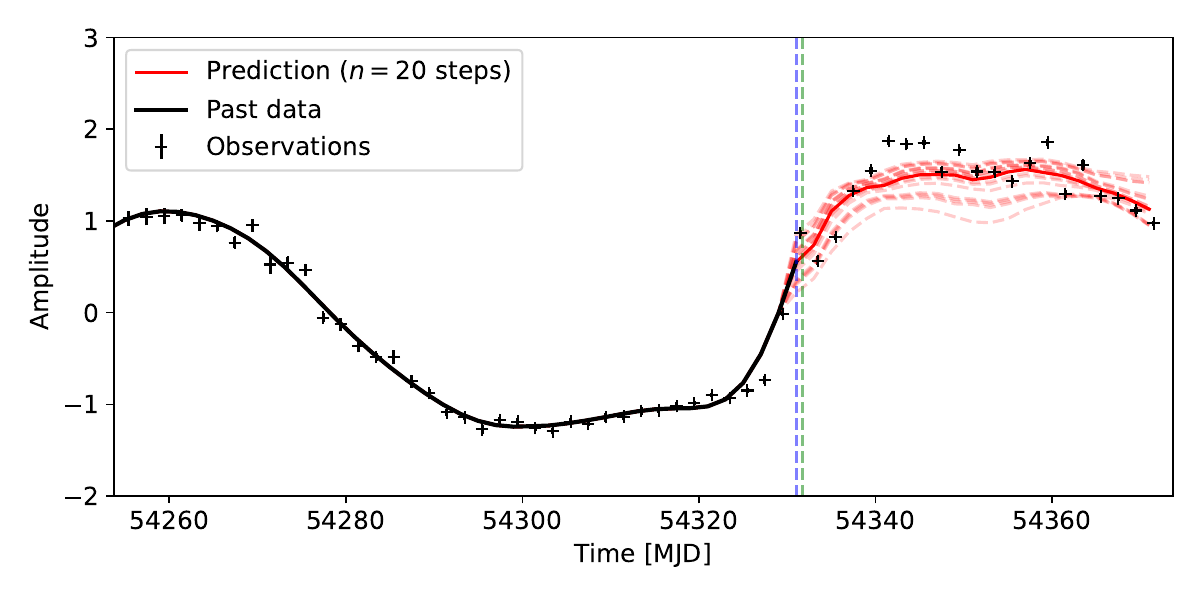}
\includegraphics[width=0.49\linewidth]{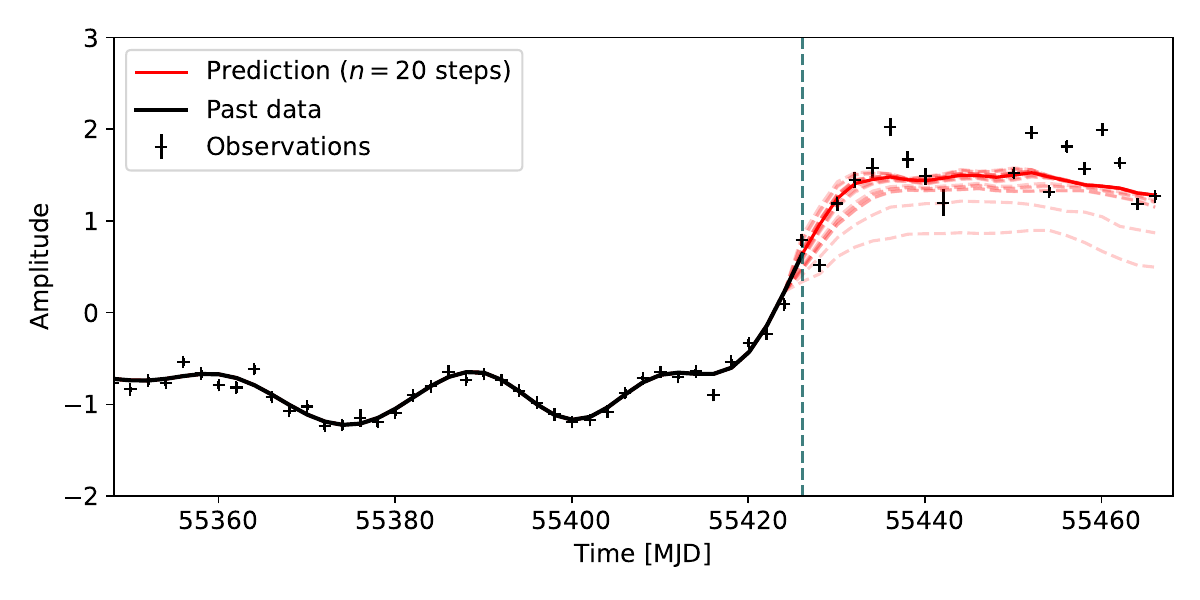} \\
\caption{
Predictions of 4U 1705-44 flux around two transition mid-points. \textbf{Left column:} Rapid-rising transition. \textbf{Right column:} Slower-rising transition. \textbf{Top to bottom:} Past data (black) ending 8, 4, and 0 days before the transition mid-point. Blue vertical lines indicate the end of past data; green vertical lines indicate transition mid-points (moments of largest flux variance within a 20-day window). Red solid curves show nominal predictions; red dashed curves show 20 alternative predictions generated by varying the last flux point, serving as uncertainty indicators. Note that predictions beginning earlier (top rows), where light curve curvature is large, show greater deviation from true future data than predictions beginning at the transition mid-point (bottom row).
}
\label{fig:predict_light_curve_4u1705}
\end{figure*}

Figure \ref{fig:predict_light_curve_4u1705} shows the predictions for the future time-series data of 4U 1705-44 around two transition mid-points between low-flux and high-flux states.
These transition mid-points are identified as the moments with the largest flux variances within a 20-day window.
The key finding is that prediction accuracy depends strongly on the curvature of the light curve at the prediction start point, as expected from Equation \ref{eq:measurement_noise}.
Predictions beginning at the onset of a rapid transition, where the curvature is large, do not agree well with the true future data.
However, predictions beginning at the mid-point of a transition, where the curvature is smaller, are significantly more accurate.
Similarly, slower-rising transitions with smaller curvature allow accurate predictions to be made earlier (e.g., 4 days before the transition mid-point).

We note that Equation \ref{eq:measurement_noise} describes only the error induced by measurement noise. Even without noise, predictions contain errors because the finite dictionary only approximately spans a Koopman-invariant subspace, and this approximation error is also expected to be largest where the dynamics change most rapidly. The observed curvature dependence reflects both effects.

\begin{figure*}[h]
\centering
\includegraphics[width=0.49\linewidth]{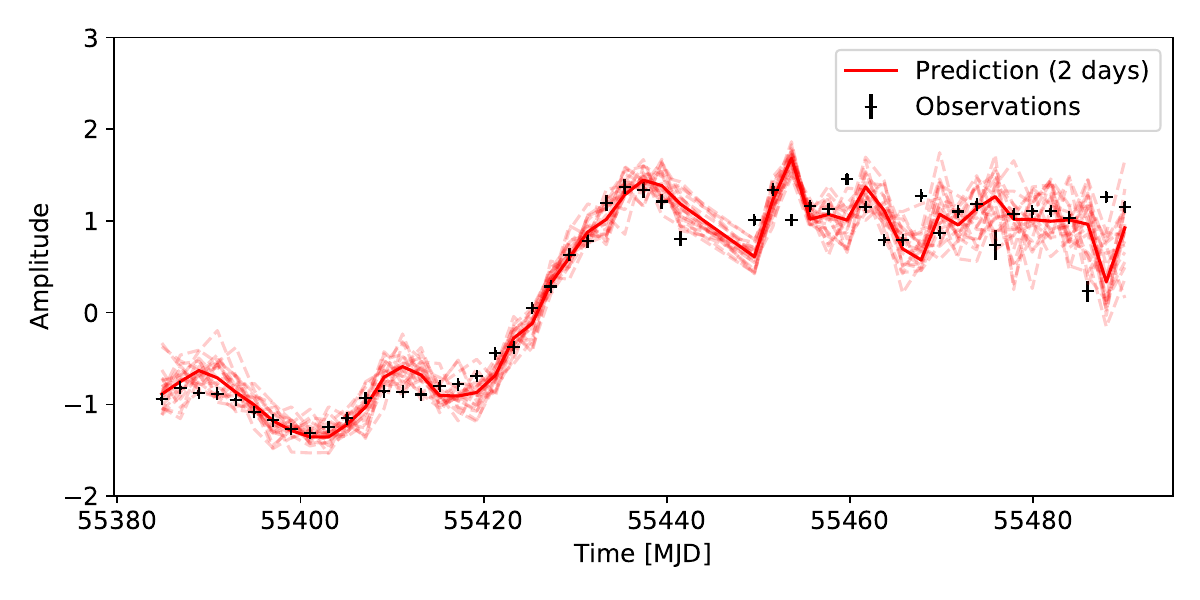}
\includegraphics[width=0.49\linewidth]{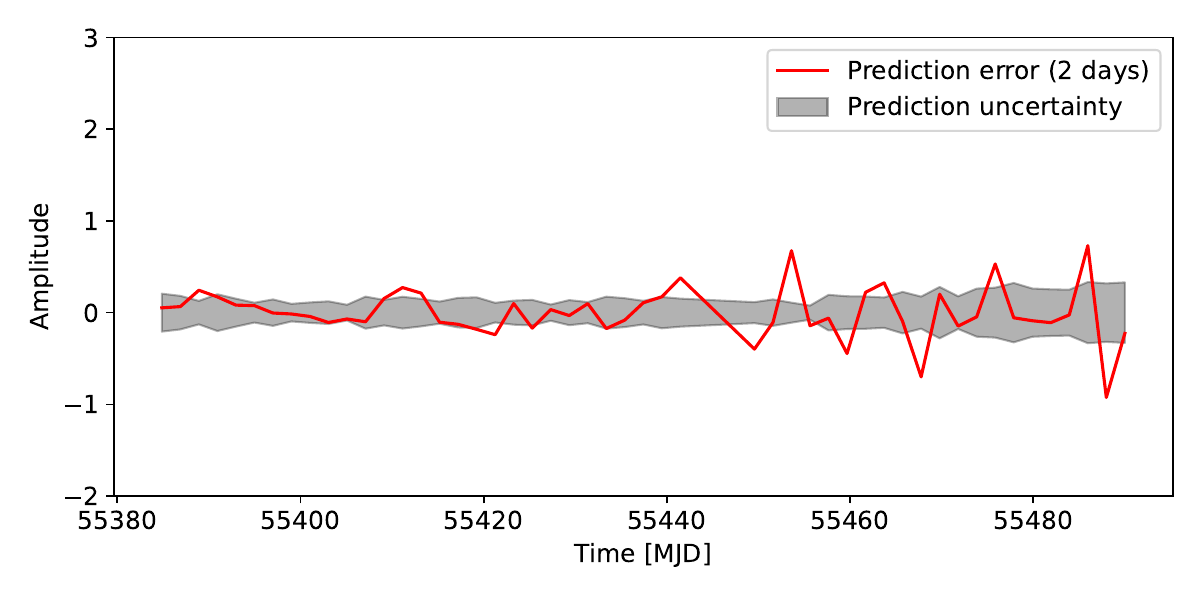}\\
\includegraphics[width=0.49\linewidth]{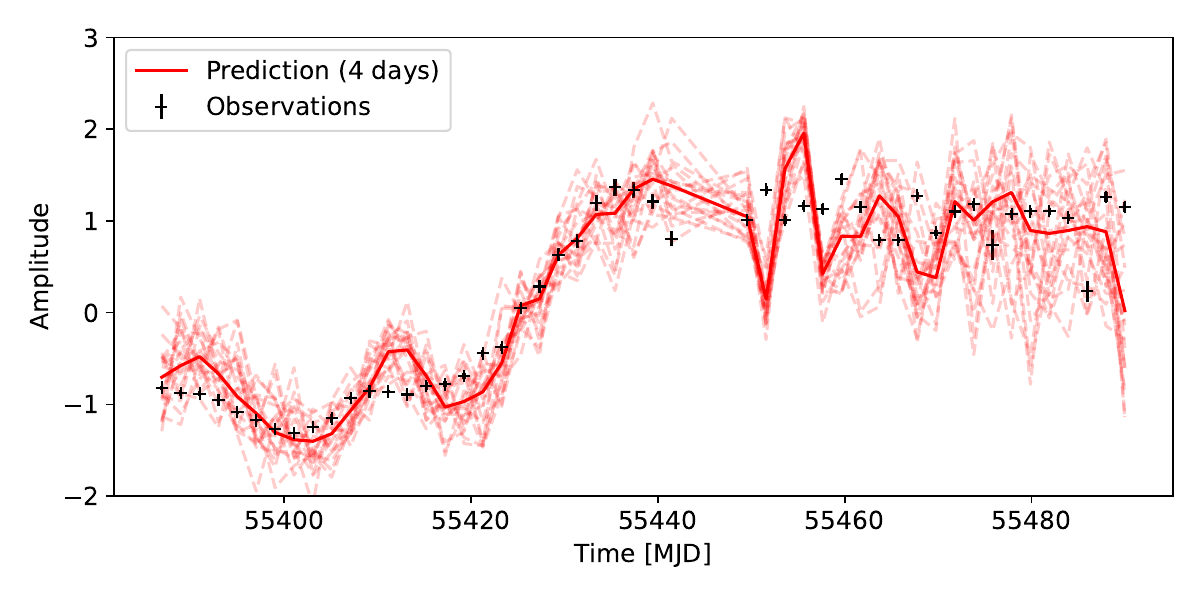}
\includegraphics[width=0.49\linewidth]{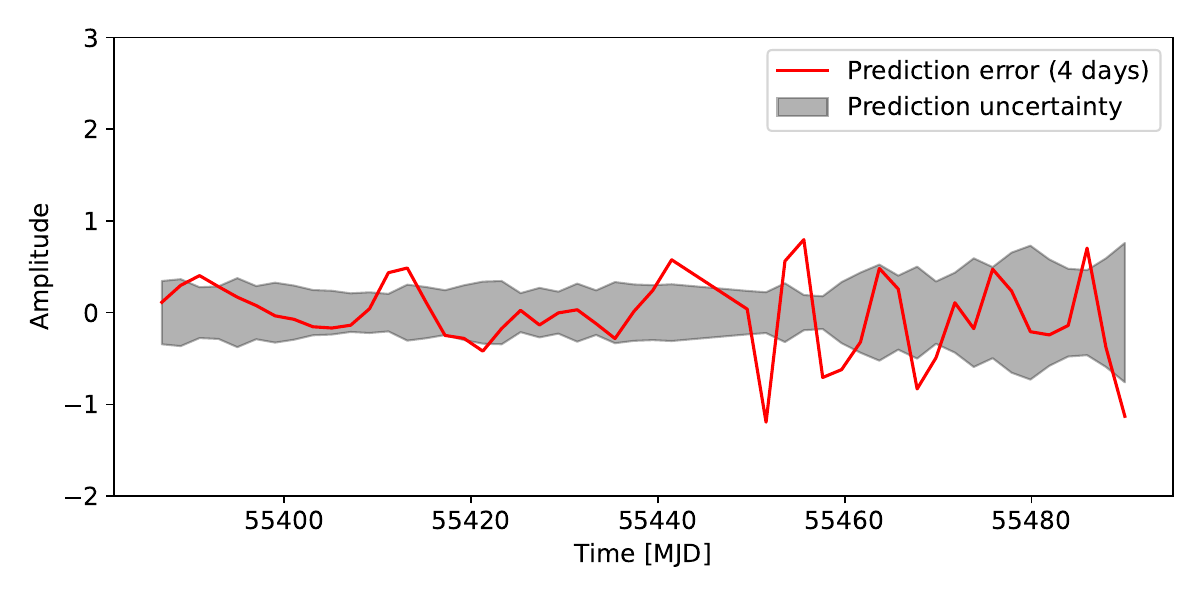}\\
\includegraphics[width=0.49\linewidth]{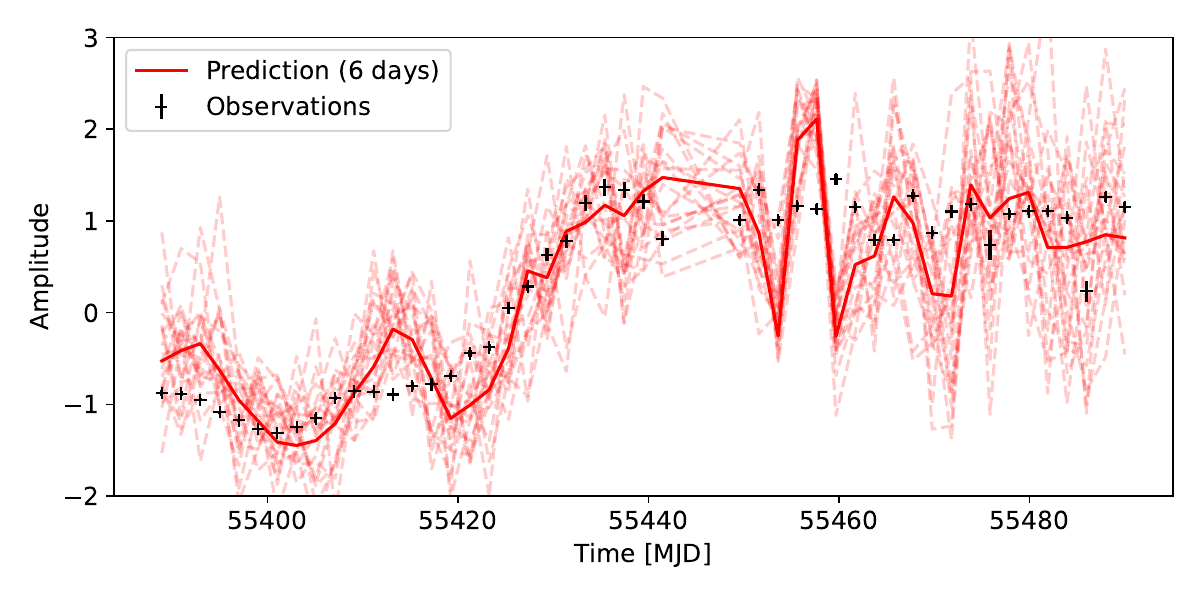}
\includegraphics[width=0.49\linewidth]{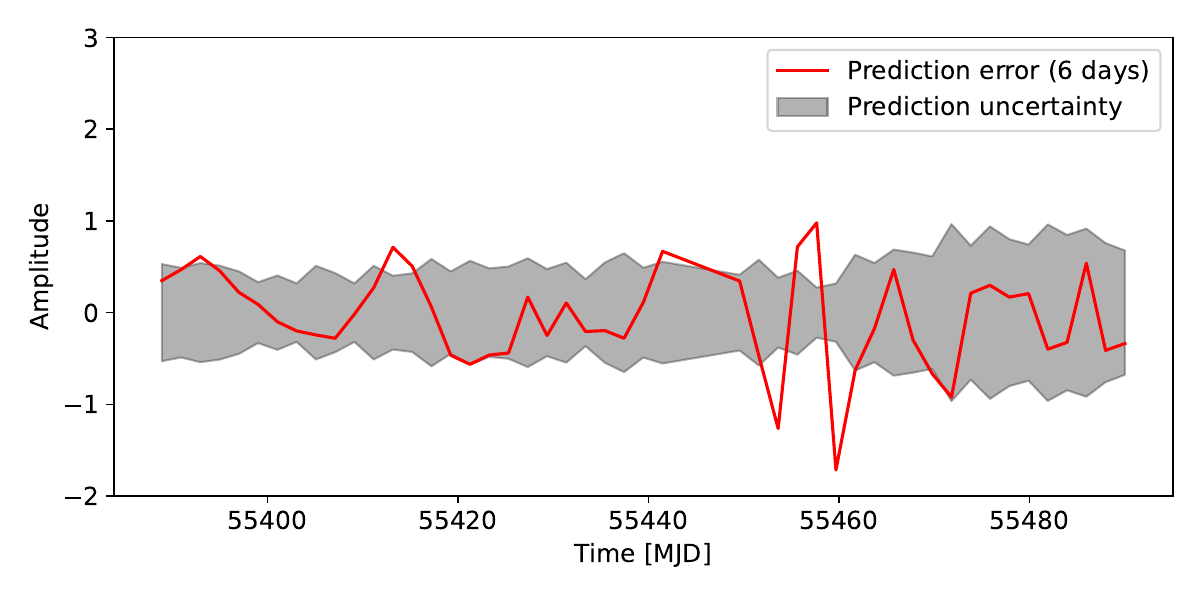}\\
\includegraphics[width=0.49\linewidth]{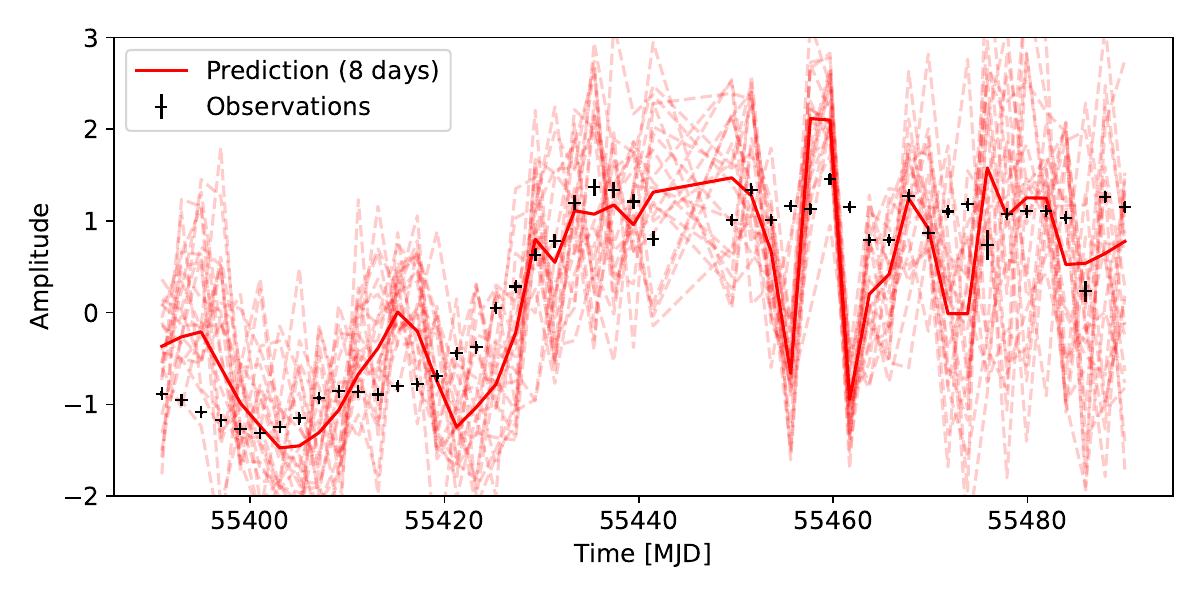}
\includegraphics[width=0.49\linewidth]{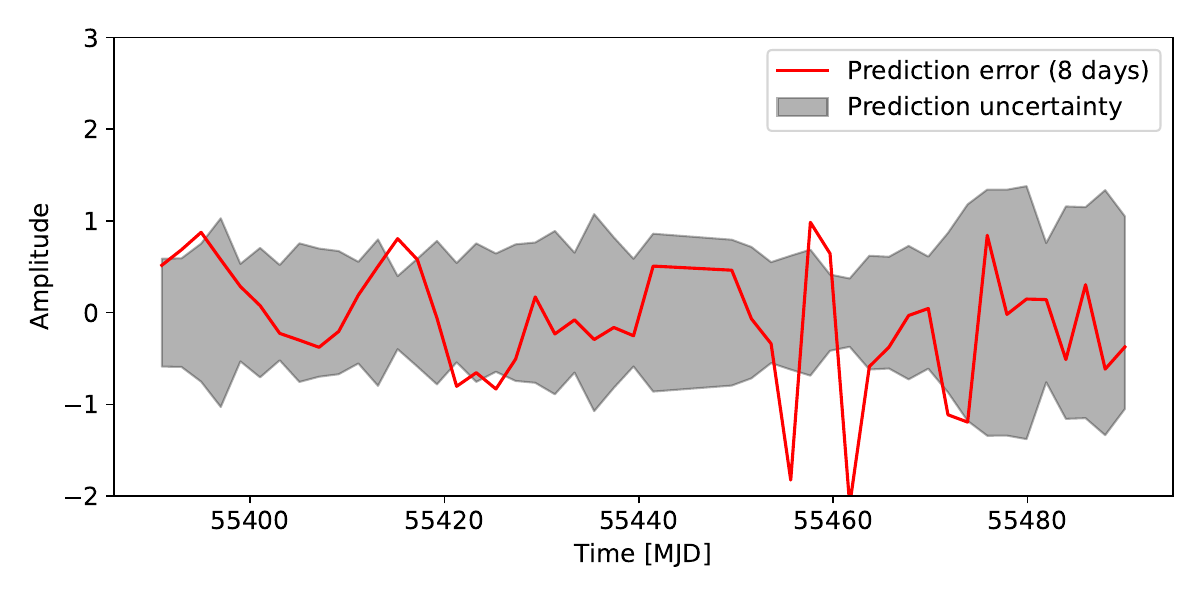}\\
\caption{
Prediction errors and standard deviations of 4U 1705-44 flux around the transition mid-point on MJD 55430. 
\textbf{Left column:} Various prediction horizons (2, 4, 6, 8 days) represented by the red curves and the observed flux represented by the black points. Red solid curves show nominal predictions; red dashed curves show 20 alternative predictions generated by varying the last flux point, serving as uncertainty indicators. 
\textbf{Right column:} Prediction standard deviations computed from the spread of the red dashed curves in the left panels are represented by the gray bands, and the prediction errors (the differences between the nominal predictions and the real observations) are represented by the red curve.
We use 20 trajectories so that individual predictions remain visible in the left panels; standard deviations estimated from 20 realizations are approximate.
}
\label{fig:predict_light_curve_4u1705_variance}
\end{figure*}

To test the stability of the predictions, we generate trajectories from 20 random flux values drawn from a normal distribution with mean and variance from the final GP-smoothed point in the past data. 
Wider spreads indicate greater uncertainty in predictions, either due to measurement noise in the last data point or to unknown future dynamics.
As the prediction horizon (number of forward prediction steps) increases, the prediction uncertainty is expected to increase.
This is demonstrated in Figure \ref{fig:predict_light_curve_4u1705_variance}, where the prediction uncertainties and errors are shown for various horizons (2, 4, 6, 8 days).
The prediction uncertainties are computed as the standard deviations of the 20 alternative predictions, and we can see that the uncertainties agree well with the actual prediction errors (the differences between the nominal prediction and the observed fluxes).
Appendix \ref{sec:longterm_predictions_appendix} further showcases predictions starting at 20 transition points between high-flux and low-flux states in the range from MJD 54000 to MJD 60400 (Figures \ref{fig:flux_prediction_points_4u1705}--\ref{fig:predict_light_curve_4u1705_longterm2}).

From these predictions, we see that reliable forecasting of the future light curve requires small measurement noise and high sampling rates to reduce its curvature.
However, a higher sampling rate also means larger measurement noise, and an optimal trade-off between the two is needed.
Several Koopman operator-based algorithms have been developed specifically to better handle measurement noise; we discuss these as future work in Section \ref{sec:conclusion}.

\section{Conclusion}
\label{sec:conclusion}

In this paper, we demonstrate that extended dynamic mode decomposition (EDMD), a data-driven approximation of Koopman operator theory, is an effective tool for studying time-series data from X-ray binaries.
Unlike Fourier methods, which decompose a signal into sinusoidal frequency components, or SSA, which focuses on identifying patterns and denoising a time-series signal, EDMD focuses on identifying the governing operator of the system.
By lifting the original state space (e.g., X-ray flux) to a high-dimensional observable space, Koopman operator theory transforms a nonlinear dynamical system into a linear one, using Koopman eigenvalues, eigenfunctions, and modes to form the Koopman mode decomposition, which allows both interpretation and prediction of dynamics.

Using Duffing oscillator simulations, we showed that EDMD can learn the timescales of variability and corresponding dynamical structures.
Crucially, we demonstrated that Koopman eigenfunctions can partition state space into distinct regimes, with sign changes of slow-varying eigenfunctions \textit{preceding} visible state transitions by days to weeks, indicating predictive power for regime changes.
We also connected Koopman eigenvalues to quasi-periodic oscillations, deriving that each eigenfunction contributes a Lorentzian peak to the power spectral density at the frequency given by the imaginary part of its eigenvalue.
Process noise induces damping to Koopman modes, broadening QPO spectral widths and shortening coherence times, providing a dynamical systems interpretation of QPO phenomenology.
We further demonstrated that the Koopman operator can advance past time-series data into the future, and we analytically showed how measurement noise corrupts the Koopman operator and degrades prediction accuracy, with errors scaling with noise variance and light curve curvature.
Finally, we applied these methods to $\sim$30 years of combined RXTE ASM and MAXI observations of 4U~1705-44, demonstrating both predictive state partitioning and direct flux forecasting on real astrophysical data.
Beyond prediction, our method is intrinsically interpretable, as timescale separation, dynamical reading of QPO widths, and eigenfunction partitioning are derived directly from the learned spectral objects. This makes our application of EDMD an instance of interpretable machine learning for scientific discovery \citep{allen2024interpretable}, in which understanding and prediction come from the same learned decomposition.

This work opens several directions for future research:
\begin{enumerate}
\item Uncertainty quantification: Developing rigorous uncertainty estimates for Koopman-based predictions.
\item Physics-informed dictionaries: Designing dictionary functions motivated by accretion physics (e.g., disk instability modes, corona dynamics) that improve predictive power and interpretability beyond generic Fourier and time-delay bases. 
\item Physical interpretation of eigenmodes: Connecting Koopman eigenmodes to the underlying physical processes driving variability, establishing a new tool for studying nonlinearity in X-ray binaries beyond conventional methods such as QPO analysis and the rms-flux relation.
\item Multi-wavelength integration: Incorporating simultaneous radio, optical, X-ray, and $\gamma$-ray observations into EDMD could construct a more complete view of the coupled dynamics across emission regions and help disentangle cause from effect during state transitions.
\item Noise-robust algorithms: Several Koopman-based methods have been developed specifically to handle measurement noise, including Total Least Squares DMD \citep{hemati2017biasing}, Optimal DMD \citep{askham2018variable}, Subspace DMD \citep{takeishi2017subspace}, Kernel DMD \citep{williams2014kernel}, and Kalman Filter--based approaches \citep{nonomura2019extended}. In addition, new methods like Symmetric Subspace Decomposition \citep{haseli_symmetric_subspace_2022} and Recursive Forward-Backward EDMD \citep{haseli_recursive_2025} offer guarantees for Koopman invariance. Applying these to astrophysical light curves may significantly improve the capture of dynamics and, therefore, prediction performance in the presence of noisy telescope data.
\item Verification of assumptions: The present analysis assumes that the delay embedding satisfies the conditions of the Takens theorem, that the finite matrix $\mathbf{K}$ is an adequate approximation of $\mathcal{K}$, that the underlying system is stationary, and that measurement errors are uncorrelated and homoscedastic. None of these are guaranteed for the data analyzed here. Verifying the embedding conditions, quantifying the operator approximation error \citep[e.g., via residual-based diagnostics;][]{colbrook2023residual,conradie2026trustworthy}, and extending the framework to non-stationary time series and to correlated, heteroscedastic measurement errors are necessary steps toward a statistically rigorous Koopman pipeline for astronomical data.
\end{enumerate}

This paper presents the first application of Koopman operator theory to X-ray binary systems, demonstrated on a single well-studied source as a pilot study. Future work will apply this framework to higher-cadence NICER observations, as well as to Fermi-LAT gamma-ray light curves of blazars and microquasars, testing its generalizability across wavelengths and source classes.

\section*{Acknowledgment}
We would like to thank the anonymous referee for critical and thoughtful comments on early versions of the paper. 
RS was supported by NSF grant PHY-2110497 at Barnard College. This work was additionally supported by NICER Grant 80NSSC26K0949.

\appendix

\section{Effect of measurement noise}
\label{sec:measurement_noise}

In this section, we discuss how measurement noise perturbs the Koopman operator and leads to biased errors in the prediction of the light curve.
Interested readers can find a more detailed discussion on how measurement noise affects the Koopman operator in \cite{dawson2016characterizing,wanner2022robust}.

Let $\vec{x}_{t+1} = \vec{F}(\vec{x}_{t})$, where $\vec{x}_{t} \in \mathbb{R}^{d}$ is the observed system state (i.e., the time-delay vector of X-ray flux), and $\vec{F}$ is the propagator that evolves $\vec{x}_{t}$.
The observed state is corrupted by the measurement noise, $\vec{z}_{t} = \vec{x}_{t} + \vec{\epsilon}_{t}$, where $\vec{z}_{t}$ is the noisy observation and $\vec{\epsilon}_{t} \sim \mathcal{N}(\vec{0},\,\sigma^{2}\mathbf{I}_{d})$ is Gaussian noise, i.e., the errors on the $d$ components of $\vec{x}_{t}$ are assumed to be identically distributed and mutually uncorrelated for simplicity of the discussion.

We further assume that $\vec{\epsilon}_{t}$ is uncorrelated in time. In practice, the GP smoothing of Section \ref{sec:rebinning_smoothing_gp} combines measurement errors from neighboring observations and induces autocorrelation in the effective errors of the final light curve, so the treatment in this appendix is a simplified model; correlated and heteroscedastic errors are left to future work (Section \ref{sec:conclusion}).

Let the dictionary $\vec{\psi}(\vec{x}_{t}): \mathbb{R}^{d} \rightarrow \mathbb{R}^{p}$ transform $\vec{x}_{t}$ to the high-dimensional lifted state.
Without measurement noise, the evolution of the lifted state is governed by the true Koopman operator
\begin{equation}
\mathcal{K} \vec{\psi}(\vec{x}_{t}) = \vec{\psi}(\vec{x}_{t+1})=\vec{\psi}\left(\vec{F}(\vec{x}_{t})\right).
\end{equation}
In the case where the lifted state is corrupted by noise,
\begin{equation}
\tilde{\mathcal{K}} \vec{\psi}(\vec{z}_{t}) = \mathbb{E} \left[ \vec{\psi}(\vec{z}_{t+1}) \right],
\end{equation}
where $\tilde{\mathcal{K}}=\mathcal{K}+\delta\mathcal{K}$ is the noisy Koopman operator.
If we replace $\vec{z}_{t+1}$ with $\vec{z}_{t+1}=\vec{F}(\vec{x}_{t})+\vec{\epsilon}_{t+1}=\vec{F}(\vec{z}_{t}-\vec{\epsilon}_{t})+\vec{\epsilon}_{t+1}$, the noisy Koopman operator equation becomes
\begin{equation}
\tilde{\mathcal{K}} \vec{\psi}(\vec{z}_{t}) = \mathbb{E} \left[
\vec{\psi}\left( \vec{F}(\vec{z}_{t}-\vec{\epsilon}_{t})+\vec{\epsilon}_{t+1} \right)
\right],
\end{equation}
and we want to Taylor expand $\vec{\psi}\left( \vec{F}(\vec{z}_{t}-\vec{\epsilon}_{t})+\vec{\epsilon}_{t+1} \right)$ to see how the measurement noise $\vec{\epsilon}$ changes the Koopman operator.

First, we fix $\vec{\epsilon}_{t}$ and expand $\vec{\psi}$ in terms of $\vec{\epsilon}_{t+1}$.
Taylor expansion (up to the second order) gives
\begin{equation}
\psi_{i}(\vec{z}_{t+1})
=
\psi_{i}\left(\vec{F}(\vec{z}_{t}-\vec{\epsilon}_{t})\right)
+ \sum_{j} J^{\psi}_{ij}\left(\vec{F}(\vec{z}_{t}-\vec{\epsilon}_{t})\right) \epsilon_{t+1,j}
+ \sum_{jk} \frac{1}{2} H^{\psi}_{ijk}\left(\vec{F}(\vec{z}_{t}-\vec{\epsilon}_{t})\right) \epsilon_{t+1,j} \epsilon_{t+1,k},
\end{equation}
where $J^{\psi}_{ij}(\vec{x}) = \partial \psi_{i}(\vec{x}) / \partial x_{j}$ is the Jacobian matrix, and $H^{\psi}_{ijk}(\vec{x}) = \partial^{2} \psi_{i}(\vec{x}) / \partial x_{j}\partial x_{k}$ is a tensor of Hessian matrices that measure the curvature of the dictionary functions.
Using the fact that $\mathbb{E}[\epsilon_{i}]=0$ and $\mathbb{E}[\epsilon_{i}\epsilon_{j}]=\sigma^{2}\delta_{ij}$, we find the expectation
\begin{equation}
\mathbb{E}\left[\psi_{i}(\vec{z}_{t+1})\right]
=
\psi_{i}\left(\vec{F}(\vec{z}_{t}-\vec{\epsilon}_{t})\right)
+ \frac{1}{2}\sigma^{2} \sum_{jk} \delta_{jk} H^{\psi}_{ijk}\left(\vec{F}(\vec{z}_{t}-\vec{\epsilon}_{t})\right).
\label{eq:expect_psi_1}
\end{equation}
Then we continue to expand $\vec{\psi}$ in terms of $\vec{\epsilon}_{t}$,
\begin{align}
\psi_{i}\left(\vec{F}(\vec{z}_{t}-\vec{\epsilon}_{t})\right) ={} &
\psi_{i}\left(\vec{F}(\vec{z}_{t})\right)
- \sum_{jk} J^{\psi}_{ij}\left(\vec{F}(\vec{z}_{t})\right) J^{F}_{jk}(\vec{z}_{t}) \epsilon_{t,k}
\nonumber \\
&+ \frac{1}{2} \sum_{jk} H^{\psi}_{ijk}\left(\vec{F}(\vec{z}_{t})\right) \sum_{uv} J^{F}_{ju}(\vec{z}_{t}) J^{F}_{kv}(\vec{z}_{t}) \epsilon_{t,u} \epsilon_{t,v}
\nonumber \\
&+ \frac{1}{2} \sum_{j} J^{\psi}_{ij}\left(\vec{F}(\vec{z}_{t})\right) \sum_{uv} H^{F}_{juv}(\vec{z}_{t}) \epsilon_{t,u} \epsilon_{t,v},
\label{eq:expand_psi}
\end{align}
where $J^{F}_{ju}(\vec{x}) = \partial F_{j}(\vec{x}) / \partial x_{u}$ is the Jacobian of the propagator, and $H^{F}_{juv}(\vec{x}) = \partial^{2} F_{j}(\vec{x}) / \partial x_{u}\partial x_{v}$ is its Hessian, which measures how strongly the trajectory bends.

Combining Equation \ref{eq:expect_psi_1} and Equation \ref{eq:expand_psi}, we have
\begin{align}
\mathbb{E}\left[\psi_{i}(\vec{z}_{t+1})\right] &=
\psi_{i}\left(\vec{F}(\vec{z}_{t})\right)
+\frac{1}{2}\sigma^{2} \sum_{jk} \delta_{jk} H^{\psi}_{ijk}\left(\vec{F}(\vec{z}_{t})\right)
+\frac{1}{2}\sigma^{2} \sum_{jk} H^{\psi}_{ijk}\left(\vec{F}(\vec{z}_{t})\right) \sum_{uv} \delta_{uv} J^{F}_{ju}(\vec{z}_{t}) J^{F}_{kv}(\vec{z}_{t}) \nonumber \\
&\quad +\frac{1}{2}\sigma^{2} \sum_{j} J^{\psi}_{ij}\left(\vec{F}(\vec{z}_{t})\right) \sum_{uv} \delta_{uv} H^{F}_{juv}(\vec{z}_{t}) \nonumber \\
&= \left(\mathcal{K}+\delta\mathcal{K}\right) \psi_{i}(\vec{z}_{t}).
\end{align}
We can see that the noisy Koopman operator perturbation $\delta \mathcal{K}$ is scaled by
the noise variance $\sigma^{2}$, and the error is amplified at large curvature of the dictionary functions $\vec{\psi}$ by $H^{\psi}$, at strong local stretching of the dynamics by $J^{F}$, and at large curvature of the trajectory by $H^{F}$.

Furthermore, let the probability measure $\nu(\vec{x})$ be the probability of the system visiting the state $\vec{x}$, and if we ask that $\vec{F}$ is invertible and preserves the probability measure, $\nu(\vec{x})=\nu\left(\vec{F}(\vec{x})\right)$, then $\mathcal{K}$ is unitary and has all eigenvalues $\vert \lambda \vert =1$.
To see that this is true, we compute the $L^{2}$-norm
\begin{align}
\left\| \mathcal{K} \psi_{i}(\vec{x}) \right\|_{L^{2}}^{2}
&=
\left\| \psi_{i}\left( \vec{F}(\vec{x}) \right) \right\|_{L^{2}}^{2} \\
&=
\int \left\vert \psi_{i}\left( \vec{F}(\vec{x}) \right) \right\vert^{2} d \nu(\vec{x}) \\
&=
\int \left\vert \psi_{i}\left( \vec{F}(\vec{x}) \right) \right\vert^{2} d \nu\left( \vec{F}(\vec{x}) \right) \\
&=
\int \left\vert \psi_{i}(\vec{y}) \right\vert^{2} d \nu(\vec{y}) \\
&=
\left\| \psi_{i}(\vec{x}) \right\|_{L^{2}}^{2},
\end{align}
which means that all eigenvalues of $\mathcal{K}$ lie on the unit circle \citep{budivsic2012applied}.
The presence of noise destroys the invertibility and probability measure preservation, and the noisy Koopman operator $\tilde{\mathcal{K}}$ is no longer unitary, making the eigenvalues leave the unit circle.

\section{Lorentzian form of QPO spikes}
\label{sec:process_noise}

Consider a stochastic system,
\begin{equation}
\frac{d}{dt} \vec{s}
= \vec{f}(\vec{s}) + \sqrt{2D} \vec{\zeta},
\label{eq:SDE}
\end{equation}
where $\vec{s}$ is the high-dimensional accretion flow state (e.g., density, temperature, magnetic fields), $\vec{f}$ is a deterministic drift operator that acts on the accretion flow state and 
$\vec{\zeta}=d \vec{W}/dt$ is stochastic process noise (e.g., turbulence), which is assumed to be uncorrelated for a simplified discussion.

Let $\phi(\vec{s})$ be a projected observable function of the accretion flow state (e.g. X-ray flux $g_i(\vec{s})$) that is governed by the stochastic Koopman generator $\mathcal{L} \phi(\vec{s}) = \mathbb{E}\left[ d\phi(\vec{s})/dt \right]$.
The Taylor expansion of $\phi(\vec{s})$ up to the second order is
\begin{equation}
d \phi(\vec{s})
= \sum_{i} ds_{i} \frac{\partial}{\partial s_{i}} \phi(\vec{s})
+ \frac{1}{2} \sum_{i} \sum_{j} ds_{i} ds_{j} \frac{\partial^{2}}{\partial s_{i} \partial s_{j}} \phi(\vec{s}).
\end{equation}
Using Equation \ref{eq:SDE}, we have
\begin{equation}
d \phi(\vec{s})
= \sum_{i} \left( f_{i}(\vec{s})dt + \sqrt{2D} dW_{i} \right) \frac{\partial}{\partial s_{i}} \phi(\vec{s})
+ \frac{1}{2} \sum_{i} \sum_{j} \left( f_{i}(\vec{s})dt + \sqrt{2D} dW_{i} \right) \left( f_{j}(\vec{s})dt + \sqrt{2D} dW_{j} \right) \frac{\partial^{2}}{\partial s_{i} \partial s_{j}} \phi(\vec{s}).
\end{equation}
Using It\^{o}'s rules \citep[e.g.,][]{gardiner2004handbook} ($dt \cdot dt = 0$, $dt \cdot dW_{i} = 0$, $dW_{i} \cdot dW_{j}=\delta_{ij} dt$, and $\mathbb{E}\left[dW_{i}\right]=0$), we find the expectation
\begin{equation}
\mathbb{E} \left[ \frac{d}{dt} \phi(\vec{s}) \right]
= \sum_{i} f_{i}(\vec{s}) \frac{\partial}{\partial s_{i}} \phi(\vec{s})
+ D \sum_{i} \sum_{j} \delta_{ij} \frac{\partial^{2}}{\partial s_{i} \partial s_{j}} \phi(\vec{s})
= \left( \vec{f}(\vec{s}) \cdot \vec{\nabla} + D \nabla^{2} \right) \phi(\vec{s}).
\end{equation}
So, the stochastic Koopman generator becomes
\begin{equation}
\mathcal{L} = \vec{f}(\vec{s}) \cdot \vec{\nabla} + D \nabla^{2},
\end{equation}
where $\mathcal{L}_{0} = \vec{f}(\vec{s}) \cdot \vec{\nabla}$ is the deterministic Koopman generator.

Letting $\varphi_{0}(\vec{s})$ be an eigenfunction that satisfies $\mathcal{L}_{0} \varphi_{0}(\vec{s}) = \mu_{0} \varphi_{0}(\vec{s})$, we can estimate the change of the eigenvalue due to the process noise using first-order perturbation theory,
\begin{equation}
\mu \approx \mu_{0} + D\, \frac{\left< \varphi_{0}(\vec{s}), \nabla^{2} \varphi_{0}(\vec{s}) \right>}{\left< \varphi_{0}(\vec{s}), \varphi_{0}(\vec{s}) \right>}.
\end{equation}
Since the inner product $\left< \varphi_{0}(\vec{s}), \nabla^{2} \varphi_{0}(\vec{s})  \right> = -\left\| \nabla \varphi_{0}(\vec{s}) \right\|^{2} \leq 0$, this means that the presence of process noise always makes $\mathrm{Re}(\mu) \leq \mathrm{Re}(\mu_{0})$ and leads to contraction of the eigenfunction (decay of its amplitude).

We now continue to examine how the process noise can change the power spectral density (PSD) of the observable function.
We define PSD \citep{merrifield1994estimating} as $S(\omega) = \int_{-\infty}^{\infty} e^{i\omega t} C(t)\, dt$, where $C(t) = \mathbb{E}\left[ \phi(\vec{s}_{t+\tau}) \phi(\vec{s}_{\tau}) \right]$ is the autocorrelation function of the observable, where the expectation is over noise realizations with $\vec{s}_{\tau}$ drawn from the stationary distribution.
Since $C(t)=C(-t)$, we can rewrite the PSD as
\begin{equation}
S(\omega) = 2 \mathrm{Re} \left[ \int_{0}^{\infty} e^{-i\omega t} C(t) dt \right].
\end{equation}
We recall that the conditional expectation of the observable is evolved by the Koopman semigroup, $\mathbb{E}\left[ \phi(\vec{s}_{t+\tau}) \,\middle|\, \vec{s}_{\tau} \right] = e^{t \mathcal{L}} \phi(\vec{s}_{\tau})$, so the PSD becomes
\begin{align}
S(\omega) &= 2 \mathrm{Re}
\left[
\int_{0}^{\infty} e^{-i\omega t}
\mathbb{E}\left[ \phi(\vec{s}_{\tau}) e^{t\mathcal{L}} \phi(\vec{s}_{\tau}) \right] dt
\right] \\
&= 2 \mathrm{Re}
\left[
\mathbb{E}\left[ \phi(\vec{s}_{\tau})
\int_{0}^{\infty} e^{-i\omega t + t\mathcal{L}} dt
\phi(\vec{s}_{\tau}) \right]
\right] \\
&= 2 \mathrm{Re}
\left[
\mathbb{E}\left[ \phi(\vec{s}_{\tau})
\left( i\omega I - \mathcal{L} \right)^{-1}
\phi(\vec{s}_{\tau}) \right]
\right],
\end{align}
where we have used the Laplace transform $\int_{0}^{\infty} e^{-i\omega t + t\mathcal{L}} dt = \left( i\omega I - \mathcal{L} \right)^{-1}$.
Let $\varphi_{k}(\vec{s})$ be an eigenfunction that satisfies $\mathcal{L}\varphi_{k}(\vec{s})=\mu_{k} \varphi_{k}(\vec{s})$, then $\left( i\omega I - \mathcal{L} \right)^{-1} \varphi_{k}(\vec{s}) = \varphi_{k}(\vec{s}) / (i\omega - \mu_{k})$.
Finally, substituting the resolvent expression into the PSD yields the Lorentzian form given in Equation \ref{eq:qpo_spike}.

\section{Data processing}
\label{sec:data_processing_appendix}

The combination of MAXI and RXTE ASM light curves for 4U~1705-44 proceeds in two stages: energy band alignment and flux cross-calibration. RXTE ASM provides integrated count rates in the 2--10~keV band, while MAXI reports fluxes in three separate channels: 2--4~keV (soft), 4--10~keV (medium), and 10--20~keV (hard).
To match the RXTE energy range, we sum the MAXI soft and medium channels, appropriately combining uncertainties.  

During the overlap period (MJD 55045--55434), we bin both light curves onto a common 4-day grid using inverse-variance weighting.
We then fit a linear model relating the MAXI and RXTE count rates:
\begin{equation}
F_{\mathrm{RXTE}} = \beta_{0} + \beta_{1} \, F_{\mathrm{MAXI}},
\end{equation}
using Orthogonal Distance Regression (ODR), which accounts for measurement uncertainties in both the predictor ($F_{\mathrm{MAXI}}$) and response ($F_{\mathrm{RXTE}}$) variables.
By contrast, Ordinary Least Squares (OLS) effectively assumes equal errors on all response-variable measurements, and Weighted Least Squares (WLS) accounts for unequal errors only in the response variable; neither is appropriate here, where both flux measurements carry significant uncertainties.
The best-fit parameters are $\beta_{0} = -0.90 \pm 0.24$ and $\beta_{1} = 27.4 \pm 1.4$.
The large scaling factor $\beta_{1}$ reflects differences in effective area and detector sensitivity between the two instruments, while the intercept factor $\beta_{0}$ could represent differences in baseline noise or background subtraction. 

\begin{figure*}
\centering
\includegraphics[width=0.64\linewidth]
{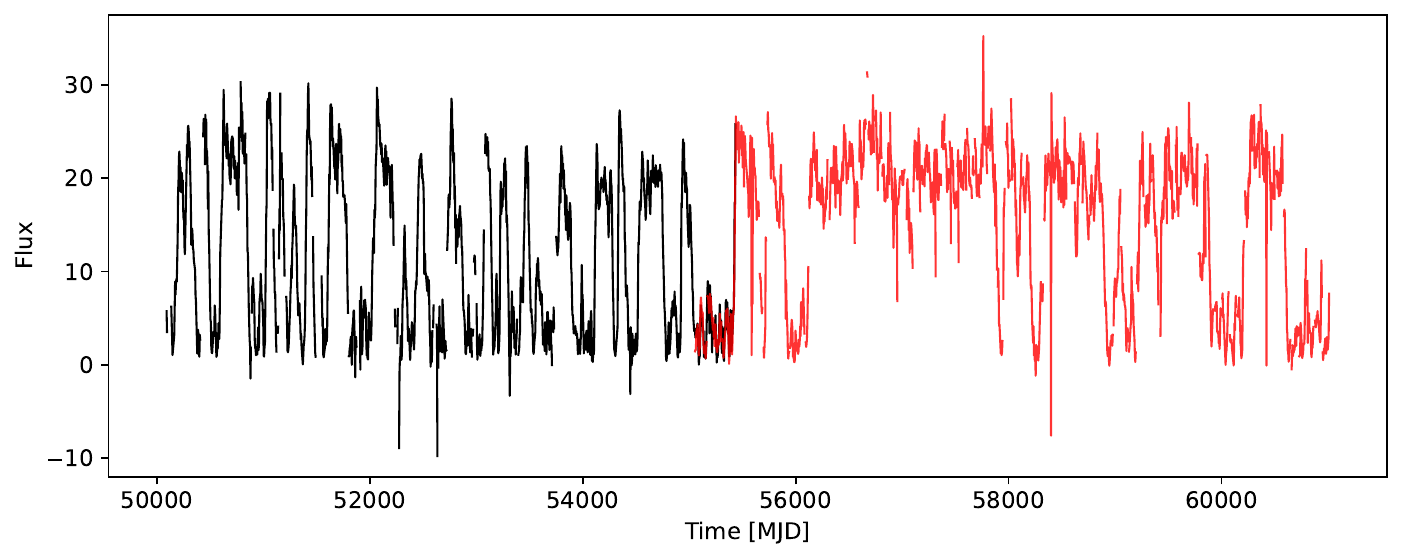}
\includegraphics[width=0.35\linewidth]
{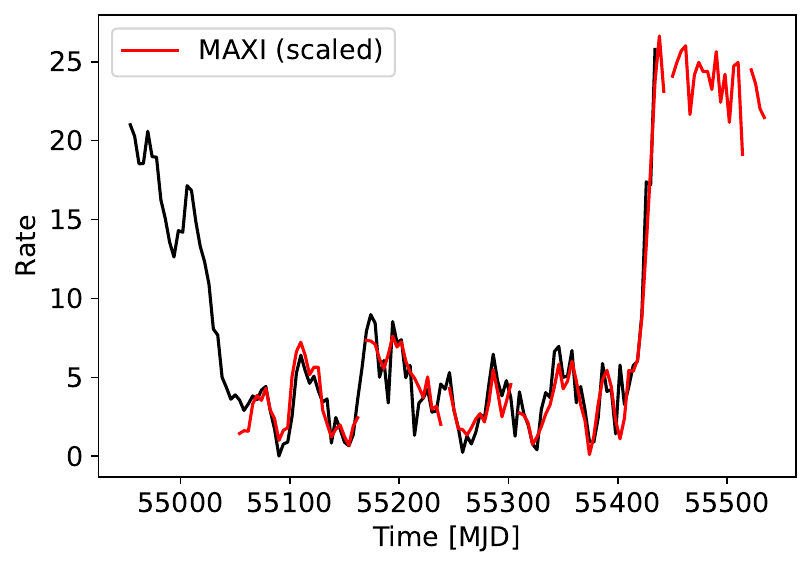} 
\caption{
Combining RXTE and MAXI data of 4U 1705-44. \textbf{Top:} The ``overlap'' section where both RXTE and MAXI are active (MJD 55045--55434). MAXI data (red) has been scaled to best match the RXTE data (black), taking into account the flux and errors of both instruments using Orthogonal Distance Regression. \textbf{Bottom:} Combined and scaled dataset, spanning the entire range of RXTE and MAXI.
}
\label{fig:appendix_combine_datasets}
\end{figure*}

To construct the combined light curve, we apply this transformation to all MAXI flux measurements, then scale the uncertainties accordingly: $\sigma_{\mathrm{scaled}} = |\beta_{1}| \, \sigma_{\mathrm{MAXI}}$.
The resulting dataset seamlessly extends the RXTE baseline from 1996 through December 2, 2025.
Figure \ref{fig:appendix_combine_datasets} shows the overlap section after the scaling and the combined dataset. The gaps and outliers clearly visible here will be treated in the next steps. While this dataset scaling is done once, outlier removal and Gaussian Process (GP) smoothing are done for each prediction point, providing a causality-enforced, simulated prediction environment. 

\begin{figure*}
\centering
\includegraphics[width=1\linewidth]
{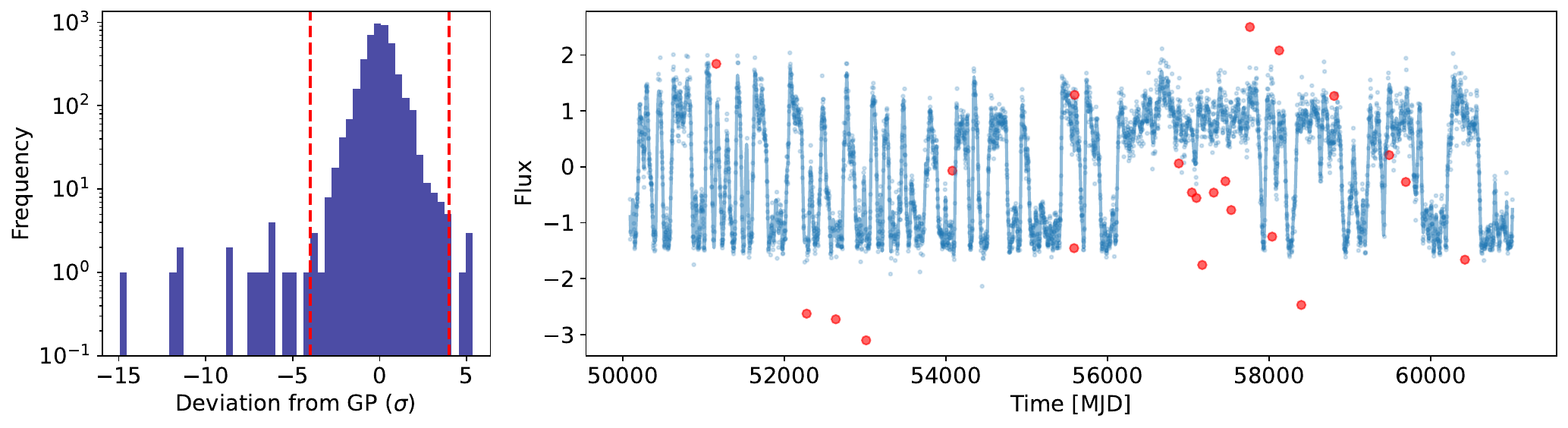}
\caption{
Outlier selection and removal for 4U 1705-44. \textbf{Left:} histogram of residuals after the first pass of GP smoothing. We choose the cutoff (red bands) at $4\sigma$, where residuals begin to significantly deviate from the main distribution. \textbf{Right:} Rebinned data points overlaid with the initial GP-smoothed light curve. Outlier points are marked in red. Outliers comprise 22 of the 4345 points ($0.5\%$).}
\label{fig:appendix_outliers}
\end{figure*}

After combining datasets, we identify and remove outliers iteratively. First, we smooth the curve using the sliding-window Gaussian Process procedure (next paragraph), and then calculate the residuals for each point relative to this smoothed curve. 
We then identify the data points whose residuals lie in the extreme tails of the residual distribution (Figure \ref{fig:appendix_outliers}) and remove these points from all subsequent analysis. A typical threshold is $4\sigma$: a point is excluded when its standardized residual exceeds four standard deviations.
After all outliers are removed, we recalculate the smoothed curve and repeat this process until convergence (no additional outliers). Convergence is typically achieved after a single iteration. This light curve is finally used for training $\mathbf{K}$. We have seen improvements in prediction performance after the removal of outliers. 
Figure \ref{fig:appendix_outliers} shows the residual distribution for 4U 1705-44 data in the first iteration, with outliers determined as data points with residuals more than $4\sigma$ deviation from zero, corresponding to about $0.5\%$ of data points. This $\sigma$ value is not static, and may change depending on source and bin size. 

We use GPs to calculate a smoothed, uniformly gridded light curve with uncertainties. The implementation uses the \texttt{scikit-learn} Python library \citep{scikit-learn}, specifically the \texttt{GaussianProcessRegressor} class with a Radial Basis Function (RBF) kernel combined with a white noise kernel.
Rather than fitting a single GP to the entire light curve (which becomes computationally expensive for long time series), we divide the data into overlapping windows. Each window contains 150 time steps, and successive windows advance by 30 steps. This means each point in the interior of the light curve is covered by approximately 5 different GP fits.

For each window, we fit a GP using an RBF kernel, which enforces smoothness over a characteristic length scale $\ell$. The kernel measures the similarity between two time points:
\begin{equation}
k(t, t') = \sigma_f^2 \exp\left( -\frac{(t-t')^2}{2\ell^2} \right),
\end{equation}
where $\sigma_f^2$ controls the amplitude of variations and $\ell$ sets the correlation timescale. We use $\ell = 10$ time steps as the initial length scale, chosen empirically to capture the characteristic variability while avoiding overfitting. The individual measurement uncertainties are incorporated directly into the GP noise model. The kernel hyperparameters are optimized automatically by \texttt{scikit-learn} using maximum likelihood estimation with 5 random restarts.

Each fitted GP provides a mean prediction $f_w(t)$ and uncertainty $\sigma_w(t)$ at any evaluation time $t$. Since windows overlap, we must combine multiple predictions at each point. We use a weighted average that favors predictions near the center of each window, where the GP is best constrained by data on both sides, ensuring that there are no discontinuities in the final light curve at window edges. The weighting also favors predictions with lower uncertainty, ensuring that high-variance sections, such as those at the window edges, are weighted less.  The weight for window $w$ centered at $c_w$ is
\begin{equation}
\pi_w(t) = \frac{1}{\sigma_w^2(t)} \cdot \exp\left( -\frac{(t - c_w)^2}{2\lambda_{w}^2} \right),
\end{equation}
where $\lambda_{w} = 0.15 \times (\text{window width})$ controls how quickly the weight decays from the center. Windows only contribute within their temporal boundaries.

The final smoothed flux and uncertainty are then computed as weighted averages:
\begin{equation}
\hat{x}(t) = \frac{\sum_{w} \pi_w(t) \, f_w(t)}{\sum_{w} \pi_w(t)}, \qquad
\mathrm{SD}(\hat{x}(t)) = \frac{\sum_{w} \pi_w(t) \, \sigma_w(t)}{\sum_{w} \pi_w(t)}.
\end{equation}
We combine the window standard deviations linearly rather than in quadrature because the overlapping windows are fit to largely the same data, so their errors are strongly correlated; the linear average is exact in the limit of perfect correlation, whereas a quadrature combination assumes independent estimates and would underestimate the uncertainty.
This procedure produces smooth transitions between adjacent windows and naturally assigns larger uncertainties in data gaps where the GP must extrapolate.

\begin{figure*}[h]
    \centering
\includegraphics[width=0.8\linewidth]
{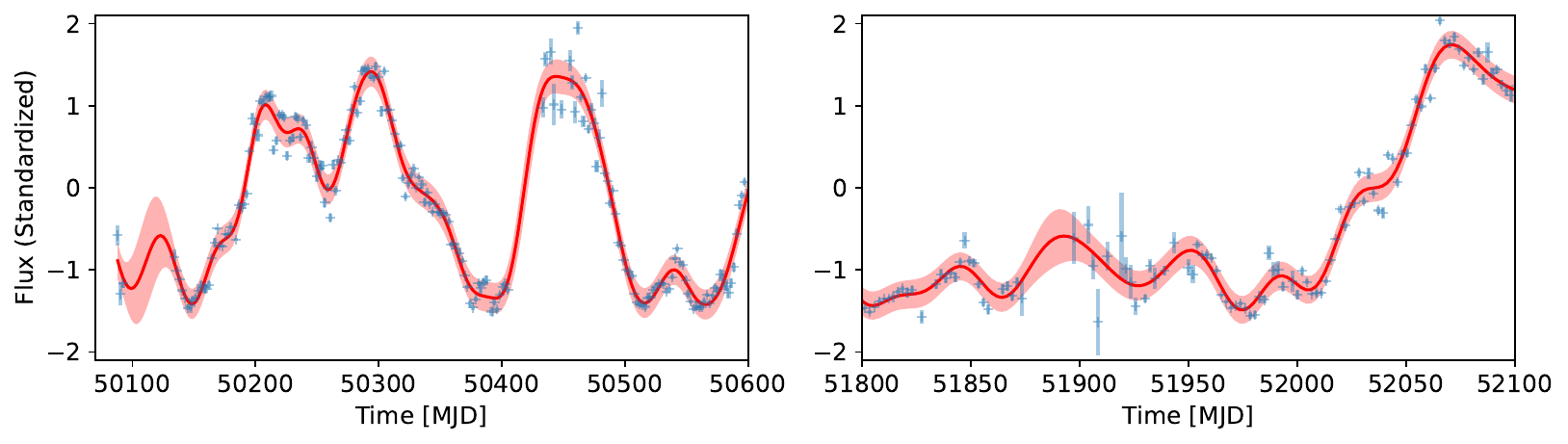}
\caption{Two segments of rebinned 4U 1705-44 data (blue) overlaid with the smoothed light curve and uncertainties. Sections with gaps and large measurement uncertainties (MJD 50100, 50450, 51900) have significantly larger smoothed uncertainties (red band), while sections with better data have smaller uncertainties.}
\label{fig:appendix_gp}
\end{figure*}

Figure \ref{fig:appendix_gp} shows the final GP smoothed curve with uncertainties at two sections in the light curve of 4U 1705-44. These sections contain significant gaps, as well as a mix of high- and low-uncertainty measurements. The GP smoothing process handles all these issues, producing a uniformly binned smooth curve that simultaneously captures the dynamics of interest, extrapolates across gaps, and appropriately treats uncertainties both across gaps and from measurements.

\section{Flux predictions at all 20 transition points}
\label{sec:longterm_predictions_appendix}

\begin{figure*}[h]
\centering
\includegraphics[width=0.9\linewidth]{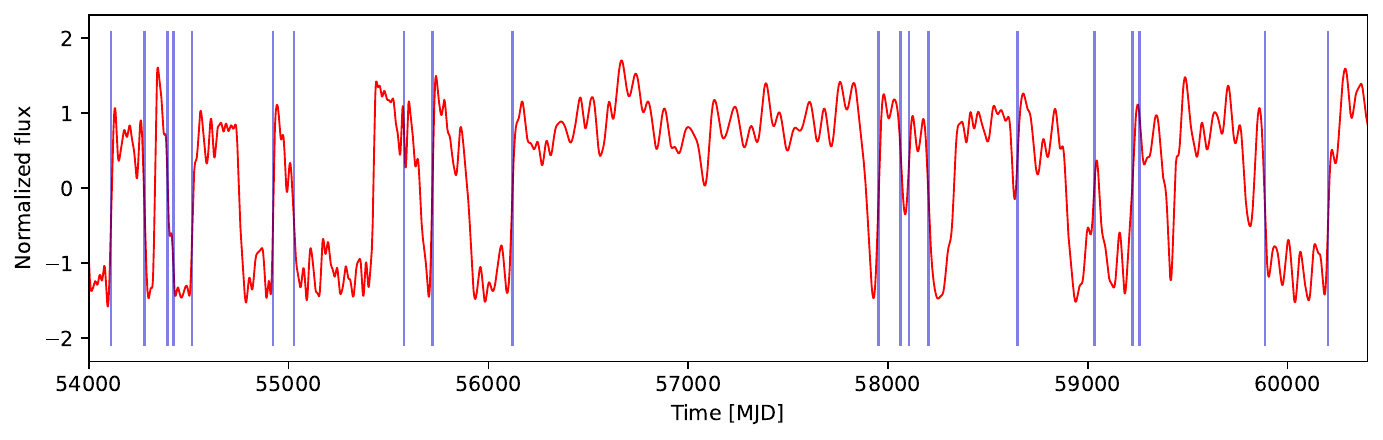}
\caption{
The 20 transition points of 4U 1705-44 in the range MJD 54000 to MJD 60400, used to demonstrate flux predictions using $\mathbf{K}$.
}
\label{fig:flux_prediction_points_4u1705}
\end{figure*}

\begin{figure*}
\centering
\includegraphics[width=0.49\linewidth]{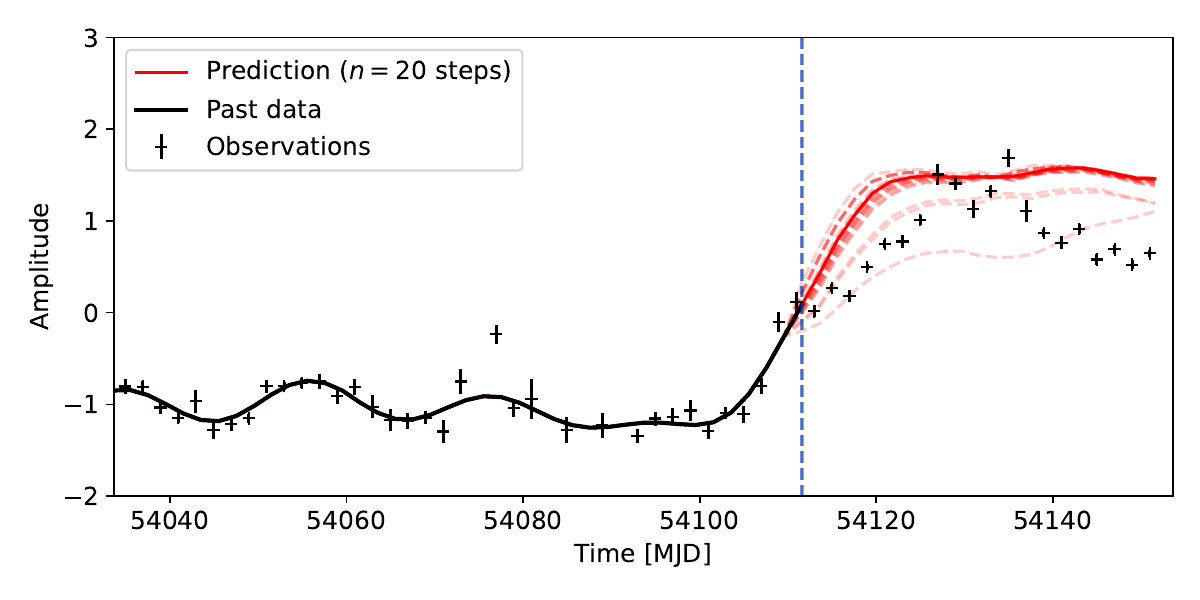}
\includegraphics[width=0.49\linewidth]{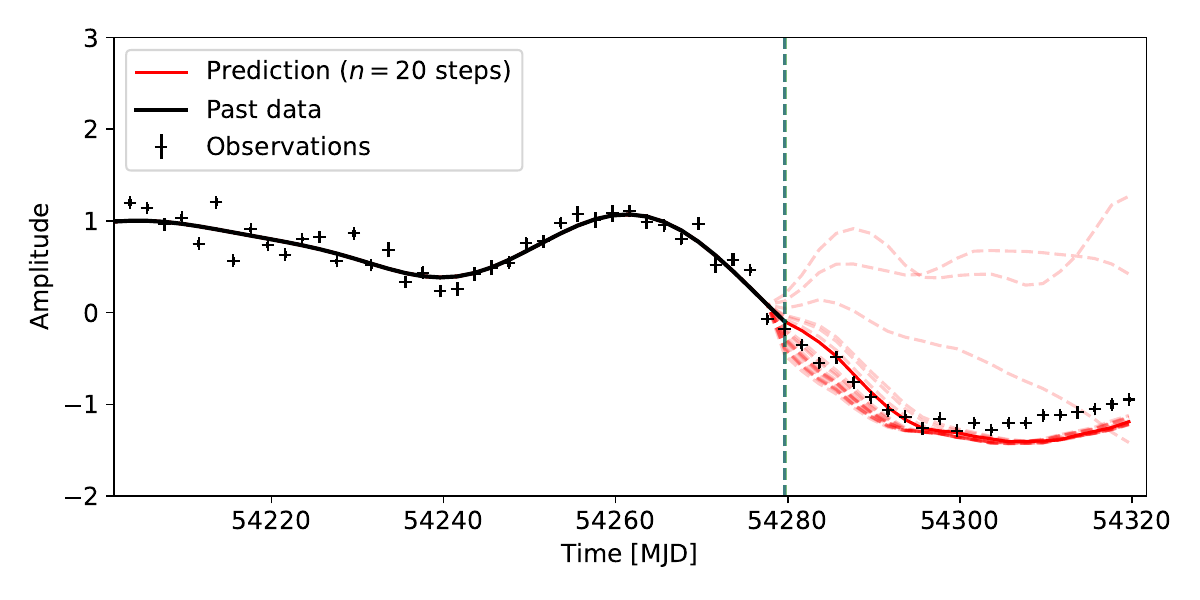}\\
\includegraphics[width=0.49\linewidth]{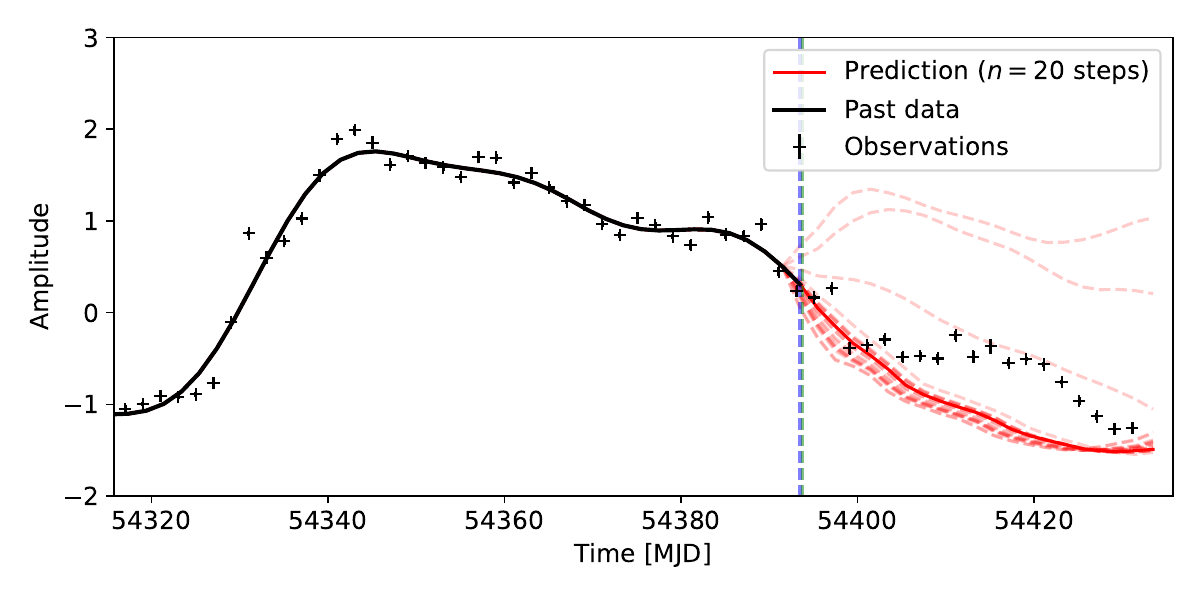}
\includegraphics[width=0.49\linewidth]{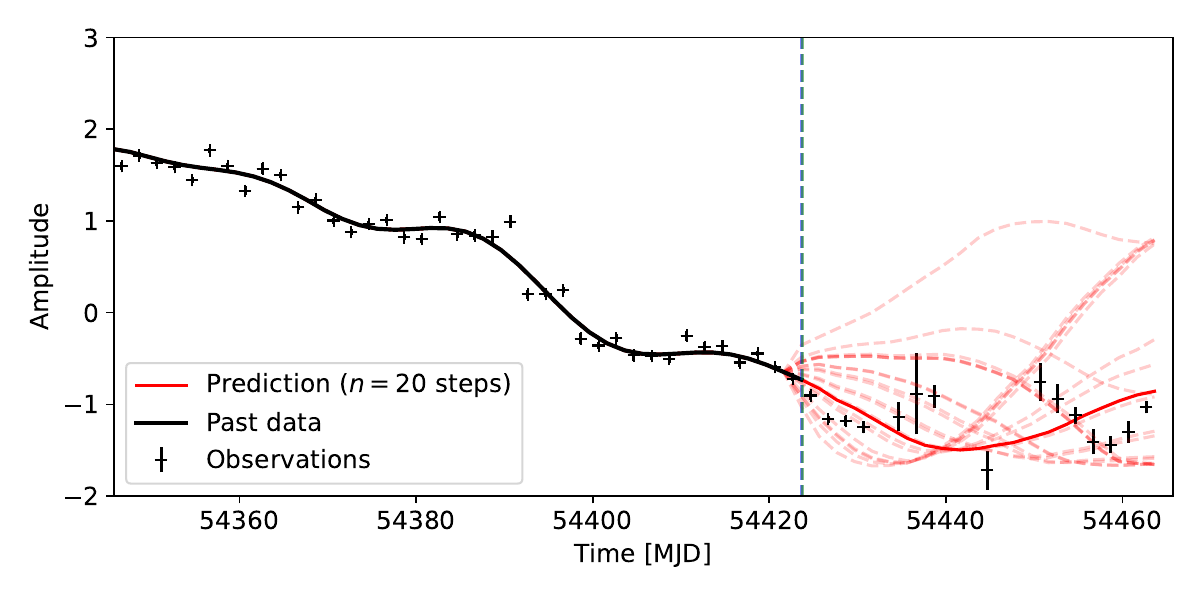}\\
\includegraphics[width=0.49\linewidth]{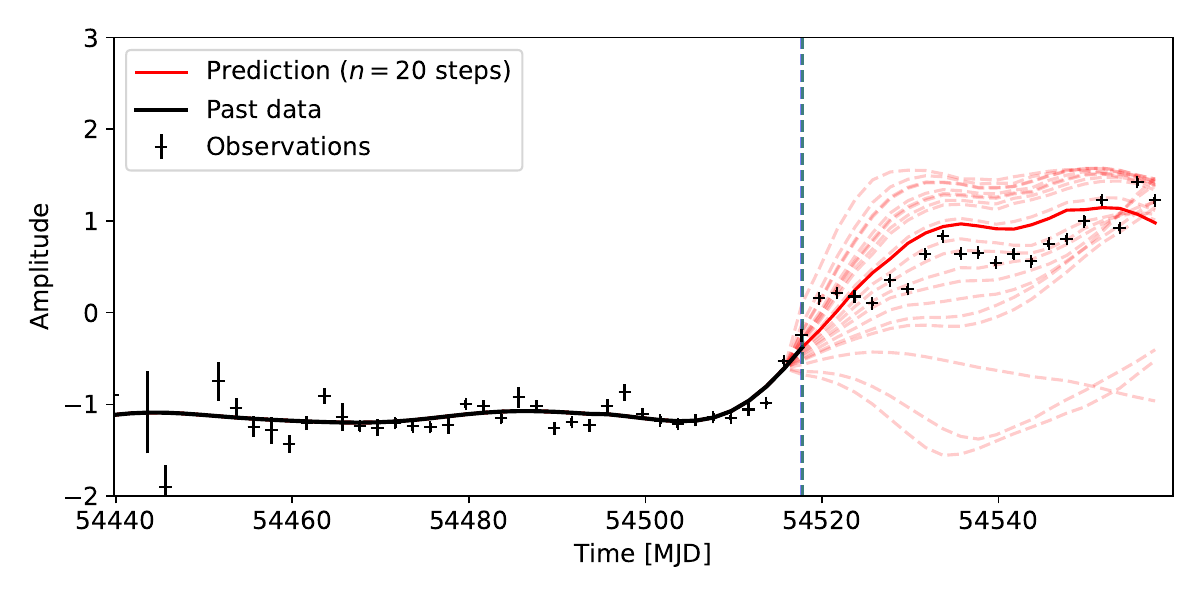}
\includegraphics[width=0.49\linewidth]{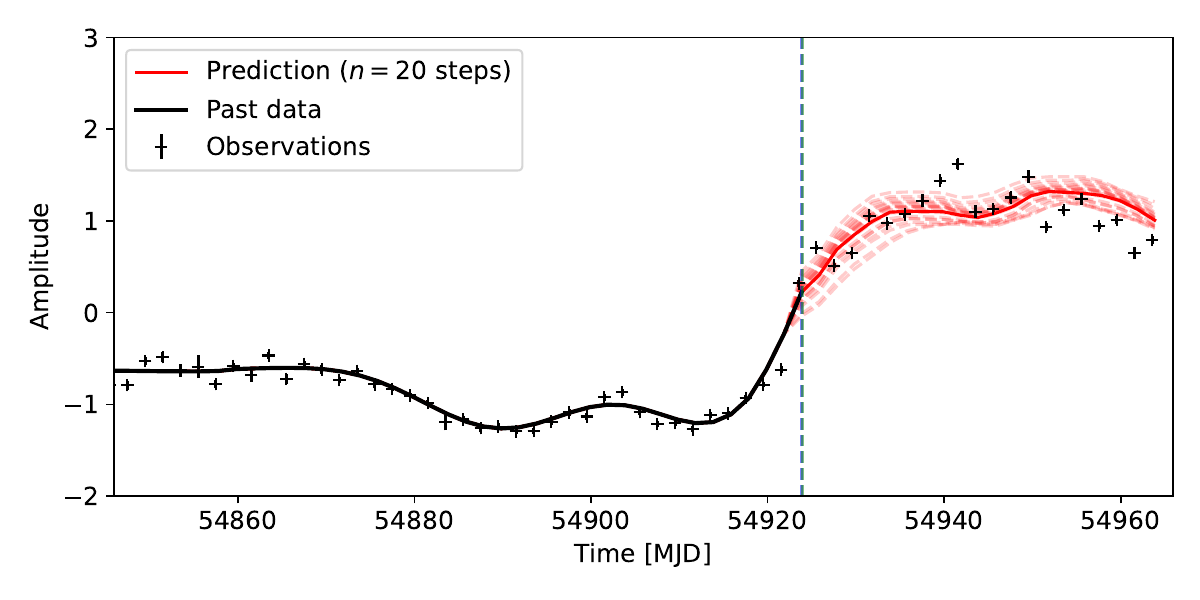} \\
\includegraphics[width=0.49\linewidth]{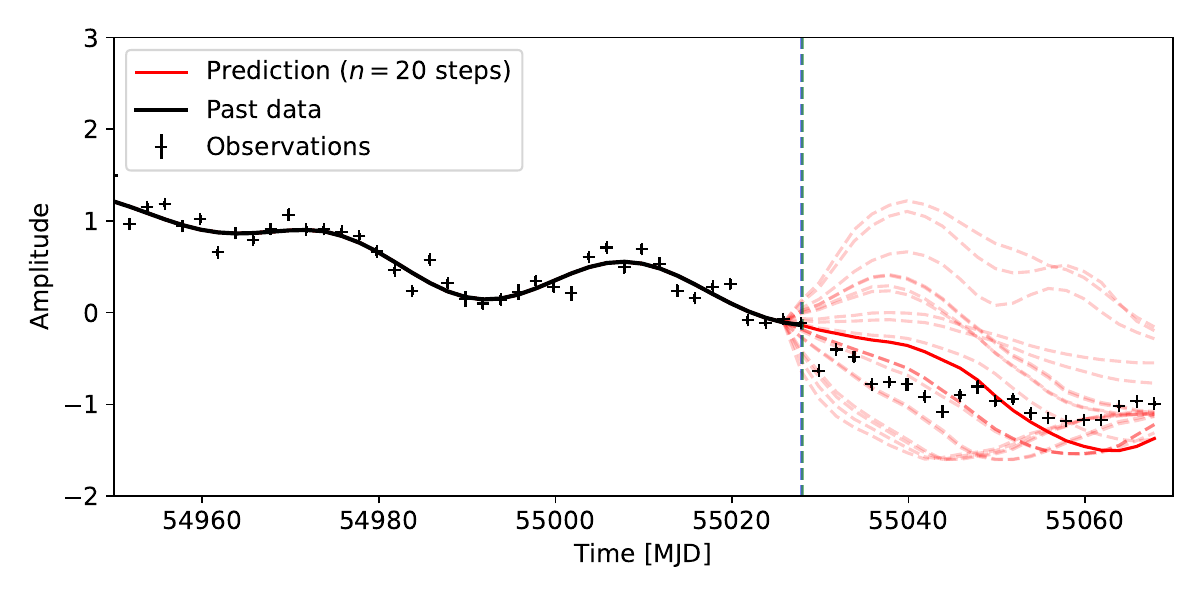}
\includegraphics[width=0.49\linewidth]{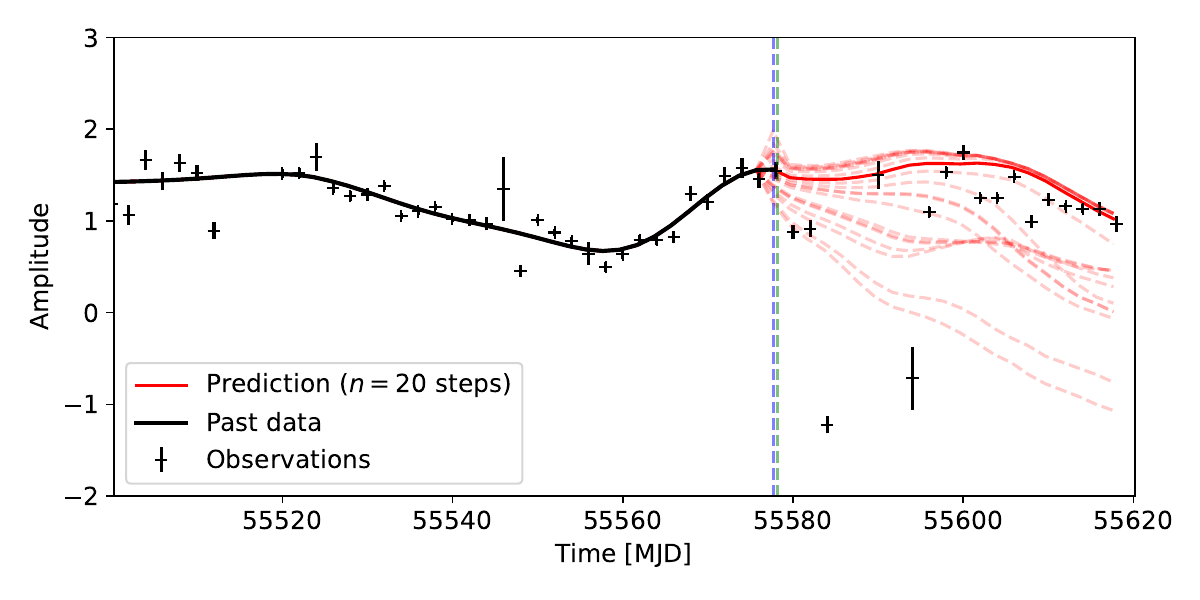} \\
\includegraphics[width=0.49\linewidth]{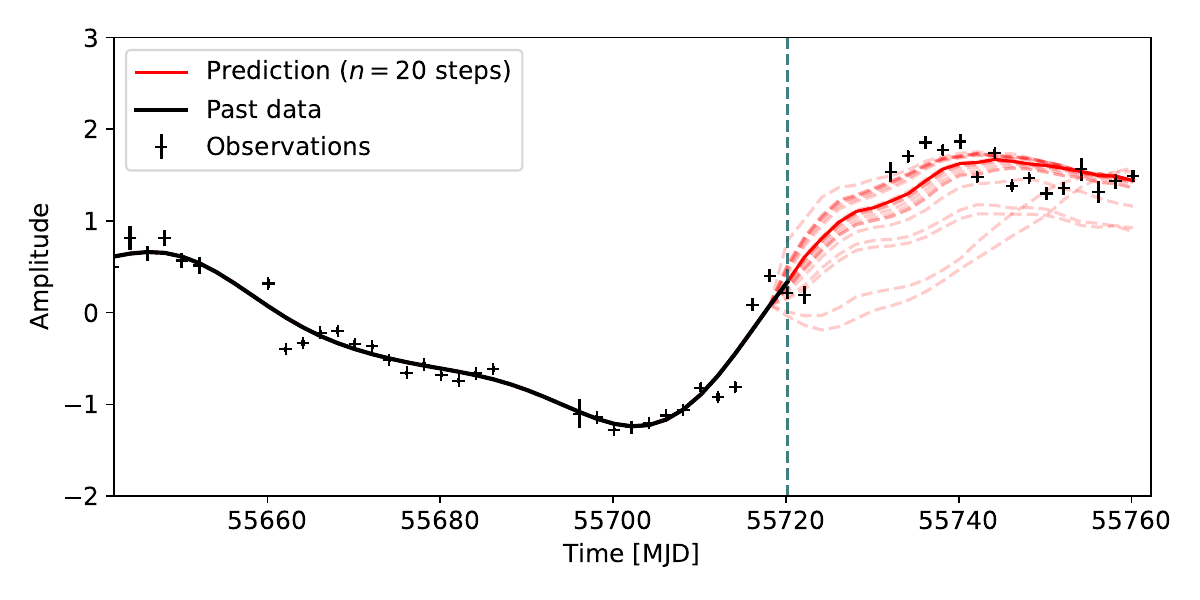}
\includegraphics[width=0.49\linewidth]{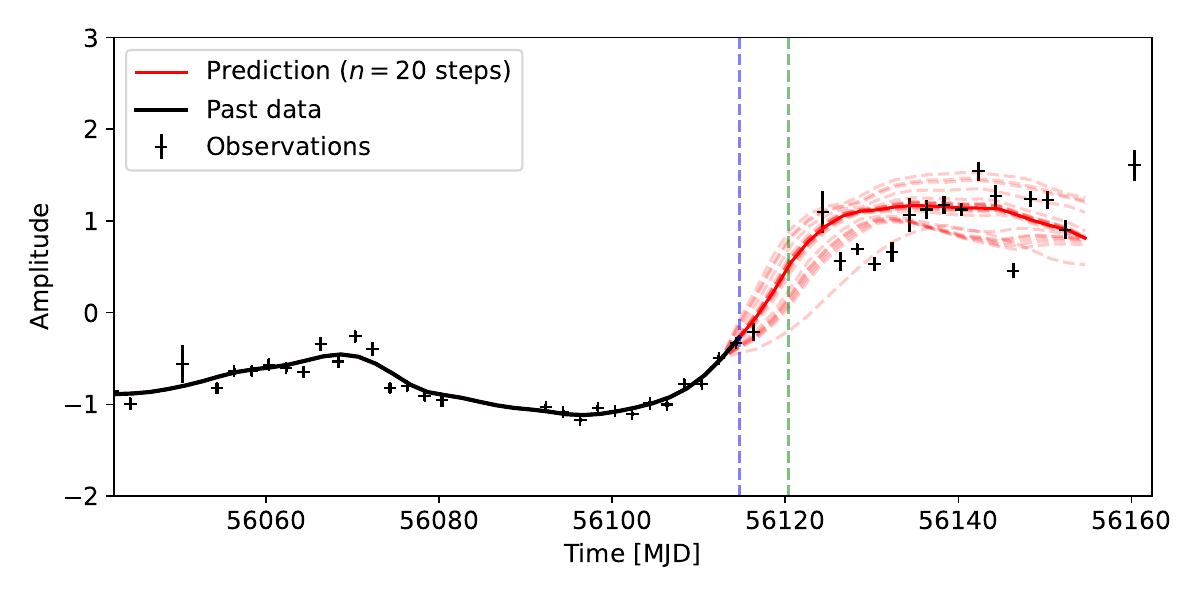} \\
\caption{
Long-term predictions of 4U 1705-44 flux at 10 transition mid-points (MJD 54000--56200). Each panel shows past data (black curve) ending at the blue vertical line, with predictions extending into the future. Red solid curves show nominal predictions; red dashed curves show alternative predictions generated by varying the last flux point, indicating forecast uncertainty. The spread of alternative predictions reflects confidence in the forecast.
}
\label{fig:predict_light_curve_4u1705_longterm1}
\end{figure*}

\begin{figure*}
\centering
\includegraphics[width=0.49\linewidth]{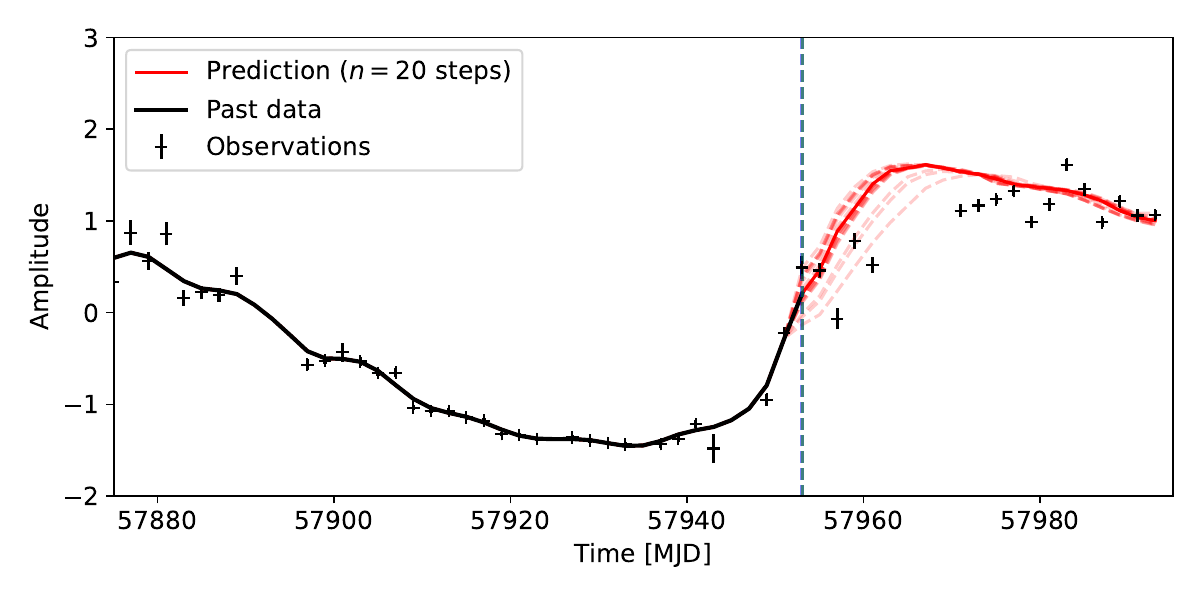}
\includegraphics[width=0.49\linewidth]{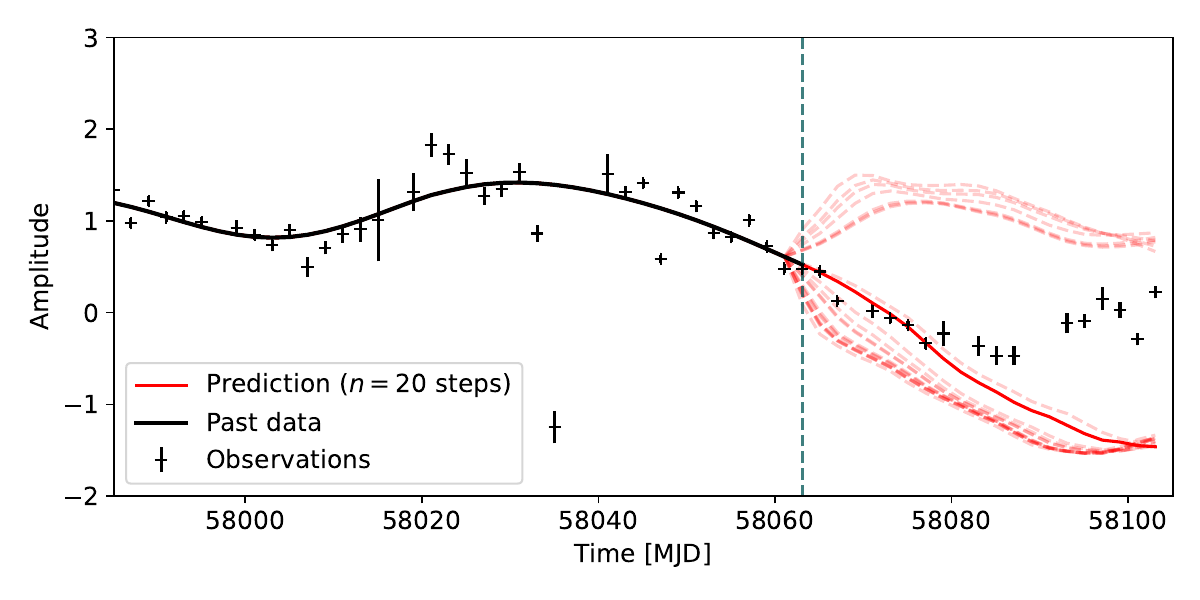}\\
\includegraphics[width=0.49\linewidth]{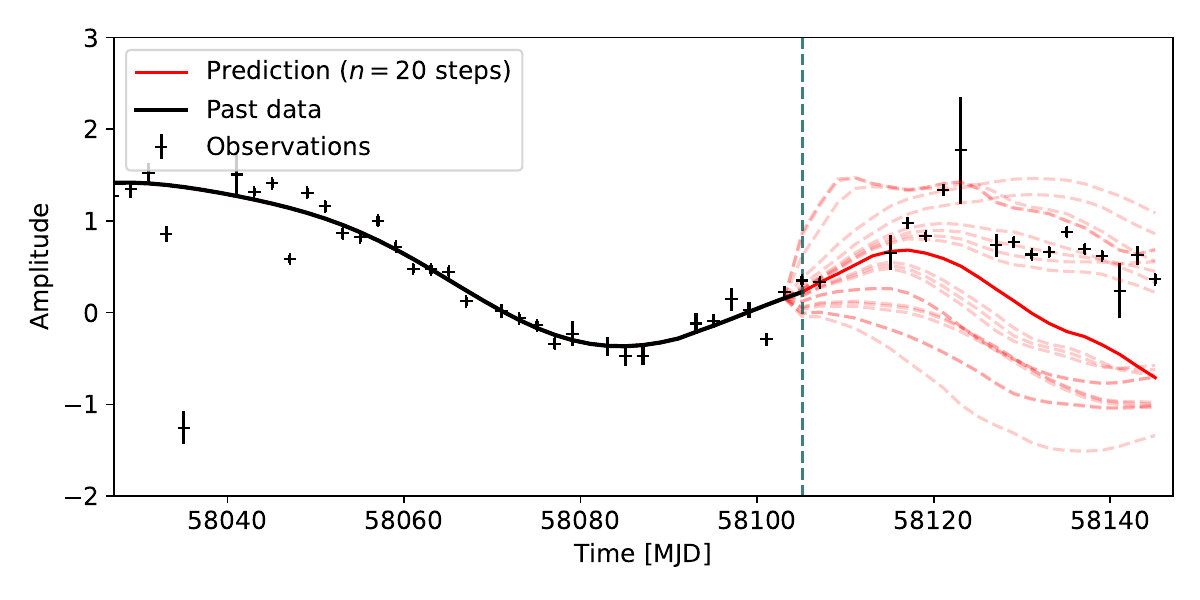}
\includegraphics[width=0.49\linewidth]{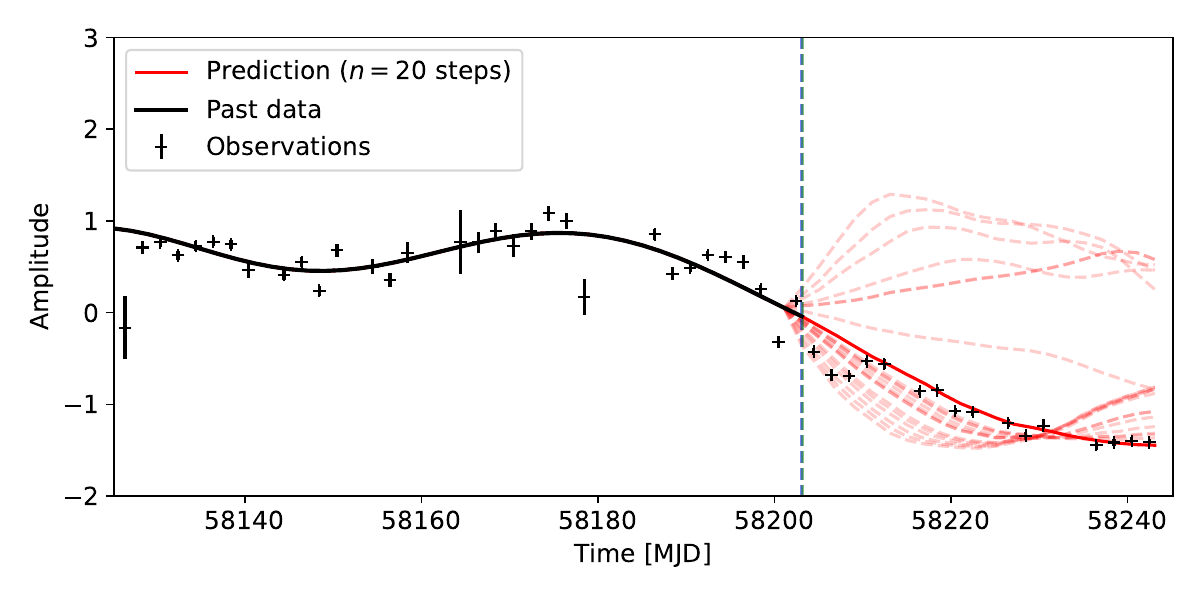}\\
\includegraphics[width=0.49\linewidth]{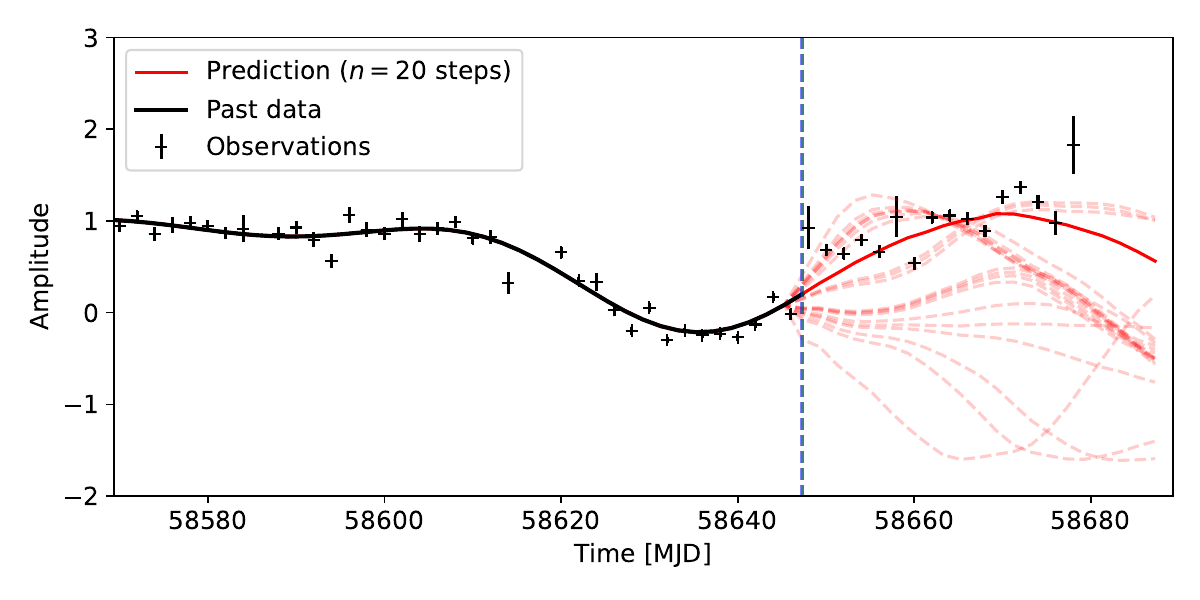}
\includegraphics[width=0.49\linewidth]{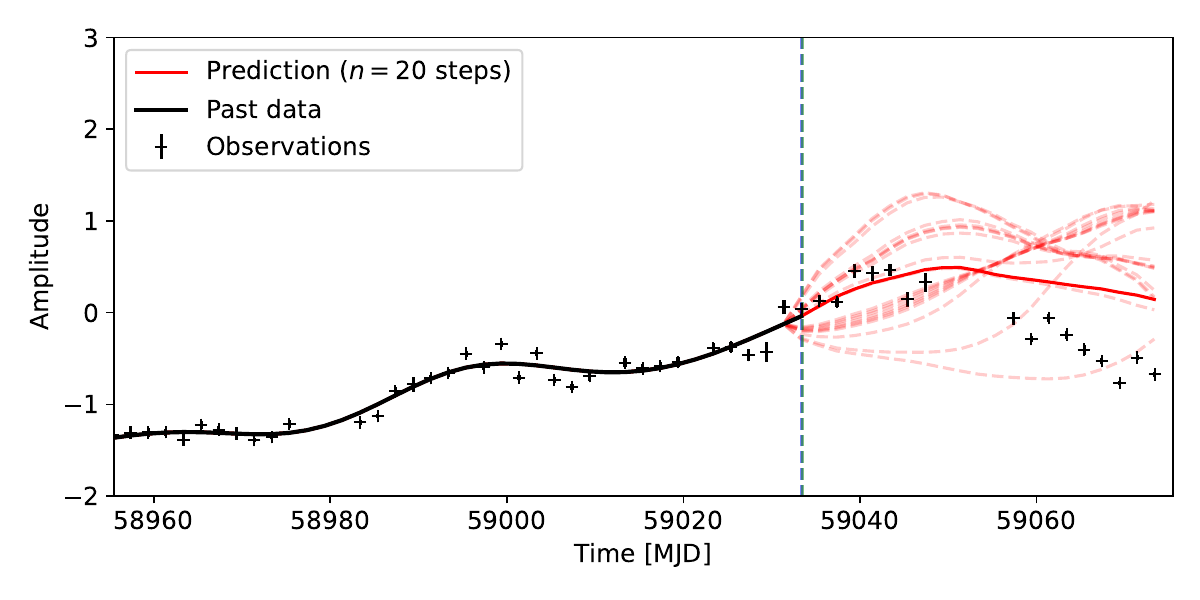} \\
\includegraphics[width=0.49\linewidth]{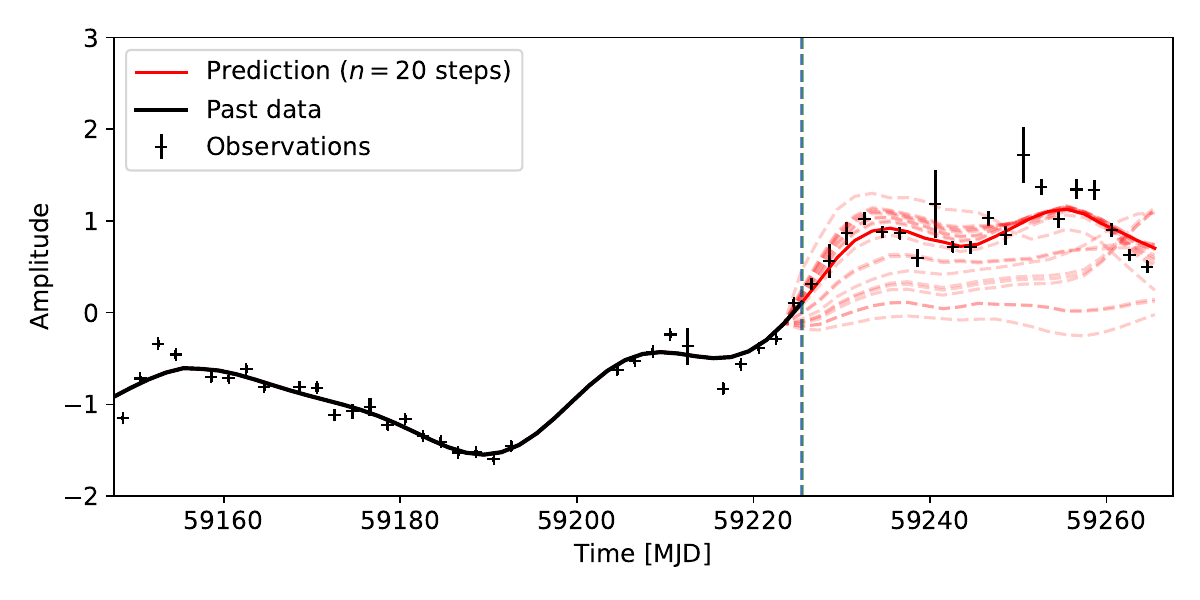}
\includegraphics[width=0.49\linewidth]{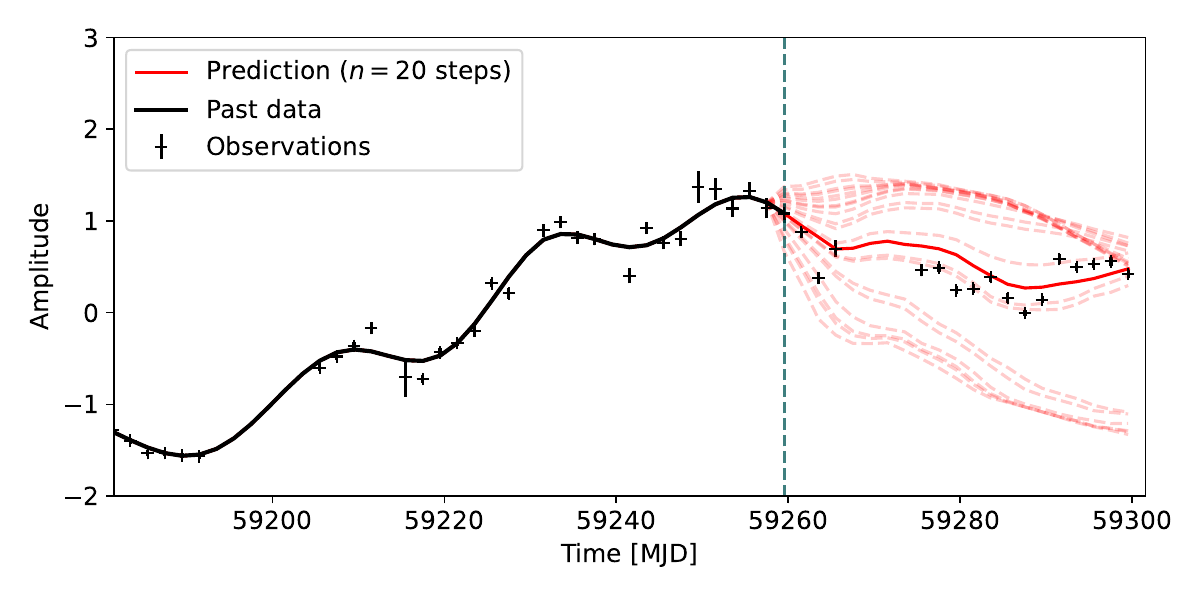} \\
\includegraphics[width=0.49\linewidth]{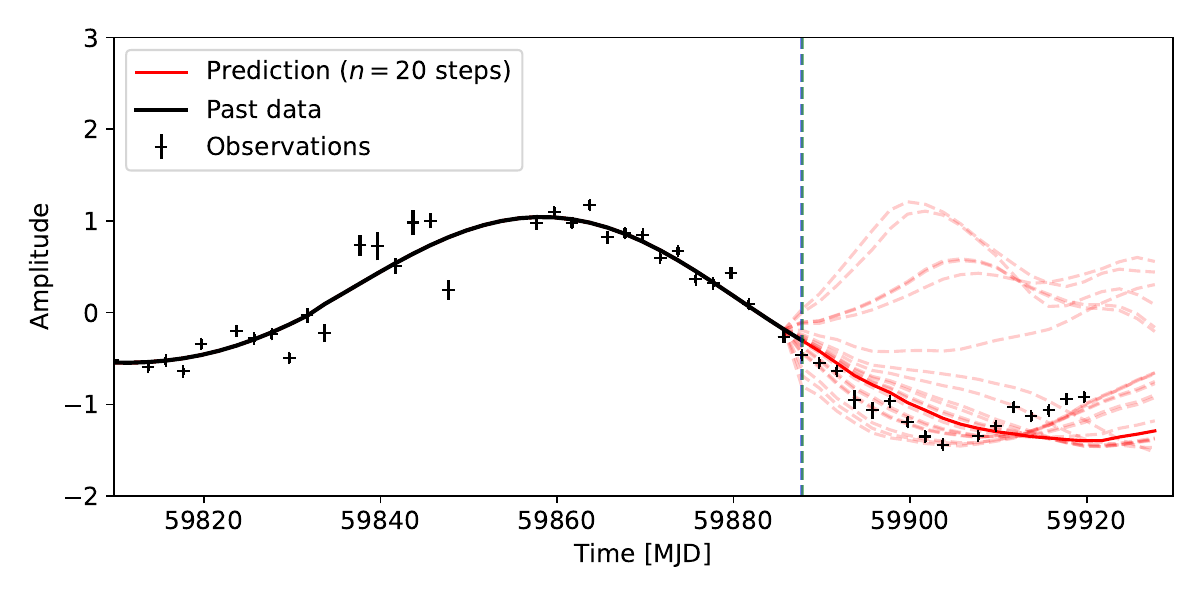}
\includegraphics[width=0.49\linewidth]{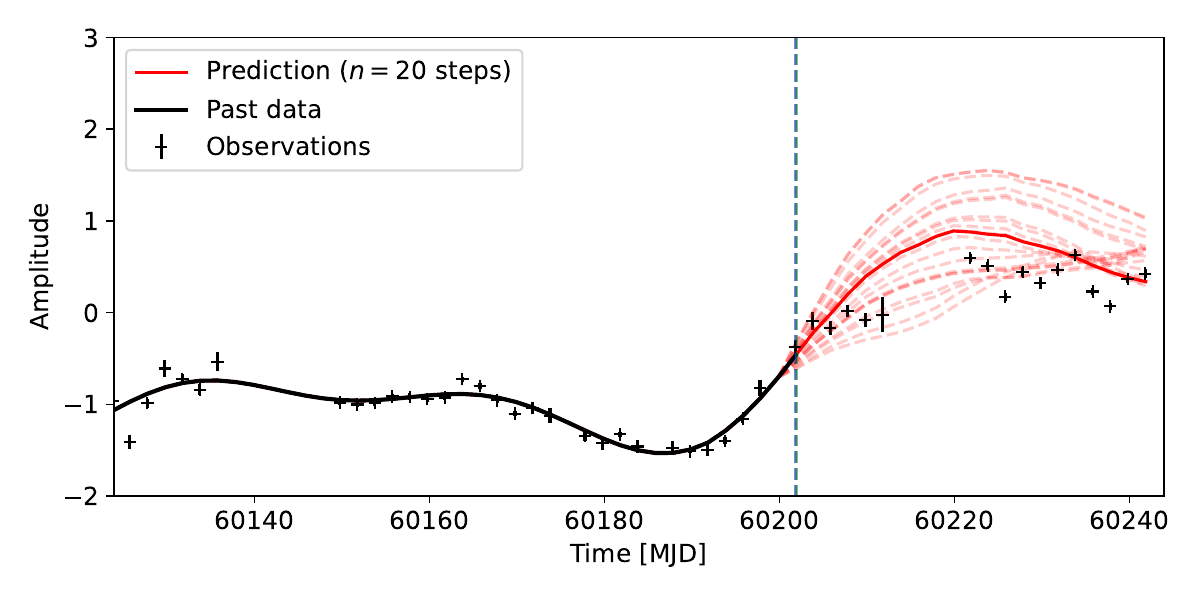} \\
\caption{
Long-term predictions of 4U 1705-44 flux at 10 transition mid-points (MJD 57900--60400). Each panel shows past data (black curve) ending at the blue vertical line, with predictions extending into the future. Red solid curves show nominal predictions; red dashed curves show alternative predictions generated by varying the last flux point, indicating forecast uncertainty. Together with Figure \ref{fig:predict_light_curve_4u1705_longterm1}, these 20 predictions span the full observation range past MJD 54000.
}
\label{fig:predict_light_curve_4u1705_longterm2}
\end{figure*}

This appendix presents the complete set of long-term flux predictions referenced in Section \ref{sec:real_data}. Figure \ref{fig:flux_prediction_points_4u1705} marks the 20 transition points between high-flux and low-flux states in the range MJD 54000--60400, and Figures \ref{fig:predict_light_curve_4u1705_longterm1} and \ref{fig:predict_light_curve_4u1705_longterm2} show the predictions starting at each point.

\newpage

\bibliographystyle{aasjournal}
\bibliography{main}

\begin{thebibliography}{}
\expandafter\ifx\csname natexlab\endcsname\relax\def\natexlab#1{#1}\fi
\providecommand{\url}[1]{\href{#1}{#1}}
\providecommand{\dodoi}[1]{doi:~\href{http://doi.org/#1}{\nolinkurl{#1}}}
\providecommand{\doeprint}[1]{\href{http://ascl.net/#1}{\nolinkurl{http://ascl.net/#1}}}
\providecommand{\doarXiv}[1]{\href{https://arxiv.org/abs/#1}{\nolinkurl{https://arxiv.org/abs/#1}}}

\bibitem[{Albidah {et~al.}(2021)Albidah, Brevis, Fedun, Ballai, Jess,
  Stangalini, Higham, \& Verth}]{albidah2021proper}
Albidah, A., Brevis, W., Fedun, V., {et~al.} 2021, Philosophical Transactions
  of the Royal Society A, 379, 20200181

\bibitem[{Allen {et~al.}(2024)Allen, Gan, \& Zheng}]{allen2024interpretable}
Allen, G.~I., Gan, L., \& Zheng, L. 2024, Annual Review of Statistics and Its
  Application, 11, 97

\bibitem[{Askham \& Kutz(2018)}]{askham2018variable}
Askham, T., \& Kutz, J.~N. 2018, SIAM Journal on Applied Dynamical Systems, 17,
  380

\bibitem[{Belloni {et~al.}(2002)Belloni, Psaltis, \& van~der
  Klis}]{belloni2002unified}
Belloni, T., Psaltis, D., \& van~der Klis, M. 2002, The Astrophysical Journal,
  572, 392

\bibitem[{Bergmeir \& Ben{\'\i}tez(2012)}]{bergmeir2012use}
Bergmeir, C., \& Ben{\'\i}tez, J.~M. 2012, Information Sciences, 191, 192

\bibitem[{Brunton {et~al.}(2017)Brunton, Brunton, Proctor, Kaiser, \&
  Kutz}]{brunton_HAVOK_2017}
Brunton, S.~L., Brunton, B.~W., Proctor, J.~L., Kaiser, E., \& Kutz, J.~N.
  2017, Nature Communications, 8, 19

\bibitem[{Brunton {et~al.}(2022)Brunton, Budišić, Kaiser, \&
  Kutz}]{brunton_modern_2022}
Brunton, S.~L., Budišić, M., Kaiser, E., \& Kutz, J.~N. 2022, SIAM Review,
  64, 229

\bibitem[{Brunton {et~al.}(2016{\natexlab{a}})Brunton, Proctor, \&
  Kutz}]{brunton2016discovering}
Brunton, S.~L., Proctor, J.~L., \& Kutz, J.~N. 2016{\natexlab{a}}, Proceedings
  of the national academy of sciences, 113, 3932

\bibitem[{Brunton {et~al.}(2016{\natexlab{b}})Brunton, Proctor, \&
  Kutz}]{brunton2016sparse}
---. 2016{\natexlab{b}}, IFAC-PapersOnLine, 49, 710

\bibitem[{Budi{\v{s}}i{\'c} {et~al.}(2012)Budi{\v{s}}i{\'c}, Mohr, \&
  Mezi{\'c}}]{budivsic2012applied}
Budi{\v{s}}i{\'c}, M., Mohr, R., \& Mezi{\'c}, I. 2012, Chaos: An
  Interdisciplinary Journal of Nonlinear Science, 22

\bibitem[{Camenzind \& Krockenberger(1992)}]{camenzind1992lighthouse}
Camenzind, M., \& Krockenberger, M. 1992, Astronomy and Astrophysics, 255, 59

\bibitem[{Caproni {et~al.}(2017)Caproni, Abraham, Motter, \&
  Monteiro}]{caproni2017jet}
Caproni, A., Abraham, Z., Motter, J.~C., \& Monteiro, H. 2017, The
  Astrophysical Journal Letters, 851, L39

\bibitem[{Casdagli(1989)}]{casdagli1989nonlinear}
Casdagli, M. 1989, Physica D: Nonlinear Phenomena, 35, 335

\bibitem[{Chekroun {et~al.}(2020)Chekroun, Tantet, Dijkstra, \&
  Neelin}]{chekroun2020ruelle}
Chekroun, M.~D., Tantet, A., Dijkstra, H.~A., \& Neelin, J.~D. 2020, Journal of
  Statistical Physics, 179, 1366

\bibitem[{Cho {et~al.}(2014)Cho, Van~Merri{\"e}nboer, Gul{\c{c}}ehre, Bahdanau,
  Bougares, Schwenk, \& Bengio}]{cho2014learning}
Cho, K., Van~Merri{\"e}nboer, B., Gul{\c{c}}ehre, {\c{C}}., {et~al.} 2014, in
  Proceedings of the 2014 conference on empirical methods in natural language
  processing (EMNLP), 1724--1734

\bibitem[{Colbrook {et~al.}(2023)Colbrook, Ayton, \&
  Sz{\H{o}}ke}]{colbrook2023residual}
Colbrook, M.~J., Ayton, L.~J., \& Sz{\H{o}}ke, M. 2023, Journal of Fluid
  Mechanics, 955, A21

\bibitem[{Colbrook \& Townsend(2024)}]{colbrook2024rigorous}
Colbrook, M.~J., \& Townsend, A. 2024, Communications on Pure and Applied
  Mathematics, 77, 221

\bibitem[{Conradie {et~al.}(2026)Conradie, Boull{\'e}, Loiseau, Brunton, \&
  Colbrook}]{conradie2026trustworthy}
Conradie, G., Boull{\'e}, N., Loiseau, J.-C., Brunton, S.~L., \& Colbrook,
  M.~J. 2026, arXiv preprint arXiv:2603.15091

\bibitem[{{\v{C}}rnjari{\'c}-{\v{Z}}ic
  {et~al.}(2020){\v{C}}rnjari{\'c}-{\v{Z}}ic, Ma{\'c}e{\v{s}}i{\'c}, \&
  Mezi{\'c}}]{vcrnjaric2020koopman}
{\v{C}}rnjari{\'c}-{\v{Z}}ic, N., Ma{\'c}e{\v{s}}i{\'c}, S., \& Mezi{\'c}, I.
  2020, Journal of Nonlinear Science, 30, 2007

\bibitem[{Darling \& Widrow(2019)}]{darling2019eigenfunctions}
Darling, K., \& Widrow, L.~M. 2019, Monthly Notices of the Royal Astronomical
  Society, 490, 114

\bibitem[{Dawson {et~al.}(2016)Dawson, Hemati, Williams, \&
  Rowley}]{dawson2016characterizing}
Dawson, S.~T., Hemati, M.~S., Williams, M.~O., \& Rowley, C.~W. 2016,
  Experiments in Fluids, 57, 42

\bibitem[{Done {et~al.}(2007)Done, Gierli{\'n}ski, \&
  Kubota}]{done2007modelling}
Done, C., Gierli{\'n}ski, M., \& Kubota, A. 2007, The Astronomy and
  Astrophysics Review, 15, 1

\bibitem[{Farmer \& Sidorowich(1987)}]{farmer1987predicting}
Farmer, J.~D., \& Sidorowich, J.~J. 1987, Physical review letters, 59, 845

\bibitem[{Fatheddin \& Sajadian(2024)}]{fatheddin2024singular}
Fatheddin, H., \& Sajadian, S. 2024, The Astronomical Journal, 168, 71

\bibitem[{Fender(2001)}]{fender2001powerful}
Fender, R.~P. 2001, Monthly Notices of the Royal Astronomical Society, 322, 31

\bibitem[{Fender {et~al.}(2004)Fender, Belloni, \& Gallo}]{fender2004towards}
Fender, R.~P., Belloni, T.~M., \& Gallo, E. 2004, Monthly Notices of the Royal
  Astronomical Society, 355, 1105

\bibitem[{Frank {et~al.}(2002)Frank, King, \& Raine}]{frank2002accretion}
Frank, J., King, A.~R., \& Raine, D. 2002, Accretion power in astrophysics
  (Cambridge university press)

\bibitem[{Gallos {et~al.}(2024)Gallos, Lehmberg, Dietrich, \&
  Siettos}]{gallos2024data}
Gallos, I.~K., Lehmberg, D., Dietrich, F., \& Siettos, C. 2024, Chaos: An
  Interdisciplinary Journal of Nonlinear Science, 34

\bibitem[{Gao {et~al.}(2026)Gao, Williams, \& Kutz}]{gao2026sparse}
Gao, M.~L., Williams, J.~P., \& Kutz, J.~N. 2026, Proceedings of the National
  Academy of Sciences, 123, e2508144123

\bibitem[{Gardiner(2004)}]{gardiner2004handbook}
Gardiner, C.~W. 2004, Handbook of Stochastic Methods for Physics, Chemistry and
  the Natural Sciences, 3rd edn. (Berlin: Springer)

\bibitem[{Gers {et~al.}(2000)Gers, Schmidhuber, \& Cummins}]{gers2000learning}
Gers, F.~A., Schmidhuber, J., \& Cummins, F. 2000, Neural computation, 12, 2451

\bibitem[{Graves(2012)}]{graves2012long}
Graves, A. 2012, Supervised sequence labelling with recurrent neural networks,
  37

\bibitem[{Haseli \& Cort{\'e}s(2022)}]{haseli_symmetric_subspace_2022}
Haseli, M., \& Cort{\'e}s, J. 2022, IEEE Transactions on Automatic Control, 67,
  3442

\bibitem[{Haseli \& Cortés(2025)}]{haseli_recursive_2025}
Haseli, M., \& Cortés, J. 2025, IEEE Access, 13, 61006

\bibitem[{Hemati {et~al.}(2017)Hemati, Rowley, Deem, \&
  Cattafesta}]{hemati2017biasing}
Hemati, M.~S., Rowley, C.~W., Deem, E.~A., \& Cattafesta, L.~N. 2017,
  Theoretical and Computational Fluid Dynamics, 31, 349

\bibitem[{Ingram \& Motta(2019)}]{ingram2019review}
Ingram, A.~R., \& Motta, S.~E. 2019, New Astronomy Reviews, 85, 101524

\bibitem[{Jaeger(2001)}]{jaeger2001echo}
Jaeger, H. 2001, Bonn, Germany: German national research center for information
  technology gmd technical report, 148, 13

\bibitem[{Koopman(1931)}]{koopman1931hamiltonian}
Koopman, B.~O. 1931, Proceedings of the National Academy of Sciences, 17, 315

\bibitem[{Korda \& Mezić(2018)}]{korda_mezic_convergence_2018}
Korda, M., \& Mezić, I. 2018, Journal of Nonlinear Science, 28, 687

\bibitem[{Krakovna \& Doshi-Velez(2016)}]{krakovna2016increasing}
Krakovna, V., \& Doshi-Velez, F. 2016, arXiv preprint arXiv:1606.05320

\bibitem[{Krastev(2020)}]{krastev2020real}
Krastev, P.~G. 2020, Physics Letters B, 803, 135330

\bibitem[{Levine {et~al.}(1996)Levine, Bradt, Cui, Jernigan, Morgan, Remillard,
  Shirey, \& Smith}]{levine1996first}
Levine, A.~M., Bradt, H., Cui, W., {et~al.} 1996, The Astrophysical Journal,
  469, L33

\bibitem[{Lewin {et~al.}(1997)Lewin, van~den Heuvel, \& van
  Paradijs}]{lewin1997x}
Lewin, W. H.~G., van~den Heuvel, E. P.~J., \& van Paradijs, J. 1997, X-ray
  Binaries, Vol.~26 (Cambridge University Press)

\bibitem[{Li {et~al.}(2017)Li, Dietrich, Bollt, \& Kevrekidis}]{li2017extended}
Li, Q., Dietrich, F., Bollt, E.~M., \& Kevrekidis, I.~G. 2017, Chaos: An
  Interdisciplinary Journal of Nonlinear Science, 27, 103111

\bibitem[{Luko{\v{s}}evi{\v{c}}ius \&
  Jaeger(2009)}]{lukovsevivcius2009reservoir}
Luko{\v{s}}evi{\v{c}}ius, M., \& Jaeger, H. 2009, Computer science review, 3,
  127

\bibitem[{Lusch {et~al.}(2018)Lusch, Kutz, \& Brunton}]{lusch2018deep}
Lusch, B., Kutz, J.~N., \& Brunton, S.~L. 2018, Nature Communications, 9, 4950

\bibitem[{Ma{\'c}e{\v{s}}i{\'c} {et~al.}(2018)Ma{\'c}e{\v{s}}i{\'c},
  {\v{C}}rnjari{\'c}-{\v{Z}}ic, \& Mezi{\'c}}]{macesic2018koopman}
Ma{\'c}e{\v{s}}i{\'c}, S., {\v{C}}rnjari{\'c}-{\v{Z}}ic, N., \& Mezi{\'c}, I.
  2018, SIAM Journal on Applied Dynamical Systems, 17, 2478

\bibitem[{Mahabal {et~al.}(2017)Mahabal, Sheth, Gieseke, Pai, Djorgovski,
  Drake, \& Graham}]{mahabal2017deep}
Mahabal, A., Sheth, K., Gieseke, F., {et~al.} 2017, in 2017 IEEE symposium
  series on computational intelligence (SSCI), IEEE, 1--8

\bibitem[{Matsuoka {et~al.}(2009)Matsuoka, Kawasaki, Ueno, Tomida, Kohama,
  Suzuki, Adachi, Ishikawa, Mihara, Sugizaki, {et~al.}}]{matsuoka2009maxi}
Matsuoka, M., Kawasaki, K., Ueno, S., {et~al.} 2009, Publications of the
  Astronomical Society of Japan, 61, 999

\bibitem[{Mekha{\"e}l {et~al.}(2024)Mekha{\"e}l, Pasquato, Carenini, Braga,
  Trevisan, Bono, \& Hezaveh}]{mekhael2024koopman}
Mekha{\"e}l, N., Pasquato, M., Carenini, G., {et~al.} 2024, arXiv preprint
  arXiv:2407.16868

\bibitem[{Merrifield \& McHardy(1994)}]{merrifield1994estimating}
Merrifield, M.~R., \& McHardy, I.~M. 1994, Monthly Notices of the Royal
  Astronomical Society, 271, 899

\bibitem[{{Mezi{\'c}}(2005)}]{mezic_kmd_2005}
{Mezi{\'c}}, I. 2005, Nonlinear Dynamics, 41, 309

\bibitem[{Mezi{\'c} {et~al.}(2026)Mezi{\'c}, Cort{\'e}s, Worthmann, Lazar, \&
  Lederer}]{mezic2026koopman}
Mezi{\'c}, I., Cort{\'e}s, J., Worthmann, K., Lazar, M., \& Lederer, A. 2026,
  arXiv preprint arXiv:2607.01819

\bibitem[{Miller {et~al.}(2001)Miller, Wijnands, Homan, Belloni, Pooley,
  Corbel, Kouveliotou, van~der Klis, \& Lewin}]{miller2001high}
Miller, J., Wijnands, R., Homan, J., {et~al.} 2001, The Astrophysical Journal,
  563, 928

\bibitem[{Mirabel \& Rodriguez(1999)}]{mirabel1999sources}
Mirabel, I.~F., \& Rodriguez, L.~F. 1999, Annual Review of Astronomy and
  Astrophysics, 37, 409

\bibitem[{Monsalves {et~al.}(2024)Monsalves, Arancibia, Bayo,
  S{\'a}nchez-S{\'a}ez, Angeloni, Damke, \& Van~de
  Perre}]{monsalves2024application}
Monsalves, N., Arancibia, M.~J., Bayo, A., {et~al.} 2024, Astronomy \&
  Astrophysics, 691, A106

\bibitem[{Motta(2016)}]{motta2016quasi}
Motta, S.~E. 2016, Astronomische Nachrichten, 337, 398

\bibitem[{Nagdi {et~al.}(2026)Nagdi, Nikolados, Yermakov, Gao, Kutz, \&
  Menolascina}]{nagdi2026learning}
Nagdi, M., Nikolados, E.-M., Yermakov, A., {et~al.} 2026, arXiv preprint
  arXiv:2606.23957

\bibitem[{Neilsen {et~al.}(2012)Neilsen, Petschek, \&
  Lee}]{neilsen2012accretion}
Neilsen, J., Petschek, A.~J., \& Lee, J.~C. 2012, Monthly Notices of the Royal
  Astronomical Society, 421, 502

\bibitem[{Nonomura {et~al.}(2019)Nonomura, Shibata, \&
  Takaki}]{nonomura2019extended}
Nonomura, T., Shibata, H., \& Takaki, R. 2019, PloS one, 14, e0209836

\bibitem[{Pedregosa {et~al.}(2011)Pedregosa, Varoquaux, Gramfort, Michel,
  Thirion, Grisel, Blondel, Prettenhofer, Weiss, Dubourg, Vanderplas, Passos,
  Cournapeau, Brucher, Perrot, \& Duchesnay}]{scikit-learn}
Pedregosa, F., Varoquaux, G., Gramfort, A., {et~al.} 2011, Journal of Machine
  Learning Research, 12, 2825

\bibitem[{Peng {et~al.}(2024)Peng, Risti{\'c}, Kedia, O'Shaughnessy, Fontes,
  Fryer, Korobkin, Mumpower, Villar, \& Wollaeger}]{peng2024kilonova}
Peng, Y., Risti{\'c}, M., Kedia, A., {et~al.} 2024, Physical Review Research,
  6, 033078

\bibitem[{Phillipson {et~al.}(2018)Phillipson, Boyd, \&
  Smale}]{phillipson2018chaotic}
Phillipson, R.~A., Boyd, P.~T., \& Smale, A.~P. 2018, Monthly Notices of the
  Royal Astronomical Society, 477, 5220

\bibitem[{Poore {et~al.}(2024)Poore, Carini, Dingler, Wehrle, \&
  Wiita}]{poore2024comparative}
Poore, E., Carini, M., Dingler, R., Wehrle, A.~E., \& Wiita, P.~J. 2024, The
  Astrophysical Journal, 966, 158

\bibitem[{Pottschmidt {et~al.}(2003)Pottschmidt, Wilms, Nowak, Pooley,
  Gleissner, Heindl, Smith, Remillard, \& Staubert}]{pottschmidt2003long}
Pottschmidt, K., Wilms, J., Nowak, M., {et~al.} 2003, Astronomy \&
  Astrophysics, 407, 1039

\bibitem[{Rico {et~al.}(2025)Rico, Dom{\'\i}nguez, Pe{\~n}il, Ajello, Buson,
  Adhikari, \& Movahedifar}]{rico2025singular}
Rico, A., Dom{\'\i}nguez, A., Pe{\~n}il, P., {et~al.} 2025, Astronomy \&
  Astrophysics, 697, A35

\bibitem[{Rowley {et~al.}(2009)Rowley, Mezi{\'c}, Bagheri, Schlatter, \&
  Henningson}]{rowley2009spectral}
Rowley, C.~W., Mezi{\'c}, I., Bagheri, S., Schlatter, P., \& Henningson, D.~S.
  2009, Journal of fluid mechanics, 641, 115

\bibitem[{Rudin(2019)}]{rudin2019stop}
Rudin, C. 2019, Nature Machine Intelligence, 1, 206

\bibitem[{Sakata \& Kawahara(2024)}]{sakata2024enhancing}
Sakata, I., \& Kawahara, Y. 2024, Scientific Reports, 14, 19276

\bibitem[{Sandrinelli {et~al.}(2016)Sandrinelli, Covino, Dotti, \&
  Treves}]{sandrinelli2016quasi}
Sandrinelli, A., Covino, S., Dotti, M., \& Treves, A. 2016, The Astronomical
  Journal, 151, 54

\bibitem[{Scargle(2020)}]{scargle2020studies}
Scargle, J.~D. 2020, The Astrophysical Journal, 895, 90

\bibitem[{Shakura \& Sunyaev(1973)}]{shakura1973black}
Shakura, N.~I., \& Sunyaev, R.~A. 1973, Astronomy and Astrophysics, 24, 337

\bibitem[{Sobacchi {et~al.}(2016)Sobacchi, Sormani, \&
  Stamerra}]{sobacchi2016model}
Sobacchi, E., Sormani, M.~C., \& Stamerra, A. 2016, Monthly Notices of the
  Royal Astronomical Society, stw2684

\bibitem[{Sobolewska \& {\.Z}ycki(2003)}]{sobolewska2003spectral}
Sobolewska, M., \& {\.Z}ycki, P. 2003, Astronomy \& Astrophysics, 400, 553

\bibitem[{Stella \& Vietri(1998)}]{stella1998lense}
Stella, L., \& Vietri, M. 1998, The Astrophysical Journal Letters, 492, L59

\bibitem[{Stella \& Vietri(1999)}]{stella1999khz}
---. 1999, Physical Review Letters, 82, 17

\bibitem[{Stella {et~al.}(1999)Stella, Vietri, \&
  Morsink}]{stella1999correlations}
Stella, L., Vietri, M., \& Morsink, S.~M. 1999, The Astrophysical Journal
  Letters, 524, L63

\bibitem[{Sugihara \& May(1990)}]{sugihara1990nonlinear}
Sugihara, G., \& May, R.~M. 1990, Nature, 344, 734

\bibitem[{Susuki {et~al.}(2016)Susuki, Mezi{\'c}, Raak, \&
  Hikihara}]{susuki2016applied}
Susuki, Y., Mezi{\'c}, I., Raak, F., \& Hikihara, T. 2016, Nonlinear Theory and
  Its Applications, IEICE, 7, 430

\bibitem[{Takeishi {et~al.}(2017)Takeishi, Kawahara, \&
  Yairi}]{takeishi2017subspace}
Takeishi, N., Kawahara, Y., \& Yairi, T. 2017, Physical Review E, 96, 033310

\bibitem[{Takens(1981)}]{takens_embedding_1981}
Takens, F. 1981, in Dynamical {Systems} and {Turbulence}, {Warwick} 1980, ed.
  D.~Rand \& L.-S. Young, Vol. 898 (Berlin, Heidelberg: Springer Berlin
  Heidelberg), 366--381

\bibitem[{Thekkeppattu {et~al.}(2023)Thekkeppattu, Trott, \&
  McKinley}]{thekkeppattu2023singular}
Thekkeppattu, J.~N., Trott, C.~M., \& McKinley, B. 2023, Monthly Notices of the
  Royal Astronomical Society, 520, 6040

\bibitem[{Trevisan(2023)}]{trevisan2023case}
Trevisan, P. 2023, PhD thesis, Sapienza Universit\`a di Roma

\bibitem[{Trushkin {et~al.}(2017)Trushkin, McCollough, Nizhelskij, \&
  Tsybulev}]{trushkin2017giant}
Trushkin, S., McCollough, M., Nizhelskij, N., \& Tsybulev, P. 2017, Galaxies,
  5, 86

\bibitem[{Uttley {et~al.}(2005)Uttley, McHardy, \& Vaughan}]{uttley2005non}
Uttley, P., McHardy, I., \& Vaughan, S. 2005, Monthly Notices of the Royal
  Astronomical Society, 359, 345

\bibitem[{Uttley \& McHardy(2001)}]{uttley2001flux}
Uttley, P., \& McHardy, I.~M. 2001, Monthly Notices of the Royal Astronomical
  Society, 323, L26

\bibitem[{Uttley {et~al.}(2017)Uttley, McHardy, \& Vaughan}]{uttley2017rms}
Uttley, P., McHardy, I.~M., \& Vaughan, S. 2017, Astronomy \& Astrophysics,
  601, L1

\bibitem[{Van~der Klis(1989)}]{van1989fourier}
Van~der Klis, M. 1989, in Timing neutron stars (Springer), 27--69

\bibitem[{Wanner \& Mezi{\'c}(2022)}]{wanner2022robust}
Wanner, M., \& Mezi{\'c}, I. 2022, SIAM Journal on Applied Dynamical Systems,
  21, 1930

\bibitem[{Wei {et~al.}(2021)Wei, Huerta, Yun, Loutrel, Shaikh, Kumar, Haas, \&
  Kindratenko}]{wei2021deep}
Wei, W., Huerta, E., Yun, M., {et~al.} 2021, The Astrophysical Journal, 919, 82

\bibitem[{Williams {et~al.}(2016)Williams, Hemati, Dawson, Kevrekidis, \&
  Rowley}]{williams2016extending}
Williams, M.~O., Hemati, M.~S., Dawson, S. T.~M., Kevrekidis, I.~G., \& Rowley,
  C.~W. 2016, IFAC-PapersOnLine, 49, 704

\bibitem[{Williams {et~al.}(2015)Williams, Kevrekidis, \&
  Rowley}]{williams2015data}
Williams, M.~O., Kevrekidis, I.~G., \& Rowley, C.~W. 2015, Journal of Nonlinear
  Science, 25, 1307

\bibitem[{Williams {et~al.}(2014)Williams, Rowley, \&
  Kevrekidis}]{williams2014kernel}
Williams, M.~O., Rowley, C.~W., \& Kevrekidis, I.~G. 2014, arXiv preprint
  arXiv:1411.2260

\bibitem[{Zdziarski \& Gierli{\'n}ski(2004)}]{zdziarski2004radiative}
Zdziarski, A.~A., \& Gierli{\'n}ski, M. 2004, Progress of Theoretical Physics
  Supplement, 155, 99

\bibitem[{Zhang {et~al.}(2019)Zhang, Rowley, Deem, \&
  Cattafesta}]{zhang2019online}
Zhang, H., Rowley, C.~W., Deem, E.~A., \& Cattafesta, L.~N. 2019, SIAM Journal
  on Applied Dynamical Systems, 18, 1586

\end{thebibliography}

\end{document}